\documentclass[journal,comsoc]{IEEEtran}

\usepackage{cite}
\usepackage{textcomp}
\usepackage[applemac]{inputenc}
\usepackage{fontenc}
\usepackage{caption}
\usepackage{subcaption}
\usepackage[ruled,vlined]{algorithm2e}
\usepackage{placeins} 
\ifCLASSINFOpdf
  \usepackage[pdftex]{graphicx}
\else
   \usepackage[dvips]{graphicx}
\fi

\usepackage{graphicx}
\usepackage{caption}
\usepackage{subcaption}

\usepackage[cmex10]{amsmath}

\usepackage{url}
\usepackage{pdfsync}
\usepackage{dsfont}
\usepackage{float}
\usepackage{amssymb}
\usepackage{acronym,xcolor}
\usepackage[squaren,Gray]{SIunits}
\usepackage{amsmath}
\usepackage{comment}

\usepackage[export]{adjustbox}

\begin{document}

\title{\huge Design of Reconfigurable Intelligent Surfaces\\ by Using $S$-Parameter Multiport Network Theory \\-- Optimization and Full-Wave Validation}

\author{Andrea Abrardo,~\IEEEmembership{Senior Member,~IEEE,}
Alberto Toccafondi,~\IEEEmembership{Senior Member,~IEEE,} and 
Marco~Di~Renzo,~\IEEEmembership{Fellow,~IEEE} \vspace{-0.75cm}

\thanks{Manuscript received Nov. 11, 2023; revised Apr. 20, 2024, and July 29, 2024. Parts of this paper were presented at IEEE EuCAP 2024 \cite{abrardo2023analysis}.} 
\thanks{A. Abrardo and A. Toccafondi are with the University of Siena, Siena, Italy. (abrardo@unisi.it, albertot@unisi.it). M. Di Renzo is with Universit\'e Paris-Saclay, CNRS, CentraleSup\'elec, Laboratoire des Signaux et Syst\'emes, 91192 Gif-sur-Yvette, France. (marco.di-renzo@universite-paris-saclay.fr).}
}

\maketitle

\begin{abstract}
Multiport network theory has been proved to be a suitable abstraction model for analyzing and optimizing reconfigurable intelligent surfaces (RISs) \textcolor{black}{in an electromagnetically consistent manner, especially for studying the impact of the electromagnetic mutual coupling among radiating elements that are spaced less than half of the wavelength apart and for considering the interrelation between the amplitude and phase of the reflection coefficients}. Both representations in terms of $Z$-parameter (impedance) and $S$-parameter (scattering) matrices are widely utilized. In this paper, we embrace multiport network theory for analyzing and optimizing the reradiation properties of RIS-aided channels, and provide four new contributions. (i) First, we offer a thorough comparison between the $Z$-parameter and $S$-parameter representations. This comparison allows us to unveil that typical scattering models utilized for RIS-aided channels ignore the structural scattering from an RIS,  \textcolor{black}{which is well documented in antenna theory. We show that the structural scattering results in an unwanted specular reflection}. (ii) Then, we develop an iterative algorithm for optimizing, in the presence of electromagnetic mutual coupling, the tunable loads of an RIS based on the $S$-parameters representation. We prove that small perturbations of the step size of the algorithm result in larger variations of the $S$-parameter matrix compared with the $Z$-parameter matrix, resulting in a faster convergence rate. (iii) Subsequently, we generalize the proposed algorithm to suppress the specular reflection due to the structural scattering, while maximizing the received power towards the direction of interest, and analyze the effectiveness and tradeoffs of the proposed approach. (iv) Finally, we validate the theoretical findings and algorithms with numerical simulations and a commercial full-wave electromagnetic  simulator based on the method of moments. 
\end{abstract} \vspace{-0.5cm}

 \vspace{-0.5cm}

\section{Introduction}
Reconfigurable intelligent surface (RIS) is an emerging technology for future wireless networks \cite{RenzoZDAYRT20}, \cite{WuZZYZ21}. The distinctive feature of RISs compared with other technologies lies in their nearly passive implementation, since no power amplifiers, RF chains, and digital signal processing units are needed for their operation \cite{RenzoNSDQLRPSZD20}. Recent system-level simulation studies have confirmed the performance benefits of nearly passive RISs when deployed in large-scale cellular networks, provided that the physical size of the surface is sufficiently large \cite{SihlbomPR23}. 

In wireless communications, either for system optimization \cite{WuZZYZ21} or signal processing design \cite{PanZZHWPRRSZZ22}, the RIS elements are often modeled as ideal scatterers, which are capable of introducing any phase shift to the impinging electromagnetic waves, without attenuation or unwanted reradiations. These models are not always electromagnetically consistent, since they do not consider several aspects that play an important role in characterizing the operation of a realistic RIS,  \textcolor{black}{including the electromagnetic mutual coupling between the scattering elements and the interrelation between the phase and amplitude of the reflection coefficient} \cite{RenzoDT22}, \cite{direnzo2022digital}, \cite{AbeywickramaZWY20}. Recent results highlight the critical need of using realistic reradiation models \cite{Galdi2023}, which result in a strong interplay between the optimization of an RIS from the surface-level standpoint and the design of the scattering elements from the element-level standpoint \cite{RafiqueHZNMRDY23}.

Motivated by these considerations, major efforts are currently devoted to develop electromagnetically consistent models for RISs, which are sufficiently tractable and suitable for performance evaluation, and to the design of efficient optimization and signal processing algorithms \cite{RenzoDT22}. In this context, multiport network theory constitutes a suitable approach for modeling and optimizing RIS-aided channels \cite{IvrlacN10}. The motivation for using multiport network theory lies in its inherent capability of modeling the electromagnetic mutual coupling between closely-spaced radiating elements (unit cells), which is a distinguishing feature of an RIS, \textcolor{black}{since a metasurface is usually characterized by sub-wavelength unit cells spaced at sub-wavelength inter-distances \cite{RenzoZDAYRT20}. Also, the coupling between the unit cells is often exploited to realize advanced wave transformations at a high power efficiency \cite{10213362}, \cite{li2023tunable}}.

The first multiport model for RIS-aided channels has been introduced in \cite{DR1}, under the assumption of minimum scattering radiating elements, and \textcolor{black}{by considering, for analytical tractability and in agreement with the discrete dipole approximation \cite{math10173049}, thin wire dipoles as radiating elements}. \textcolor{black}{A similar model based on the coupled dipoles formalism is presented in \cite{FaqiriSASIH23}, and has recently been further analyzed and experimentally validated in \cite{R1}, \cite{R3}, \cite{R8}. The analytical analogies between the multiport network theory model and the coupled dipoles formalism are apparent by comparing \cite{DR1} with \cite{FaqiriSASIH23} and the series representations in \cite{DR2} and  \cite{R1}.}

\textcolor{black}{As far the multiport network model is concerned, specifically,} the authors of \cite{DR1} model an RIS as a collection of scattering elements that are loaded with tunable impedances, which determine the reradiation properties of the RIS. An end-to-end communication model for the analysis and optimization of RIS-aided channels is formulated and is expressed in terms of the $Z$-parameter (impedance) matrix. A closed-form expression for the entries of the impedance matrix is available in \cite{10133771}. The main feature that distinguishes the communication model in \cite{DR1} from conventional scattering models utilized for RIS-aided channels is the non-linearity, as a function of the tunable impedances, of the end-to-end system response (the channel) due to the presence of electromagnetic mutual coupling between closely spaced RIS elements. \textcolor{black}{A similar non-linear response is found in the coupled dipoles model \cite{FaqiriSASIH23}}.

The model in \cite{DR1} has been subsequently utilized in several research works, including the following. In \cite{DR2}, the authors develop an iterative algorithm for optimizing the tunable impedances of an RIS-aided single-input single-output (SISO) channel, in order to maximize the end-to-end received power. The approach leverages the Neumann series approximation for the inversion of non-diagonal matrices. The Neumann series approach is utilized and generalized in several subsequent research works. In \cite{ABR}, the authors apply it for optimizing the sum-rate in multiple-input multiple-output (MIMO) interference networks. In \cite{SARIS}, the authors generalize the RIS-aided multiport channel model in \cite{DR1} to account for the presence of natural scatterers, i.e., multipath propagation, and propose, using the Neumann series approximation, an optimization algorithm for application to multi-user multiple-input single-output networks. The authors of \cite{SARIS} unveil that the multipath generated by natural scatterers cannot be modeled as an additive term, due to the non-linearity, as a function of the tunable impedances, of the reradiatiated electromagnetic field from the RIS. \textcolor{black}{This finding is obtained by using the coupled dipoles model as well \cite{FaqiriSASIH23}, which further corroborates the analysis and the similarity between the multiport network theory and coupled dipoles models.} More recently, the authors of \cite{DR3} have proposed an iterative algorithm for optimizing the tunable loads in RIS-aided MIMO channels, which overcomes the Neumann series approximation and leverages the Gram-Schmidt orthogonalization method for optimizing one tunable impedance at a time, which results in a simple and computationally efficient method. In addition, the authors of \cite{abs-2305-12735} introduce an optimization algorithm, for application to RIS-aided SISO and MIMO channels, that accounts for the power mismatch at the transmitter and receiver; the authors of \cite{akrout2023physically} apply the multiport network model in \cite{DR1} by modeling the RIS scattering elements as canonical minimum scattering Chu's antennas; the authors of \cite{abs-2306-03761} validate the model in \cite{DR1} by using a simulator based on the method of moments (MoM) and considering a general unit cell structure; the authors of \cite{zheng2023impact} utilize the model in \cite{DR1} for analyzing the impact of the electromagnetic mutual coupling for channel estimation; and, finally, the authors of \cite{10032536} utilize the model in \cite{DR1} for analyzing the impact of the electromagnetic mutual coupling in terms of energy efficiency optimization. \textcolor{black}{More recently, the multiport network theory model has been experimentally validated in \cite{R4}, \cite{KAUST} and in \cite{CEA} to model multipath channels}.

The aforementioned research works utilize a multiport network model for RIS-aided channels that is based on the $Z$-parameter representation. An alternative formulation can be expressed in terms of $S$-parameter representation (scattering matrix) \cite{PozarBook}. This is considered in \cite{ShenCM22} for application to RISs whose scattering properties can be controlled through individual (independent) tunable loads or through a network of interconnected loads, with the objective of enhancing the control of the reradiated electromagnetic field in the presence of electromagnetic mutual coupling \cite{10042168}. RISs whose elements are controlled by a network of connected tunable loads are characterized by an impedance matrix and a scattering matrix that are not diagonal. Therefore, they are referred to as beyond-diagonal RISs \cite{abs-2301-03288}. The information-theoretic optimality of non-diagonal scattering matrices for nearly passive RISs is proved in \cite{sil2.12195}. Recent contributions on beyond diagonal RISs include \cite{LiEHSMCH22}, \cite{LiSC23}, \cite{10155675}, \cite{li2023diagonal}. These research works are focused on evaluating the tradeoff between the performance and the implementation complexity of beyond-diagonal and diagonal RISs. There exist no research works on the optimization of the scattering matrix $S$ in the presence of electromagnetic mutual coupling, either for diagonal or for beyond-diagonal RISs. In this context, the closest research contribution to the present paper is \cite{li2023diagonal}. Therein, however, the optimization algorithm is developed based on the $Z$-parameter representation, and, more importantly, the impact of the structural scattering (introduced and discussed next) \textcolor{black}{\cite{R7}} is neither discussed nor considered in the proposed optimization algorithm. Recently, the authors of \cite{nossek2023physically} have conducted a comparative study of the multiport network model for RISs formulated in terms of $Z$-parameter and $S$-parameter representations. The analysis ignores the impact of the electromagnetic mutual coupling. The authors prove the equivalence between the two representations, but they unveil major differences in terms of the physical meaning of the individual matrices that constitute the end-to-end channel, especially in terms of the direct transmitter-receiver links. We have conducted a similar study in the companion conference paper \cite{abrardo2023analysis}, by considering the impact of mutual coupling. 

The present paper provides a major extension of the three-page companion conference paper \cite{abrardo2023analysis}, with focus on understanding, optimizing, and validating with a full-wave electromagnetic simulator the $S$-parameter multiport network model for RISs in the presence of electromagnetic mutual coupling. Specifically, the analysis of the aforementioned state-of-the-art shows that (i) the physical understanding of the multiport network model for RISs is often misunderstood, especially when the scattering parameters $S$ are utilized; (ii) as far as the $S$-parameter representation is concerned, there exists no algorithm that optimizes the RIS elements in the presence of electromagnetic mutual coupling. Also, it is unclear whether any benefits with respect to  currently available optimization algorithms based on the $Z$-parameter representation exist; and (iii) \textcolor{black}{as far as modeling the structural scattering is concerned, recent works have accounted for it and have performed optimizations considering on it \cite{R3}, \cite{R4}. However, the structural scattering was not explicitly singled out from other background scattering components. In the context of multiport network modeling, there exist no prior works in which the structural scattering is mathematically and explicitly formulated, optimization algorithms are developed to mitigate it, and full-wave simulations are used to validate the accuracy of the models and the effectiveness of the optimization algorithms}.

Against this background, the present paper provides the following four main and novel contributions:

$\bullet$ We consider the $Z$-parameter and $S$-parameter representations for RIS-aided channels, and offer a thorough comparison between them. \textcolor{black}{Even though these two representations are known to be equivalent from the mathematical point of view \cite{PozarBook}, we highlight that the physical interpretation of their constituent terms is different}. Specifically, we prove that the impedance matrix between the transmitter and the receiver is always zero if this link is physically blocked by the presence of obstacles. We prove, however, that the corresponding matrix expressed in terms of scattering parameters is not equal to zero, as it accounts for the structural scattering from the RIS \cite{AntennaTheory}. This comparison allows us to show that the $Z$-parameter representation for RISs inherently accounts for the structural scattering from the RIS when the direct link is ignored, as customary done in the literature. The $S$-parameter representation, on the other hand, accounts for the structural scattering from the RIS provided that the direct link is correctly modeled. We show that the structural scattering from the RIS results in a specular reflection, which is usually ignored when considering the ideal scattering models utilized in communication theory. \textcolor{black}{Even though the structural scattering is well documented in antenna theory \cite{R7}, it has not been well recognized and appropriately modeled in wireless communications, especially in the context of multiport network models for RISs. This is remarked and recognized in recently published papers too \cite{R8}}.

$\bullet$ We develop the first iterative algorithm for optimizing the tunable loads of the RIS, based on the $S$-parameter representation, in the presence of electromagnetic mutual coupling. To this end, we utilize the Neumann series approximation, similar to \cite{DR2}. \textcolor{black}{We prove that optimizing the scattering matrix instead of the impedance matrix results in a faster convergence rate. The reason is that small perturbations of the step size of the algorithm result in larger variations of the $S$-parameter matrix compared with the variations of the $Z$-parameter matrix. Since the step size needs to be small for ensuring convergence, and the considered optimization problem is not convex, the use of a different multiport representation is shown to significantly influence the convergence properties of the optimization algorithm and the final optimized variables that it returns}. 

$\bullet$ To tackle the specular component that originates from the structural scattering from the RIS, we propose an optimization algorithm that considers the twofold objective of maximizing the scattered signal towards the desired direction of reradiation and suppressing the scattered signal towards the specular direction. To this end, we formulate a weighted optimization problem, whose weight allows us to control the intensity of the desired and unwanted reradiated components of the signal. The performance tradeoffs of the proposed approach are analyzed with the aid of numerical simulations, and the suppression of the undesired specular reflection is validated.

$\bullet$ With the aid of a commercial full-wave simulator that implements the MoM, we validate the correctness of the theoretical findings and the effectiveness of the proposed optimization algorithms for maximizing the received power towards the direction of interest while suppressing the undesired specular reflection. To this end, we use a three-step approach: (i) first, we compute the $S$-parameters matrix with the full-wave simulator; (ii) then, we input the obtained matrix into the proposed optimization algorithms and compute the optimal tunable loads; and (iii) finally, we simulate the entire RIS structure connected to the obtained loads, and obtain the reradiation pattern. \textcolor{black}{By using this approach, we prove the existence of the specular reflection and we show that the multiport network model provides estimates of the reradiation pattern from the RIS that are consistent with those obtained with the full-wave simulator. Specifically, the full-wave simulations are conducted for two practical implementations for the RIS elements: (i) free-standing loaded thin wire dipoles and (ii) loaded rectangular patch antennas on a grounded substrate}.

The paper is organized as follows. In Sec. II, we introduce the system model, and compare the $Z$-parameter and $S$-parameter representations. In Sec. III, we introduce the optimization algorithms based on the $S$-parameter representation. In Sec. IV, we illustrate numerical results and compare them against full-wave simulations. Sec. V concludes the paper.

\emph{Notation}.
Scalars are denoted by italic letters, and vectors and matrices are denoted by bold-face lower-case and bold-face upper-case letters. $\odot$ is the Hadamard product. $\Re\{\cdot\}$ and $\Im\{\cdot\}$ are the real and imaginary parts of complex numbers. $\|\mathbf{X}\|$ is the Frobenius norm of $\mathbf{X}$, i.e., $\|\mathbf{X}\|^2 = \text{trace}\left(\mathbf{X}\mathbf{X}^H\right)$ with $\mathbf{X}^H$ being the Hermitian of $\mathbf{X}$. The transpose of $\mathbf{X}$ is denoted by $\mathbf{X}^T$. ${\rm{diag}}\left( \mathbf{X} \right)$ is a column vector given by the main diagonal of $\mathbf{X}$. ${\rm{diag}}\left({\rm{diag}}\left( \mathbf{X} \right)\right)$ is a diagonal matrix whose diagonal is the main diagonal of $\mathbf{X}$. The distribution of a complex Gaussian random vector with mean ${x}$ and variance $\sigma^2$ is denoted by $\mathcal{CN}({x},\sigma^2)$, and $\sim$ stands for ``distributed as''. $\mathbf{U}$ is the identity matrix. $j$ is the imaginary unit, $(\cdot)^*$ is the complex conjugate. $\left|  \cdot  \right|$ and $\angle{\cdot}$ are the modulus and phase of complex numbers, $\odot$ is the Hadamard (element-wise) product.

\begin{figure}[!t]
	\centering
\includegraphics[width=0.6\columnwidth]{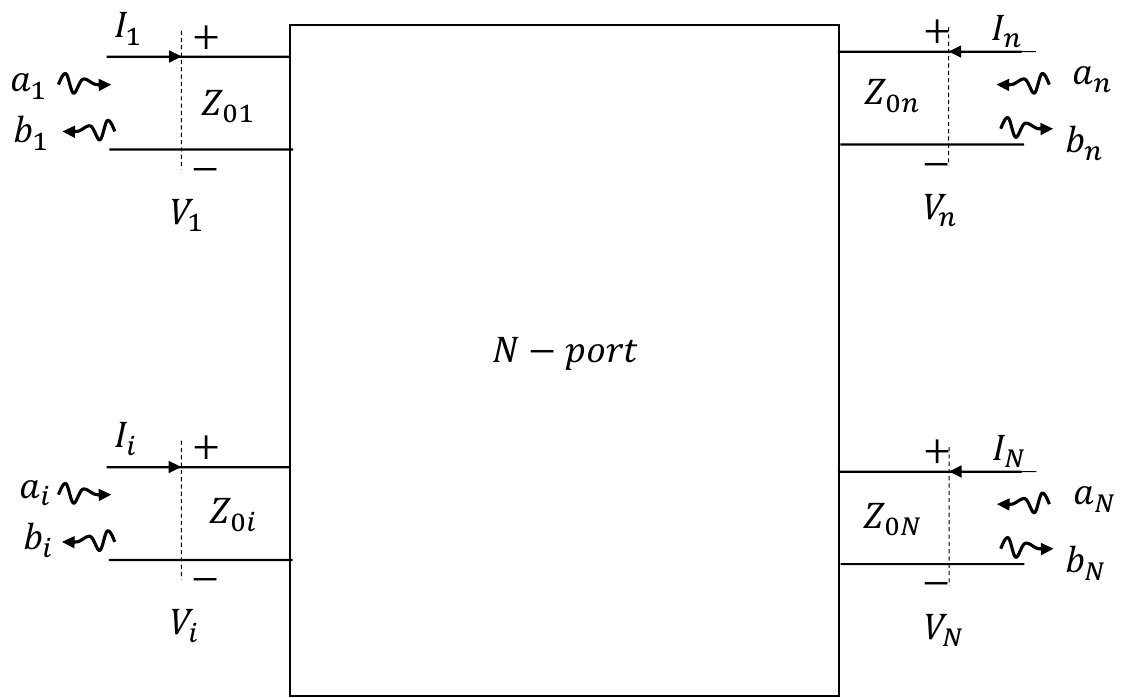}
	\caption{Considered $N$-port network representation.}
	\label{Smodel} \vspace{-0.75cm}
\end{figure}
\vspace{-0.25cm}
\section{System Model}\label{System model}
In this section, we introduce the multiport network theory model for RIS-aided channels, and compare the $Z$-parameter and $S$-parameter representations. Also, we elaborate on the electromagnetic differences between the proposed models and those typically utilized in communication theory. \textcolor{black}{For ease of description, we consider a free space scenario in the absence of multipath propagation. The generalization to multipath channels is elaborated in Sec. III-D based on \cite{SARIS}.}

\vspace{-0.25cm}
\subsection{Multiport Network Theory Model}
We consider a general RIS-aided communication channel that consists of a transmitter with $N_T$ antennas, a receiver with $N_R$ antennas, and an RIS with $K$ scattering elements. Each antenna at the transmitter and receiver, and each RIS element are modeled as a port of an $N$-port network with $N=N_T+K+N_R$, as illustrated in Fig. \ref{Smodel}. The whole $N$-port network model is characterized by an $N \times N$ scattering matrix $\mathbf{S}$, which relates the vector $\mathbf{a}$ of (forward) traveling waves that are incident on the ports to the vector $\mathbf{b}$ of scattered (reverse) traveling waves that exit from the ports. In mathematical terms, the $N$-port network model can be formulated as \cite[Eq. (4.40)]{PozarBook}
\vspace{-0.1cm}
\begin{equation}
    \mathbf{b}=\mathbf{S}\mathbf{a}
    \label{eq:bSa} \vspace{-0.1cm}
\end{equation}

Specifically, the vectors $\mathbf{a}$ and $\mathbf{b}$ of the incident and scattered electromagnetic waves, respectively, are related to the total voltages ${\bf{V}}$ and currents ${\bf{I}}$ at the $N$ ports, as follows \cite{AntennaTheory}:
\vspace{-0.1cm}
\begin{equation} \label{Eq:ab_VI}
    {\bf{V}} = {\sqrt {{Z_0}} } \left( {\bf{a}} + {\bf{b}}\right),\quad \quad {\bf{I}} = \frac{{{\bf{a}} - {\bf{b}}}}{{{{\sqrt {{Z_0}} }}}} \vspace{-0.1cm}
\end{equation} 
where $Z_0$ is a reference impedance that is assumed, for simplicity, to be the same for all the $N$ ports. Usually, $Z_0=\ 50\,\Omega$. 

The $\mathbf{S}$ matrix is typically decomposed into block matrices, in order to identify the ports of the transmitter, RIS, and receiver. Specifically, we obtain the following: \vspace{-0.1cm}
\begin{equation}
    \left[\begin{array}{c}\mathbf{b}_T \\\mathbf{b}_S \\\mathbf{b}_R\end{array}\right]=
    \left[\begin{array}{ccc}\mathbf{S}_{TT} & \mathbf{S}_{TS}& \mathbf{S}_{TR}\\
    \mathbf{S}_{ST} & \mathbf{S}_{SS}& \mathbf{S}_{SR}\\
    \mathbf{S}_{RT}& \mathbf{S}_{RS}& \mathbf{S}_{RR}
    \end{array}\right] 
    \left[\begin{array}{c}\mathbf{a}_T \\\mathbf{a}_S \\\mathbf{a}_R\end{array}\right] 
    \label{eq:absub} \vspace{-0.1cm}
\end{equation}
where $\mathbf{a}_x$ and $\mathbf{b}_x$ for $x \in \{T,S,R\}$ denote the incident and scattered electromagnetic waves at the ports of the transmitter ($T$), RIS ($S$) and receiver ($R$). Accordingly,  $\mathbf{S}_{xy}$ for $x,y \in \{T,S,R\}$ is the scattering sub-matrix that relates the vector $\mathbf{b}_x$ of the scattered waves at the ports of $x$ with the vector $\mathbf{a}_y$ of the incident waves at the ports of $y$.

In a general network configuration, the $N_T$ ports of the transmitter are connected to a vector of voltage generators $\mathbf{V}_g=\left[V_{g,1},\ldots,V_{g,N_T} \right]^T$ with internal impedances given by the diagonal matrix $\mathbf{Z}_g$ whose $i$th diagonal element is $Z_{g,i}$. The source wave vector associated with the voltage generators is denoted by $\mathbf{a}_g=\left[a_{g,1},\ldots,a_{g,N_T} \right]^T$. The relation between $\mathbf{a}_g$ and $\mathbf{V}_g$ is given next. The $N_R$ ports of the receiver are connected to load impedances given by the diagonal matrix $\mathbf{Z}_R$ whose $i$th diagonal element is $Z_{R,i}$. Also, the $K$ ports of the RIS are connected to tunable load impendances given by the matrix $\mathbf{Z}_S$. The matrix of impedances connected to the ports of the RIS, $\mathbf{Z}_S$, is an optimization variable, while the matrices $\mathbf{Z}_g$ and $\mathbf{Z}_R$ at the transmitter and receiver are given. Given the voltages and the loads connected to the ports, the sub-vectors in \eqref{eq:absub} are related through the following expressions: \vspace{-0.1cm}
\begin{equation} \label{eq:atasar}
\mathbf{a}_T= \mathbf{a}_g+\mathbf{\Gamma}_T\mathbf{b}_T, \quad 
\mathbf{a}_S= \mathbf{\Gamma}_S\mathbf{b}_S, \quad \mathbf{a}_R= \mathbf{\Gamma}_R\mathbf{b}_R \vspace{-0.1cm}
\end{equation}
where $\mathbf{\Gamma}_x$ for $x \in \{T,R\}$ is the diagonal matrix of reflection coefficients at the ports of $x$, whose entries are $\Gamma_{x,i}=(Z_{x,i}-Z_{0})/(Z_{x,i}+Z_{0})$. If $x = T$, $Z_{T,i} = Z_{g,i}$ is the internal impedance of the $i$th voltage generator. In conventional (diagonal) RISs \cite{DR1}, $\mathbf{Z}_S$ is a diagonal matrix, and it holds $\Gamma_{S,i}=(Z_{S,i}-Z_{0})/(Z_{S,i}+Z_{0})$, with $Z_{S,i}$ being the impedance connected to the $i$th port of the RIS. In non-diagonal, in general fully-connected, RISs \cite{ShenCM22}, $\mathbf{Z}_S$ is a full matrix and the reflection coefficient is given by ${{\bf{\Gamma }}_S} = {\left( {{{\bf{Z}}_S} + {Z_0}{\bf{U}}} \right)^{ - 1}}\left( {{{\bf{Z}}_S} - {Z_0}{\bf{U}}} \right)$.

Given the multiport network model in \eqref{eq:bSa}, the end-to-end RIS-aided channel is defined as the matrix $\mathbf{{H}}_{\rm{e2e}}^{(S)}$ that expresses the vector $\mathbf{b}_R$ of the waves at the ports of the receiver as a function of the vector $\mathbf{a}_g$ of the voltage generators at the ports of the transmitter, i.e., $\mathbf{b}_R=\mathbf{{H}}_{\rm{e2e}}^{(S)}\mathbf{a}_g$. The analytical expression of $\mathbf{{H}}_{\rm{e2e}}^{(S)}$ is derived in Appendix A. The final result, without any restrictions on the parameters, is the following: \vspace{-0.1cm}
\begin{equation} \label{Eq:HE2Eexact}
{{\mathbf{{H}}}_{{\rm{e2e}}}^{(S)}} = {\left( {{\bf{U}} - {{{\bf{\tilde S}}}_{RR}}{{\bf{\Gamma }}_R}} \right)^{ - 1}}{{{\bf{\tilde S}}}_{RT}}{\left( {{\bf{U}} - {{\bf{\Gamma }}_T}{{{\bf{\bar S}}}_{TT}}} \right)^{ - 1}} \vspace{-0.1cm}
\end{equation} 
where the following definitions hold: \vspace{-0.1cm}
\begin{equation}
{{{\bf{\tilde S}}}_{xy}} = {{\bf{S}}_{xy}} + {{\bf{S}}_{xS}}{\left( {{\bf{U}} - {{\bf{\Gamma }}_S}{{\bf{S}}_{SS}}} \right)^{ - 1}}{{\bf{\Gamma }}_S}{{\bf{S}}_{Sy}}\vspace{-0.1cm}
\end{equation} 
\vspace{-0.5cm}
\begin{equation} \label{Eq:BarSTT}
{{{\bf{\bar S}}}_{TT}} = {{{\bf{\tilde S}}}_{TT}} + {{{\bf{\tilde S}}}_{TR}}{{\bf{\Gamma }}_R}{\left( {{\bf{U}} - {{{\bf{\tilde S}}}_{RR}}{{\bf{\Gamma }}_R}} \right)^{ - 1}}{{{\bf{\tilde S}}}_{RT}} \vspace{-0.1cm}
\end{equation}
with $x \in \{T,R\}$  and $y \in \{T,R\}$.
 
In the rest of this paper, we assume $\mathbf{\Gamma}_T=\mathbf{0}$ and $\mathbf{\Gamma}_R=\mathbf{0}$. This ensures that the ports of the transmitter and receiver are matched for zero reflection \cite[p. 561]{PozarBook}. This configuration is obtained by setting $Z_{g,i} = Z_{R,i} = Z_0$, i.e., the transmitter and receiver are ``match terminated''. Under this assumption, which is the usual configuration in practice, $\mathbf{{H}}_{\rm{e2e}}^{(S)}$ in \eqref{Eq:HE2Eexact} simplifies to \vspace{-0.1cm}
\begin{equation} \label{Eq:HE2Ematched}
{{\mathbf{{H}}}_{{\rm{e2e}}}^{(S)}} = {{{\bf{\tilde S}}}_{RT}} = {{\bf{S}}_{RT}} + {{\bf{S}}_{RS}}{\left( {{\bf{U}} - {{\bf{\Gamma }}_S}{{\bf{S}}_{SS}}} \right)^{ - 1}}{{\bf{\Gamma }}_S}{{\bf{S}}_{ST}} \vspace{-0.1cm}
\end{equation}

The expression in \eqref{Eq:HE2Ematched} is relatively simple and much simpler than the expression in \eqref{Eq:HE2Eexact}. Both expressions can be applied to multiple-antenna transmitters and receivers, and to single-connected (diagonal) or fully-connected (non-diagonal) RISs.

\vspace{-0.25cm}
\subsection{Comparison with the $Z$-Parameter Representation}
In \cite{DR1}, the authors first introduced a multiport network model for RISs based on the $Z$-parameter representation, which relates the voltages ${\bf{V}}$ and the current ${\bf{I}}$ at the ports of the multiport network illustrated in Fig. \ref{Smodel}. Similar to the scattering matrix $\mathbf{S}$ in \eqref{eq:absub}, the impedance matrix $\mathbf{Z}$ is usually decomposed into block matrices, in order to identify the ports of the transmitter, RIS, and receiver. Specifically, we have  \vspace{-0.1cm}
\begin{equation}
    \left[\begin{array}{c}\mathbf{V}_T \\\mathbf{V}_S \\\mathbf{V}_R\end{array}\right]=
    \left[\begin{array}{ccc}\mathbf{Z}_{TT} & \mathbf{Z}_{TS}& \mathbf{Z}_{TR}\\
    \mathbf{Z}_{ST} & \mathbf{Z}_{SS}& \mathbf{Z}_{SR}\\
    \mathbf{Z}_{RT}& \mathbf{Z}_{RS}& \mathbf{Z}_{RR}
    \end{array}\right] 
    \left[\begin{array}{c}\mathbf{I}_T \\\mathbf{I}_S \\\mathbf{I}_R\end{array}\right] 
    \label{eq:Zmat} \vspace{-0.1cm}
\end{equation}
where $\mathbf{I}_x$ and $\mathbf{V}_x$ for $x \in \{T,S,R\}$ denote the currents and voltages at the ports of the transmitter ($T$), RIS ($S$), and receiver ($R$). Accordingly,  $\mathbf{Z}_{xy}$ for $x,y \in \{T,S,R\}$ is the impedance sub-matrix that relates the vector $\mathbf{V}_x$ of the voltages at the ports of $x$ with the vector $\mathbf{I}_y$ of the currents at the ports of $y$. Specifically, the voltages and the currents at the ports are related through the following Ohm's laws: \vspace{-0.1cm}
\begin{equation} \label{Eq:VoltageCurrent}
{{\bf{V}}_T} = {{\bf{V}}_g} - {{\bf{Z}}_g}{{\bf{I}}_T},\quad {{\bf{V}}_S} =  - {{\bf{Z}}_S}{{\bf{I}}_S},\quad {{\bf{V}}_R} =  - {{\bf{Z}}_R}{{\bf{I}}_R} \vspace{-0.1cm}
\end{equation}

Thanks to the relationships in \eqref{Eq:ab_VI}, it is known that the $\mathbf{S}$ matrix in \eqref{eq:absub} and the $\mathbf{Z}$ matrix in \eqref{eq:Zmat} are related to each other through the identities \cite[Eq. (4.44) and Eq. (4.45)]{PozarBook} \vspace{-0.1cm}
\begin{align}
& {\bf{S}} = {\left( {{\bf{Z}} + {Z_0}{\bf{U}}} \right)^{ - 1}}\left( {{\bf{Z}} - {Z_0}{\bf{U}}} \right) \label{Eq:ZtoS} \\
& {\bf{Z}} = \left( {{Z_0}{\bf{U}} + {\bf{S}}} \right){\left( {{Z_0}{\bf{U}} - {\bf{S}}} \right)^{ - 1}} \label{Eq:StoZ} \vspace{-0.25cm}
\end{align}

Based on the $Z$-parameter representation in \eqref{eq:Zmat}, the end-to-end RIS-aided channel is defined as the matrix $\mathbf{{H}}_{\rm{e2e}}^{(Z)}$ that expresses the vector $\mathbf{V}_R$ of the voltages at the ports of the receiver as a function of the vector $\mathbf{V}_g$ of the voltage generators at the ports of the transmitter, i.e., $\mathbf{V}_R=\mathbf{{H}}_{\rm{e2e}}^{(Z)}\mathbf{V}_g$. By using the approach in \cite{DR1} or, equivalently, the analytical derivation briefly summarized in Appendix A, the end-to-end RIS-aided channel matrix can be formulated as follows: \vspace{-0.1cm}
\begin{equation} \label{Eq:HZexact}
{\bf{{H}}}_{{\rm{e2e}}}^{\left( Z \right)} = {{\bf{Z}}_R}{\left( {{{\bf{Z}}_R} + {{{\bf{\bar Z}}}_{RR}}} \right)^{ - 1}}{{{\bf{\tilde Z}}}_{RT}}{\left( {{{\bf{Z}}_g} + {{{\bf{\tilde Z}}}_{TT}}} \right)^{ - 1}} \vspace{-0.1cm}
\end{equation}
where the following definitions hold: \vspace{-0.1cm}
\begin{equation} \label{Eq:ZtildeXY}
{{{\bf{\tilde Z}}}_{xy}} = {{\bf{Z}}_{xy}} - {{{\bf{ Z}}}_{xS}}{\left( {{{\bf{Z}}_S} + {{\bf{Z}}_{SS}}} \right)^{ - 1}}{{{\bf{ Z}}}_{Sy}} \vspace{-0.1cm}
\end{equation}
\vspace{-0.35cm}
\begin{equation} 
{{{\bf{\bar Z}}}_{RR}} = {{{\bf{\tilde Z}}}_{RR}} - {{{\bf{\tilde Z}}}_{RT}}{\left( {{{\bf{Z}}_g} + {{{\bf{\tilde Z}}}_{TT}}} \right)^{ - 1}}{{{\bf{\tilde Z}}}_{TR}} \vspace{-0.1cm}
\end{equation}
with $x \in \{T,R\}$  and $y \in \{T,R\}$.

Besides the relationship between the matrices $\mathbf{S}$ and $\mathbf{Z}$ in \eqref{Eq:ZtoS} and  \eqref{Eq:StoZ}, it is important to clarify the relationships between the matrices ${\bf{{H}}}_{{\rm{e2e}}}^{\left( S \right)}$ and ${\bf{{H}}}_{{\rm{e2e}}}^{\left( Z \right)}$, and the different physical meaning between the sub-matrices that constitute $\mathbf{S}$ and $\mathbf{Z}$.

\textbf{Relationship between ${\bf{{H}}}_{{\rm{e2e}}}^{\left( S \right)}$ and ${\bf{{H}}}_{{\rm{e2e}}}^{\left( Z \right)}$} -- From Appendix B, we obtain the  following two identities: \vspace{-0.1cm}
\begin{align} \label{Eq:VR}
& {{\bf{V}}_R} = 2{\sqrt {{Z_0}}}{\bf{{\rm H}}}_{{\rm{e2e}}}^{\left( Z \right)}\left( {{\bf{U}} - {{\bf{\Gamma }}_T}} \right)^{ - 1}{{\bf{a}}_g}\\
& {{\bf{V}}_R} = {\sqrt {{Z_0}}}\left( {{\bf{U}} + {{\bf{\Gamma }}_R}} \right){\bf{{\rm H}}}_{{\rm{e2e}}}^{\left( S \right)}{{\bf{a}}_g} \vspace{-0.1cm}
\end{align}

Therefore, the relation between ${\bf{{H}}}_{{\rm{e2e}}}^{\left( S \right)}$ and ${\bf{{H}}}_{{\rm{e2e}}}^{\left( Z \right)}$ is as follows: \vspace{-0.1cm}
\begin{equation} \label{Eq:HStoHZ}
{\bf{{H}}}_{{\rm{e2e}}}^{\left( Z \right)} = \frac{1}{2}\left( {{\bf{U}} + {{\bf{\Gamma }}_R}} \right){\bf{{H}}}_{{\rm{e2e}}}^{\left( S \right)}\left( {{\bf{U}} - {{\bf{\Gamma }}_T}} \right) \vspace{-0.1cm}
\end{equation}

From Appendix B, the relationship between ${{\bf{a}}_g}$ and ${{\bf{V}}_g}$ is \vspace{-0.1cm}
\begin{equation} \label{Eq:AgtoVg}
{{\bf{V}}_g} = 2{\sqrt {{Z_0}}}{\left( {{\bf{U}} - {{\bf{\Gamma }}_T}} \right)^{ - 1}}{{\bf{a}}_g} \vspace{-0.1cm}
\end{equation}

It is worth mentioning that ${\bf{{H}}}_{{\rm{e2e}}}^{\left( S \right)}$ and ${\bf{{H}}}_{{\rm{e2e}}}^{\left( Z \right)}$ are not identical because of their slightly different definitions, i.e., $\mathbf{b}_R=\mathbf{{H}}_{\rm{e2e}}^{(S)}\mathbf{a}_g$ and $\mathbf{V}_R=\mathbf{{H}}_{\rm{e2e}}^{(Z)}\mathbf{V}_g$. In other words, they relate different input and output vectors, which are in turn related to one another. The definitions of ${\bf{{H}}}_{{\rm{e2e}}}^{\left( S \right)}$ and ${\bf{{H}}}_{{\rm{e2e}}}^{\left( Z \right)}$ employed in this paper are those typically utilized in multiport network theory based on the $S$-parameter and $Z$-parameter representations, respectively.

\textbf{Zero reflection at the ports of the transmitter and receiver} -- The obtained expressions for the end-to-end RIS-aided channels ${\bf{{H}}}_{{\rm{e2e}}}^{\left( S \right)}$ and ${\bf{{H}}}_{{\rm{e2e}}}^{\left( Z \right)}$, and the relationships between them, have general validity. A sensible case study is considering $\mathbf{\Gamma}_T=\mathbf{0}$ and $\mathbf{\Gamma}_R=\mathbf{0}$, which ensures that the ports of the transmitter and receiver are matched for zero reflection \cite[p. 561]{PozarBook}. In this case, the matrix ${\bf{{H}}}_{{\rm{e2e}}}^{\left( S \right)}$ greatly simplifies, as given in \eqref{Eq:HE2Ematched}. By direct inspection of \eqref{Eq:HStoHZ} and \eqref{Eq:AgtoVg}, in fact, we obtain ${\bf{{ H}}}_{{\rm{e2e}}}^{\left( Z \right)} = \left( {{1 \mathord{\left/ {\vphantom {1 2}} \right.
 \kern-\nulldelimiterspace} 2}} \right){\bf{{ H}}}_{{\rm{e2e}}}^{\left( S \right)}$ and ${{\bf{V}}_g} = 2{\sqrt {{Z_0}}}{{\bf{a}}_g}$, respectively. Therefore, except for a simple scaling factor, the end-to-end channels, and the source waves and voltages are equivalent. 

It is instructive to scrutinize ${\bf{ H}}_{{\rm{e2e}}}^{\left( Z \right)}$ in \eqref{Eq:HZexact} under the assumption of zero reflection at the ports of the transmitter and receiver. By definition, the conditions $\mathbf{\Gamma}_T=\mathbf{0}$ and $\mathbf{\Gamma}_R=\mathbf{0}$ imply ${{\bf{Z}}_g} = {Z_0}{\bf{U}}$ and ${{\bf{Z}}_R} = {Z_0}{\bf{U}}$. By inserting them into \eqref{Eq:HZexact}, we see, however, that the expression of ${\bf{ H}}_{{\rm{e2e}}}^{\left( Z \right)}$ does not simplify significantly. This is in contrast with the major simplification that is obtained for ${\bf{ H}}_{{\rm{e2e}}}^{\left( S \right)}$, as apparent in \eqref{Eq:HE2Eexact} and \eqref{Eq:HE2Ematched}. To obtain an expression for ${\bf{ H}}_{{\rm{e2e}}}^{\left( Z \right)}$ that resembles ${\bf{ H}}_{{\rm{e2e}}}^{\left( S \right)}$, thus offering the same analytical tractability, it is necessary to apply some approximations, as proposed in \cite[Corollary 1]{DR1} and \cite{ABR}.

Specifically, the following approximations can be applied:
\begin{equation}
{{\bf{Z}}_g} + {{{\bf{\tilde Z}}}_{TT}} \approx {{\bf{Z}}_g} + {{\bf{Z}}_{TT}}, \quad
{{\bf{Z}}_R} + {{{\bf{\bar Z}}}_{RR}} \approx {{\bf{Z}}_R} + {{\bf{Z}}_{RR}} \vspace{-0.1cm}
\end{equation}
since the internal impedances ${{\bf{Z}}_g}$, the load impedances ${{\bf{Z}}_R}$, and the self-impedances ${{{\bf{\tilde Z}}}_{TT}}$ and ${{{\bf{\tilde Z}}}_{RR}}$ are the dominant terms in the expressions of ${{{\bf{\tilde Z}}}_{xy}}$ in \eqref{Eq:ZtildeXY}.

With this approximation at hand, ${\bf{ H}}_{{\rm{e2e}}}^{\left( Z \right)}$ simplifies as follows: \vspace{-0.2cm}
\begin{equation} \label{Eq:HZsimpler}
{\bf{{H}}}_{{\rm{e2e}}}^{\left( Z \right)} \approx {{\bf{Z}}_R}{\left( {{{\bf{Z}}_R} + {{\bf{Z}}_{RR}}} \right)^{ - 1}}{{{\bf{\tilde Z}}}_{RT}}{\left( {{{\bf{Z}}_g} + {{\bf{Z}}_{TT}}} \right)^{ - 1}} \vspace{-0.1cm}
\end{equation}
with ${{\bf{Z}}_g} = {Z_0}{\bf{U}}$ and ${{\bf{Z}}_R} = {Z_0}{\bf{U}}$ if the ports of the transmitter and receiver are matched for zero reflection. However, \eqref{Eq:HZsimpler} holds true for any values of the impedances ${{\bf{Z}}_g}$ and ${{\bf{Z}}_R}$.

Even though \eqref{Eq:HE2Ematched} and \eqref{Eq:HZsimpler} are similar to one another, it needs to be kept in mind that, if $\mathbf{\Gamma}_T=\mathbf{0}$ and $\mathbf{\Gamma}_R=\mathbf{0}$, \eqref{Eq:HE2Ematched} is exact whereas \eqref{Eq:HZsimpler} is, although tight, an approximation. An example is presented in Appendix B, moving from \eqref{Eq:ZtoS} and \eqref{Eq:StoZ}.

\textbf{Direct transmitter-receiver link} -- A subtle difference between the $S$-parameter and $Z$-parameter representations is the physical meaning of the sub-matrices ${{\bf{S}}_{xy}}$ and ${{\bf{Z}}_{xy}}$. Due to its practical relevance and mathematical tractability, we illustrate this, often overlooked aspect, by considering the setting $\mathbf{\Gamma}_T=\mathbf{0}$ and $\mathbf{\Gamma}_R=\mathbf{0}$. Thus, we compare ${\bf{ H}}_{{\rm{e2e}}}^{\left( S \right)}$ and ${\bf{ H}}_{{\rm{e2e}}}^{\left( Z \right)}$ in \eqref{Eq:HE2Ematched} and \eqref{Eq:HZsimpler} by assuming ${{\bf{Z}}_g} = {Z_0}{\bf{U}}$ and ${{\bf{Z}}_R} = {Z_0}{\bf{U}}$. Also, we focus on the setting ${{\bf{Z}}_{TT}} = {Z_0}{\bf{U}}$ and ${{\bf{Z}}_{RR}} = {Z_0}{\bf{U}}$, i.e., there is no mutual coupling at the transmitter and receiver, and their self-impedances are matched to the reference impedance $Z_0$. These assumptions simplify the analytical expressions, without jeopardizing the general findings on the impact of the RIS.

Under the considered assumptions, the relationship between the sub-matrices ${{\bf{S}}_{xy}}$ and ${{\bf{Z}}_{xy}}$ was first reported in \cite{li2023diagonal}. Specifically, the following identities hold \cite[Eq. (6)]{li2023diagonal}: \vspace{-0.1cm}
\begin{align}
& {{\bf{S}}_{{{ST}}}} = {\left( {{{\bf{Z}}_{{{SS}}}} + {Z_0}{\bf{U}}} \right)^{ - 1}}{{\bf{Z}}_{{{ST}}}} \label{Eq:S_ST}\\
& {{\bf{S}}_{{{RT}}}} = \frac{{{{\bf{Z}}_{{{RT}}}}}}{{2{Z_0}}} - \frac{{{{\bf{Z}}_{{{RS}}}}}}{{2{Z_0}}}{\left( {{{\bf{Z}}_{{{SS}}}} + {Z_0}{\bf{U}}} \right)^{ - 1}}{{\bf{Z}}_{{{ST}}}} \label{Eq:SZ_RT}  \\ 
& {{\bf{S}}_{{{SS}}}} = {\left( {{{\bf{Z}}_{{{SS}}}} + {Z_0}{\bf{U}}} \right)^{ - 1}}\left( {{{\bf{Z}}_{{{SS}}}} - {Z_0}{\bf{U}}} \right) \label{Eq:SZ_SS}  \\
& {{\bf{S}}_{{{RS}}}} = \frac{{{{\bf{Z}}_{{{RS}}}}}}{{2{Z_0}}}{\bf{U}} - \frac{{{{\bf{Z}}_{{{RS}}}}}}{{2{Z_0}}}{\left( {{{\bf{Z}}_{{{SS}}}} + {Z_0}{\bf{U}}} \right)^{ - 1}}\left( {{{\bf{Z}}_{{{SS}}}} - {Z_0}{\bf{U}}} \right)
\label{Eq:StructScatt} \vspace{-0.1cm}
\end{align}

The relationships in \eqref{Eq:S_ST}-\eqref{Eq:StructScatt} are obtained by applying \eqref{Eq:ZtoS}, and by noting that the approximation in \eqref{Eq:HZsimpler} implies ${{\bf{Z}}_{TS}} = {\bf{0}}$, ${{\bf{Z}}_{TR}} = {\bf{0}}$, and ${{\bf{Z}}_{SR}} = {\bf{0}}$. Then, the corresponding matrix $\bf{Z}$ in \eqref{Eq:ZtoS} is lower triangular and it is easier to invert.

\textcolor{black}{The RIS-aided channels in \eqref{Eq:HE2Ematched} and  \eqref{Eq:HZexact} are equivalent, except for simple scaling factors. The block sub-matrices in ${\bf{{ H}}}_{{\rm{e2e}}}^{\left( Z \right)}$ and ${\bf{{ H}}}_{{\rm{e2e}}}^{\left( S \right)}$ are, however, not one-to-one related, since the identity in \eqref{Eq:ZtoS} cannot be applied to each sub-matrix of $\bf{Z}$, but it needs to be applied to the whole matrix $\bf{Z}$. Also, \eqref{Eq:HE2Ematched} and \eqref{Eq:HZexact} are both useful, since they offer an easier understanding of different aspects that determine the scattering from an RIS}. 

A major aspect is the different physical meaning between ${{\bf{S}}_{{{RT}}}}$ in \eqref{Eq:HE2Ematched} and ${{{\bf{Z}}_{{{RT}}}}}$ in \eqref{Eq:HZexact}. The sub-matrices ${{\bf{S}}_{{{RT}}}}$ and ${{{\bf{Z}}_{{{RT}}}}}$ are usually associated with the direct link between the transmitter and receiver. It is assumed,  stated differently, that ${{\bf{S}}_{{{RT}}}} = \bf{0}$ and ${{{\bf{Z}}_{{{RT}}}}} = \bf{0}$ if the direct link is physically blocked by objects. This is not, however, in agreement with \eqref{Eq:SZ_RT}. Specifically, we see that the condition ${{{\bf{Z}}_{{{RT}}}}} = \bf{0}$ does not imply ${{{\bf{S}}_{{{RT}}}}} = \bf{0}$. The reason is that the physical direct link between the transmitter and receiver is given by the impedance matrix ${{{\bf{Z}}_{{{RT}}}}}$. Based on \cite{DR1}, in fact, the entries of the impedance matrices correspond to the electric field that is generated by the transmitter and is observed at the receiver. As a result, only the impedance sub-matrices are related to the physical links between pairs of devices. The scattering sub-matrices, on the other hand, are not one-to-one related to the physical links between pairs of devices. As far as the sub-matrix ${{{\bf{S}}_{{{RT}}}}}$ is concerned, we evince from \eqref{Eq:SZ_RT} that ${{{\bf{S}}_{{{RT}}}}} \ne \bf{0}$ if ${{{\bf{Z}}_{{{RT}}}}} = \bf{0}$, i.e., ${{{\bf{S}}_{{{RT}}}}} \ne \bf{0}$ even if the physical direct transmitter-receiver link is blocked by an obstacle. The sub-matrix ${{{\bf{S}}_{{{RT}}}}}$ includes an additional term that originates from the scattering of the RIS. It is worth mentioning that this additional term is present in ${{{\bf{Z}}_{{{RT}}}}}$ in \eqref{Eq:HZexact} and \eqref{Eq:HZsimpler} as well. However, it is hidden in the second addend of ${{{\bf{\tilde Z}}}_{RT}}$ and, hence, it is not explicitly visible. The electromagnetic meaning of the additional addend in ${{{\bf{S}}_{{{RT}}}}}$ when ${{{\bf{Z}}_{{{RT}}}}} = \bf{0}$ is discussed in the following sub-section. 

\vspace{-0.25cm}
\subsection{Comparison with Conventional Scattering Models}
In this sub-section, we elaborate on the electromagnetic differences between the scattering model for RISs that is typically utilized in communication theory, and those obtained in \eqref{Eq:HE2Ematched} and \eqref{Eq:HZexact} by utilizing multiport network theory.

\textbf{Scattering model employed in communication theory} -- The typical end-to-end RIS-aided channel model utilized in communication theory is usually formulated in terms of scattering matrices, and is expressed as follows \cite{WuZZYZ21}, \cite{Galdi2023}: \vspace{-0.1cm}
\begin{equation} \label{Eq:HCT}
{\bf{{H}}}_{{\rm{e2e}}}^{\left( CT \right)} = {{\bf{H}}_{{{RT}}}} + {{\bf{H}}_{{{RS}}}}{{\bf{\Gamma }}_{\rm{S}}}{{\bf{H}}_{{{ST}}}} \vspace{-0.1cm}
\end{equation}
\textcolor{black}{where ${{\bf{H}}_{{RT}}}$ represents the direct link between the transmitter and the receiver, and ${{\bf{H}}_{{RS}}}$ and ${{\bf{H}}_{{ST}}}$ represent the links between the receiver and the RIS, and the transmitter and the RIS, respectively. In typical communication-theoretic models, the channel matrices ${{\bf{H}}_{{RT}}}$, ${{\bf{H}}_{{RS}}}$ and ${{\bf{H}}_{{ST}}}$ represent the \textit{physical} channels between the network elements. In other words, these matrices are equal to zero when the corresponding links are physically blocked by objects. In addition, ${{\bf{H}}_{{RT}}}$ is assumed, by definition, to be independent of the RIS structure and of the RIS configuration. Similar to the multiport network theory model, ${{\bf{\Gamma }}_{\rm{S}}}$ is the matrix of RIS reflection coefficients, and is still defined as ${{\bf{\Gamma }}_S} = {\left( {{{\bf{Z}}_S} + {Z_0}{\bf{U}}} \right)^{ - 1}}\left( {{{\bf{Z}}_S} - {Z_0}{\bf{U}}} \right)$. Since \eqref{Eq:HCT} is formulated in terms of the matrix ${{\bf{\Gamma }}_{\rm{S}}}$ of RIS reflection coefficients, it is natural to compare ${\bf{{H}}}_{{\rm{e2e}}}^{\left( CT \right)}$ with ${\bf{{H}}}_{{\rm{e2e}}}^{\left( S \right)}$ in \eqref{Eq:HE2Ematched}. The following fundamental differences can be identified:}

$\bullet$ ${\bf{{H}}}_{{\rm{e2e}}}^{\left( CT \right)}$ in \eqref{Eq:HCT} ignores the mutual coupling among the RIS elements. In fact, ${\bf{{H}}}_{{\rm{e2e}}}^{\left( CT \right)}={\bf{{H}}}_{{\rm{e2e}}}^{\left( S \right)}$ if and only if ${{{\bf{S}}_{SS}}} = \bf{0}$. 

$\bullet$ In \eqref{Eq:HCT}, the self-impedances of the RIS elements are assumed to be equal to the reference impedance $Z_0$. This immediately follows from the implicit assumption ${{{\bf{S}}_{SS}}} = \bf{0}$ in \eqref{Eq:HCT}, as well as from \eqref{Eq:SZ_SS}, which yields  ${{\bf{Z}}_{SS}} = {Z_0}{\bf{U}}$.

$\bullet$ ${\bf{{H}}}_{{\rm{e2e}}}^{\left( CT \right)}$ in \eqref{Eq:HCT} ignores the interrelation between the phase and the amplitude transformation that an RIS applies to the incident waves. For illustration, consider a diagonal RIS. Based on ${\bf{{H}}}_{{\rm{e2e}}}^{\left( CT \right)}$, the condition of lossless scattering is usually imposed by setting $\left| {{\Gamma _{S,i}}} \right| = 1$, for every RIS element From the equation ${{\bf{\Gamma }}_S} = {\left( {{{\bf{Z}}_S} + {Z_0}{\bf{U}}} \right)^{ - 1}}\left( {{{\bf{Z}}_S} - {Z_0}{\bf{U}}} \right)$, this implies that the load impedances ${{{\bf{Z}}_S}}$ are purely imaginary values. Then, it is argued that an RIS is capable of changing the phase of the incident waves without altering their amplitude. Realistic scattering models, however, show that this is not easy to obtain \cite{Galdi2023}. Even assuming $\left| {{\Gamma _{S,i}}} \right| = 1$ for every RIS element, the RIS-aided channel in ${\bf{{H}}}_{{\rm{e2e}}}^{\left( S \right)}$ leads to a different conclusion. For ease of exposition, consider the simple case study in the absence of mutual coupling, i.e., ${{{\bf{S}}_{SS}}}$ is a diagonal matrix whose $(i,i)$th entry is $S_{SS, ii}$. Then, the matrix ${\left( {{\bf{U}} - {{\bf{\Gamma }}_S}{{\bf{S}}_{SS}}} \right)^{ - 1}}$ in \eqref{Eq:HE2Ematched} is diagonal and its $(i,i)$th entry is ${\left( {1 - {\Gamma _{S,i}}{S_{SS,ii}}} \right)^{ - 1}}$. Therefore, its absolute value is, in general, less than one even if $\left| {{\Gamma _{S,i}}} \right| = 1$. Also, this absolute value strongly depends on the phase of ${{\Gamma _{S,i}}}$. In other words, the matrix ${\left( {{\bf{U}} - {{\bf{\Gamma }}_S}{{\bf{S}}_{SS}}} \right)^{ - 1}}$ determines the interrelation between the amplitude and the phase of the scattered electromagnetic wave from an RIS. The amplitude-phase dependence is even more intricate in the presence of electromagnetic mutual coupling, and for non-diagonal RISs.

\textbf{Structural scattering} -- In the previous sub-section, we have analyzed the differences between ${{\bf{S}}_{{{RT}}}}$ in \eqref{Eq:HE2Ematched} and ${{{\bf{Z}}_{{{RT}}}}}$ in \eqref{Eq:HZexact} from the mathematical standpoint. It is instructive, and important for the rest of this paper, to analyze the difference between ${{\bf{S}}_{{{RT}}}}$ and ${{{\bf{Z}}_{{{RT}}}}}$ from the electromagnetic standpoint, and to put the findings in relation with ${\bf{{H}}}_{{\rm{e2e}}}^{\left( CT \right)}$ in \eqref{Eq:HCT}.

For ease of exposition, we introduce the following matrix: 
\begin{align} \label{Eq:StSc}
{{\bf{S}}_{{\rm{StSc}}}} =  - \frac{{{{\bf{Z}}_{{{RS}}}}}}{{2{Z_0}}}{\left( {{{\bf{Z}}_{{{SS}}}} + {Z_0}{\bf{U}}} \right)^{ - 1}}{{\bf{Z}}_{{{ST}}}} 
\end{align}

From \eqref{Eq:SZ_RT}, we obtain, therefore, the following: \vspace{-0.1cm}
\begin{align} \label{Eq:S_RT}
{{\bf{S}}_{{{RT}}}} = \frac{{{{\bf{Z}}_{{{RT}}}}}}{{2{Z_0}}} + {{\bf{S}}_{{\rm{StSc}}}} \vspace{-0.1cm}
\end{align}

In the previous sub-section, we have clarified that ${{\bf{S}}_{{\rm{StSc}}}}$ depends on the RIS, and that it is not related to the transmitter-receiver physical direct link. \textcolor{black}{In antenna theory, the sub-matrix ${{\bf{S}}_{{\rm{StSc}}}}$ is well-known and includes the structural scattering \cite{R7}, \cite{KnottBook}. The structural scattering is always present in RIS-aided channels, regardless of whether the transmitter-receiver direct link is physically blocked. This fundamental  electromagnetic aspect is ignored, by assumption, in conventional channel models used in communication theory, i.e., ${\bf{{H}}}_{{\rm{e2e}}}^{\left( CT \right)}$ in \eqref{Eq:HCT}}. 

To better understand the impact of the structural scattering, it is necessary to analyze the origin and meaning of ${{\bf{S}}_{{\rm{StSc}}}}$ from the electromagnetic standpoint. By direct inspection, we evince that ${{\bf{S}}_{{\rm{StSc}}}}$ is the signal scattered from an RIS when ${{\bf{Z}}_S} = {Z_0}{\bf{U}}$, i.e., when the tunable loads of the RIS are matched to the reference impedance. This follows because ${{\bf{\Gamma }}_S} = {\left( {{{\bf{Z}}_S} + {Z_0}{\bf{U}}} \right)^{ - 1}}\left( {{{\bf{Z}}_S} - {Z_0}{\bf{U}}} \right) = {\bf{0}}$ when ${{\bf{Z}}_S} = {Z_0}{\bf{U}}$, which, based on \eqref{Eq:HE2Ematched}, results in ${{\mathbf{{H}}}_{{\rm{e2e}}}^{(S)}} = {{{\bf{ S}}}_{RT}}$. Based on \eqref{Eq:SZ_RT}, in addition, we see that ${{{\bf{ S}}}_{RT}} = {{\bf{S}}_{{\rm{StSc}}}}$ when the transmitter-receiver direct link is blocked. This has a major implication: An RIS reradiates an electromagnetic wave even if the reflection coefficient ${{\bf{\Gamma }}_S}$ is set equal to zero. This is in stark contrast with the conventional channel model ${\bf{{H}}}_{{\rm{e2e}}}^{\left( CT \right)}$ in \eqref{Eq:HCT}, which provides no redadiated signal if ${{{\bf{\Gamma }}_{{S}}}} = {\bf{0}}$. \textcolor{black}{In simple terms, ${\bf{{H}}}_{{\rm{e2e}}}^{\left( CT \right)}$ completely ignores the structural scattering from an RIS. By contrast, ${{\bf{S}}_{{\rm{StSc}}}}$ is duly considered in \eqref{Eq:HE2Ematched} and \eqref{Eq:HZsimpler}}.

{\textbf{Unwanted specular reradiation}} -- The presence of the structural scattering in ${{\bf{S}}_{{\rm{StSc}}}}$ has major implications when optimizing an RIS: The term ${{\bf{S}}_{{\rm{StSc}}}}$ in \eqref{Eq:StSc} corresponds to a specular reflection, i.e., a signal reradiated according to Snell's law of reflection. This specular reflection may result in undesired/unwanted interference, which is, however, usually ignored in communication theory because the channel model ${\bf{{H}}}_{{\rm{e2e}}}^{\left( CT \right)}$ does not account for the structural scattering from the RIS, {\textcolor{black}{since, as mentioned, ${{\bf{H}}_{{RT}}}$ is, by definition, independent of the RIS and the RIS reradiates no signal if ${{{\bf{\Gamma }}_{{S}}}} = {\bf{0}}$, by definition of  ${\bf{{H}}}_{{\rm{e2e}}}^{\left( CT \right)}$ in \eqref{Eq:HCT}}}. This implies that an RIS needs to be optimized for steering the electromagnetic waves towards the desired direction of reradiation, but also for reducing its structural scattering, so as to ensure a sufficient amount of scattered power towards the intended locations and to reduce the  interference towards other locations/users. Next, we will show that the structural scattering in ${{\bf{S}}_{{\rm{StSc}}}}$ can be compensated by appropriately controlling the tunable loads ${{\bf{Z}}_S}$ of the RIS. \textcolor{black}{To the best of the authors' knowledge, this optimization problem has never been considered in the literature. It is interesting to note, in addition, that the structural scattering from an RIS is, in free space channels, known by design to the transmitter. In fact, the structural scattering in ${{\bf{S}}_{{\rm{StSc}}}}$ always corresponds to a specular reflection, which depends only on the location of the transmitter and the RIS, i.e., it is determined by the angle of incidence in the Fraunhofer far-field region. In fading channels, ${{\bf{S}}_{{\rm{StSc}}}}$ depends on the multipath components as well, as elaborated next in Sec. III-D. However, many channel estimation algorithms for RIS-aided channels are available \cite{PanZZHWPRRSZZ22}, including algorithms for the multiport network model \cite{zheng2023impact}}.

\vspace{-0.25cm}
\section{RIS Optimization}
In this section, we provide two contributions: (i) we introduce the first algorithm that optimizes an RIS based on the $S$-parameter representation, and discuss its advantages with respect to utilizing the $Z$-parameter representation; and (ii) we introduce the first algorithm that optimizes the scattered signal from an RIS, while considering the structural scattering as an optimization constraint. For simplicity, we analyze a SISO ($N_T = N_R = 1$) system and a diagonal RIS (${{\bf{\Gamma }}_S}$ is a diagonal matrix). We assume that the ports of the transmitter and receiver are matched for zero reflection (${{\bf{\Gamma }}_T} = {\bf{0}}$, ${{\bf{\Gamma }}_R} = {\bf{0}}$). Possible generalizations are postponed to future work.

\vspace{-0.25cm}
\subsection{$S$-Parameter Based Optimization}\label{Opt_uni}
In the considered setting, the optimization problem of interest is equivalent to maximize the received power. Moving from \eqref{Eq:HE2Ematched}, the optimization problem is as follows: \vspace{-0.1cm}
\begin{align}\label{Optprob1}
&\max \limits_{\mathbf{\Gamma}}  \left|{S}_{RT} + \mathbf{S}_{RS} \mathbf{\Gamma}(\mathbf{U}-\mathbf{S}_{SS}\mathbf{\Gamma})^{-1}\mathbf{S}_{ST} \right|^2  \\
&\text{s.t.} \quad \left|\Gamma_{k}\right| \le 1 \text{ for } k=1,\ldots,K \nonumber \vspace{-0.1cm}
\end{align} 
where $\mathbf{\Gamma} = \mathbf{\Gamma}_{S}$ and $\Gamma_{i} = \Gamma_{s,i}$ to simplify the notation, and the identity ${\left( {{\bf{U}} - {{\bf{\Gamma }}_S}{{\bf{S}}_{SS}}} \right)^{ - 1}}{{\bf{\Gamma }}_S} = {{\bf{\Gamma }}_S}{\left( {{\bf{U}} - {{\bf{\Gamma }}_S}{{\bf{S}}_{SS}}} \right)^{ - 1}}$ is utilized. The constraint $\left|\Gamma_{k}\right| \le 1$ accounts for the fact that the tunable impedances of the RIS introduce some losses. The impact of these losses when solving \eqref{Optprob1} is elaborated next.

The main challenge in solving \eqref{Optprob1} is that the objective function depends on the inverse of a matrix, which in turn depends on $\mathbf{\Gamma}$. This makes the problem non-convex. To circumvent this issue, we propose an iterative algorithm, which provably allows the objective function to increase at each iteration and  to converge to a locally optimal solution. 

We denote by $X_k$ the $k$th tunable reactance of the RIS and by $r_0$ a small parasitic resistance ($r_0 \ll Z_0$). Thus, the $k$th tunable impedance is $Z_{S,k} = Z_{k} = jX_k + r_0$. The parasitic resistance $r_0$ makes the absolute value of $\left|\Gamma_{k}\right|$ less than one. Thus, the reflection coefficient at the $k$th port of the RIS is \vspace{-0.1cm}
\begin{align}\label{Gamma1}
& \Gamma_{k} = \frac{jX_k-Z_0+r_0}{jX_k+Z_0+r_0} = \frac{j\tilde{X}_k-1+\varepsilon}{j\tilde{X}_k+1+\varepsilon}  \vspace{-0.1cm}
\end{align} 
where $\tilde{X}_k = X_k/Z_0$ and $\varepsilon = r_0/Z_0 \ll 1$. 

Since $\varepsilon \ll 1$, we can apply Taylor's series approximation to $\Gamma_{k}$ at $\varepsilon = 0$. By doing so, we obtain the following: \vspace{-0.1cm}
\begin{align}\label{Gamma2}
\Gamma_{k} & \approx \Gamma_k|_{\varepsilon = 0} + \varepsilon \frac{d \Gamma_{k}}{d \varepsilon}|_{\varepsilon = 0} \\
& = e^{j\phi_k} + \frac{2}{\left(j\tilde{X}_k+1\right)^2}\varepsilon = e^{j\phi_k}\left(1-\frac{2}{\tilde{X}_k^2+1}\varepsilon\right)\nonumber
& \mathop  \approx \limits^{\left( a \right)} e^{j\phi_k} \vspace{-0.1cm}
\end{align}
where $\phi_k = \angle{\frac{j\tilde{X}_k-1}{j\tilde{X}_k+1}}$, and the approximation in $(a)$ follows because $\varepsilon$ is small and $\tilde X_k^2$ is usually large.

From \eqref{Gamma2}, we evince that the phase of $\Gamma_k$ coincides, to the first-order approximation, with the phase of $\phi_k$, which is the phase of $\Gamma_{k}$ when $\varepsilon = 0$. In addition, the approximation $(a)$ in \eqref{Gamma2} implies that the impact of $\varepsilon$ on the absolute value of $\Gamma_k$ can be ignored at the optimization stage. The optimization problem in \eqref{Optprob1} can, therefore, be reformulated as follows: \vspace{-0.1cm}
\begin{align}\label{Optprob1a}
&\max \limits_{\mathbf{\Gamma}}  \left|{S}_{RT} + \mathbf{S}_{RS} \mathbf{\Gamma}(\mathbf{U}-\mathbf{S}_{SS}\mathbf{\Gamma})^{-1}\mathbf{S}_{ST} \right|^2  \\
&\text{s.t.} \quad \left|\Gamma_{k}\right| = 1 \text{ for } k=1,\ldots,K \nonumber \vspace{-0.1cm}
\end{align} 
with $\Gamma_k = e^{j\phi_k}$. Next, we will see, however, that $r_0$ is not completely ignored in the proposed optimization algorithm.

The proposed algorithm works iteratively, by adjusting $\phi_k$ or, equivalently, $\tilde{X}_k = \frac{1}{j}\frac{1+e^{j\phi_{k}}}{1-e^{j\phi_{k}}}$ at each iteration. Let us assume to be at the $m$th iteration and that the phase at this iteration is $\phi_{k}^{(m)}$, i.e., $\Gamma_{k}^{(m)} \approx e^{j\phi_k^{(m)}}$. Then, we add a small perturbation, which is denoted by $\delta_k^{(m)}$, to $\phi_{k}^{(m)}$ with $\left|\delta_k^{(m)}\right| \ll 1$, so that the solution at the $(m+1)$th iteration is $\phi_{k}^{(m+1)} = \phi_{k}^{(m)} + \delta_k^{(m)}$. Thus, the reflection coefficient at the $(m+1)$th iteration is \vspace{-0.1cm}
\begin{align}\label{approx1}
\Gamma_{k}^{(m+1)} = e^{j(\phi_{k}^{(m)}+\delta_k^{(m)})} \mathop  \approx \limits^{\left( a \right)} \Gamma_{k}^{(m)}  + j e^{j\phi_{k}^{(m)}} \delta_k^{(m)} \vspace{-0.1cm}
\end{align} 
\noindent where $(a)$ follows by applying Taylor's series approximation, since $\left|\delta_k^{(m)}\right| \ll 1$, i.e., ${e^{j\delta _k^{\left( m \right)}}} \approx 1 + j\delta _k^{\left( m \right)}$. 

The main difficulty when solving the optimization problem in \eqref{Optprob1a} is the non-linear dependence of the objective function with the reflection coefficient $\mathbf{\Gamma}$, i.e., the term $\mathbf{\Gamma}(\mathbf{U}-\mathbf{S}_{ SS }\mathbf{\Gamma})^{-1}$. This is in stark contrast with the conventional RIS-aided channel model in \eqref{Eq:HCT}, making the considered optimization problem even harder to solve. To tackle the optimization problem in \eqref{Optprob1a}, we first use the following identity: \vspace{-0.1cm}
\begin{equation} \label{Eq:MatInv}
\mathbf{\Gamma}(\mathbf{U}-\mathbf{S}_{ SS }\mathbf{\Gamma})^{-1} = (\mathbf{\Gamma}^{-1}-\mathbf{S}_{ SS })^{-1} \vspace{-0.1cm}
\end{equation}

By using \eqref{approx1}, we then obtain \vspace{-0.1cm}
\begin{align}\label{approx4p}
& \left({\Gamma}_{k}^{(m+1)}\right)^{-1} \mathop  \approx \limits^{\left( a \right)} \left({\Gamma}_{k}^{(m)}\right)^{-1} - j \left({\Gamma}_{k}^{(m)}\right)^{-2} e^{j{\phi}_k^{(m)}} {\delta}_k^{(m)} \vspace{-0.1cm}
\end{align}
where $(a)$ follows from Taylor's series approximation, since $\left|\delta_k^{(m)}\right| \ll 1$, i.e., ${\left( {a_0 + b_0\delta _k^{\left( m \right)}} \right)^{ - 1}} \approx {a_0^{ - 1}} - b_0{a_0^{ - 2}}\delta_k^{(m)}$.

Recalling that $\mathbf{\Gamma}^{(m)}$ is a $K \times K$ diagonal matrix, and introducing the $K \times K$ diagonal matrices $\boldsymbol{\phi}^{(m)}$ and $\boldsymbol{\delta}^{(m)}$, whose diagonal elements are ${\phi}_k^{(m)}$ and ${\delta}_k^{(m)}$, respectively, \eqref{Eq:MatInv} simplifies to \vspace{-0.2cm}
\begin{align}\label{approx2}
& \mathbf{\Gamma}^{(m+1)}\left(\mathbf{U}-\mathbf{S}_{SS}\mathbf{\Gamma}^{(m+1)}\right)^{-1} \nonumber \\
& \hspace{0.5cm} \approx \left(\left(\mathbf{\Gamma}^{(m)}\right)^{-1} - j \left(\mathbf{\Gamma}^{(m)}\right)^{-2} e^{j\boldsymbol{\phi}^{(m)}} \boldsymbol{\delta}^{(m)}-\mathbf{S}_{SS}\right)^{-1} \vspace{-0.2cm} 
\end{align} 

Introducing the notation $\mathbf{Q}^{(m)} = \left(\mathbf{\Gamma}^{(m)}\right)^{-1} -\mathbf{S}_{SS}$ and $\mathbf{P}^{(m)} = j (\mathbf{Q}^{(m)})^{-1}\left(\mathbf{\Gamma}^{(m)}\right)^{-2}e^{j\boldsymbol{\phi}^{(m)}}$, \eqref{approx2} can be rewritten as follows: \vspace{-0.2cm}
\begin{equation}\label{approx3}
\mathbf{\Gamma}^{(m+1)}\left(\mathbf{U}-\mathbf{S}_{SS}\mathbf{\Gamma}^{(m+1)}\right)^{-1} \approx \left(\mathbf{U}-  \mathbf{P}^{(m)} \boldsymbol{\delta}^{(m)}\right)^{-1} (\mathbf{Q}^{(m)})^{-1} \vspace{-0.2cm} 
\nonumber 
\end{equation} 

Therefore, the optimization variable at the $m$th iterations is the diagonal matrix $\boldsymbol{\delta}^{(m)}$, which is inside the inversion of a matrix. The resulting optimization problem is still not convex in $\boldsymbol{\delta}^{(m)}$. To tackle it, we apply, similar to \cite{DR2}, \cite{ABR}, Neumann's series approximation, which yields the following: 
\vspace{-0.2cm}
\begin{equation}\label{approx4s}
\mathbf{\Gamma}^{(m+1)}\left(\mathbf{U}-\mathbf{S}_{ SS }\mathbf{\Gamma}^{(m+1)}\right)^{-1} \approx (\mathbf{Q}^{(m)})^{-1} + \mathbf{P}^{(m)} \boldsymbol{\delta}^{(m)}(\mathbf{Q}^{(m)})^{-1} 
\end{equation}
\textcolor{black}{provided that the following condition is fulfilled:} 
\begin{equation}\label{condition}\textcolor{black}{\left\|\boldsymbol{\delta}^{(m)}\mathbf{P}^{(m)}\right\| \ll 1 }
\end{equation}

\textcolor{black}{A simple, but sub-optimal, approach for ensuring that the condition is satisfied consists of applying the same small increment to all the elements of $\boldsymbol{\delta}^{(m)}$. Denoting by $\delta_0^{(m)}$ the value of the small increment  at the $m$th iteration, the condition $\left\|\boldsymbol{\delta}^{(m)}\mathbf{P}^{(m)}\right\| = \bar \delta_{0} \ll 1$ is equivalent to $\left|\delta_k^{(m)} \right| = \left|\delta_0^{(m)} \right| = {\bar \delta_0}{\left\|\mathbf{P}^{(m)}\right\|^{-1}}$ for $k=1, 2, \ldots, K$ and $\bar \delta_{0} \ll 1$. This approach is referred to as $\text{S-UNI}$. The possibility of applying different increment values to the elements of $\boldsymbol{\delta}^{(m)}$ is discussed next.}

By utilizing \eqref{approx4s}, the considered optimization problem at the $(m+1)$th iteration can be formulated as follows: \vspace{-0.2cm}
\begin{align}\label{Optprob2}
&\max \limits_{{\boldsymbol{\delta}^{(m)}}}  \left| a^{(m)} +  \left(\mathbf{b}_1^{(m)}\right)^T \boldsymbol{\delta}^{(m)}  \mathbf{b}_2^{(m)} \right|^2\\
&\text{s.t.} \quad \left|\delta_k^{(m)}\right|  \le \bar \delta_0 {\left\|\mathbf{P}^{(m)}\right\|^{-1}}, \quad k=1,2, \ldots , K \nonumber \vspace{-0.2cm}
\end{align} 
where $\bar \delta_0 \ll 1$, $a^{(m)} = {S}_{ RT } + \mathbf{S}_{ RS }(\mathbf{Q}^{(m)})^{-1}\mathbf{S}_{ ST }$, and $\left(\mathbf{b}_1^{(m)}\right)^T = \mathbf{S}_{ RS }\mathbf{P}^{(m)}$ and $\mathbf{b}_2^{(m)} = (\mathbf{Q}^{(m)})^{-1} \mathbf{S}_{ ST }$ are $K \times 1$ column vectors.

Since $\boldsymbol{\delta}^{(m)}$ is a diagonal matrix, $\left(\mathbf{b}_1^{(m)}\right)^T{{\boldsymbol{\delta }}^{\left( m \right)}}{{\bf{b}}_2^{(m)}} = \sum\nolimits_{k = 1}^K {{b_{1,k}^{(m)}}\delta _k^{\left( m \right)}{b_{2,k}^{(m)}}}$, where $b_{1,k}^{(m)}$ and $b_{2,k}^{(m)}$ are the elements of $\mathbf{b}_1^{(m)}$ and $\mathbf{b}_2^{(m)}$, respectively. For ease of writing, we use the notation $c_k^{(m)} = b_{1,k}^{(m)} b_{2,k}^{(m)}$. By virtue of the constraint $\bar \delta_0 \ll 1$, the objective function in \eqref{Optprob2} can be approximated as follows: \vspace{-0.1cm}
\begin{align} \label{Eq:ApproxObj}
&{\left| {{a^{\left( m \right)}} + \left(\mathbf{b}_1^{(m)}\right)^T{{\boldsymbol{\delta }}^{\left( m \right)}}{{\bf{b}}_2^{(m)}}} \right|^2} \\ &\hspace{0.5cm} = {\left| {{a^{\left( m \right)}}} \right|^2} + 2{\mathop{\Re}\nolimits} \left\{ {{a^{\left( m \right)}}{{\left( {{\left(\mathbf{b}_1^{(m)}\right)^T}{{\boldsymbol{\delta }}^{\left( m \right)}}{{\bf{b}}_2^{(m)}}} \right)}^*}} \right\} \nonumber  \\
&\hspace{0.75cm} + {\left| {{\left(\mathbf{b}_1^{(m)}\right)^T}{{\boldsymbol{\delta }}^{\left( m \right)}}{{\bf{b}}_2^{(m)}}} \right|^2} \nonumber \\
& \hspace{0.5cm} \mathop  \approx \limits^{\left( a \right)} {\left| {{a^{\left( m \right)}}} \right|^2} + 2{\mathop{\Re}\nolimits} \left\{ {{a^{\left( m \right)}}{{\left( {{\left(\mathbf{b}_1^{(m)}\right)^T}{{\boldsymbol{\delta }}^{\left( m \right)}}{{\bf{b}}_2^{(m)}}} \right)}^*}} \right\} \nonumber\\
& \hspace{0.5cm} \mathop  \approx \limits^{\left( b \right)} {\left| {{a^{\left( m \right)}}} \right|^2} + 2\sum\nolimits_{k = 1}^K {{\mathop{\Re}\nolimits} \left\{ {{a^{\left( m \right)}}\left(c_k^{(m)}\right)^*} \right\}\delta _k^{\left( m \right)}}  \nonumber \vspace{-0.1cm}
\end{align}
where $(a)$ follows by ignoring the quadratic term in ${{\boldsymbol{\delta }}^{\left( m \right)}}$ thanks to the constraint $\bar \delta_0 \ll 1$, and $(b)$ follows because $c_k^{(m)} = b_{1,k}^{(m)} b_{2,k}^{(m)}$ and $\delta _k^{\left( m \right)}$ are real-valued variables.

By inserting \eqref{Eq:ApproxObj} into \eqref{Optprob2}, we see that the optimization problem can be separated in the variables $\delta_k^{(m)}$.  In mathematical terms, $\delta_k^{(m)}$ is the solution of the following problem: \vspace{-0.1cm}
\begin{align}\label{Optprob3}
&\max \limits_{\delta_k^{(m)}} \left\{  {\left| {{a^{\left( m \right)}}} \right|^2} + 2{\mathop{\Re}\nolimits} \left\{ {{a^{\left( m \right)}}\left(c_k^{(m)}\right)^*} \right\}\delta _k^{\left( m \right)} \right\}\\
&\text{s.t.} \quad \left|\delta_k^{(m)}\right|  \le \bar \delta_0^{(m)} \nonumber \vspace{-0.1cm}
\end{align} 
where $\bar \delta_0^{(m)} = \bar \delta_0 {\left\|\mathbf{P}^{(m)}\right\|^{-1}}$ with $\bar \delta_0 \ll 1$.

The optimal solution of the optimization problem in \eqref{Optprob3} is always at the edge of the interval $\left[-\bar \delta_0^{(m)} ,\bar \delta_0^{(m)} \right]$, as follows: \vspace{-0.1cm}
\begin{align}\label{solution1}
\delta _k^{\left( m \right)} = \left\{ {\begin{array}{*{20}{l}}
{{{\bar \delta }_0^{(m)}}\quad \quad \;\;\,{\rm{if}}\quad {\mathop{\Re}\nolimits} \left\{ {{a^{\left( m \right)}}\left(c_k^{(m)}\right)^*} \right\} \ge 0}\\
{ - {{\bar \delta }_0^{(m)}}\quad \quad {\rm{if}}\quad {\mathop{\Re}\nolimits} \left\{ {{a^{\left( m \right)}}\left(c_k^{(m)}\right)^*} \right\} < 0}
\end{array}} \right. \vspace{-0.1cm}
\end{align}

Once the small phase increments $\delta _k^{\left( m \right)}$ are obtained for $k=1,2,\ldots, K$, the phase shift at the $(m+1)$th iteration is computed as $\phi_{k}^{(m+1)} = \phi_{k}^{(m)} + \delta_k^{(m)}$. Then, the reactance and reflection coefficient are  given by 
\begin{equation}
X_k^{(m)} = \frac{Z_0}{j}\frac{1+e^{j\phi_{k}^{(m+1)}}}{1-e^{j\phi_{k}^{(m+1)}}}, \quad \Gamma_{k}^{(m)} = \frac{jX_k^{(m)}-Z_0+r_0}{jX_k^{(m)}+Z_0+r_0}
\end{equation}
respectively. The process is repeated until convergence. The complete algorithm for the proposed S-UNI approach is summarized in Algorithm \ref{Alg:S-UNI}. It is worth mentioning that the parasitic resistance $r_0$ is accounted for when updating $\Gamma_{k}^{(m)}$ at each iteration of Algorithm \ref{Alg:S-UNI}. The parasitic resistance $r_0$ is ignored only to compute the small phase increment, as a first-order approximation, according to \eqref{Gamma2}.

\begin{algorithm}[t!]
\footnotesize
\caption{S-UNI Optimization Algorithm}
\textbf{Input}: $S_{RT}$, $\mathbf{S}_{RS}$, $\mathbf{S}_{ST}$, $Z_0 =50 \, \Omega$, $r_0 > 0$\;
\textbf{Initialize}:\\
Generate the initial values of $X_k$ and evaluate $\tilde{X}_k = X_k/Z_0$\; 
$\Gamma_{k} \leftarrow \frac{jX_k-Z_0+r_0}{jX_k+Z_0+r_0}$, for $k = 1,\ldots,K$\;
Set an arbitrarily small value $\eta > 0$, $m \leftarrow 1$, $\bar \delta_0 \ll 1$\;
$\boldsymbol{\Gamma}^{(m)} \leftarrow \mathbf{0}$, ${\Gamma}^{(m)}_{k} \leftarrow  \Gamma_{k}$\;
$\phi_k = \angle{\frac{j\tilde{X}_k-1}{j\tilde{X}_k+1}}$, $\boldsymbol{\phi}^{(m)} = \mathbf{0}$, ${\phi}^{(m)}_k \leftarrow {\phi}_k $\;
\While{$\rho >\eta $}
{		
		$\mathbf{Q}^{(m)} \leftarrow \left(\mathbf{\Gamma}^{(m)}\right)^{-1} -\mathbf{S}_{SS}$\;
$\mathbf{P}^{(m)} \leftarrow -j \left(\mathbf{Q}^{(m)}\right)^{-1}\left(\mathbf{\Gamma}^{(m)}\right)^{-2}e^{j\boldsymbol{\phi}^{(m)}}$\;
		$ \delta_0^{(m)} \leftarrow {\bar\delta_0}/{\left\|\mathbf{P}^{(m)}\right\|}$\;
		$a^{(m)} \leftarrow {S}_{ RT } + \mathbf{S}_{ RS }\left(\mathbf{Q}^{(m)}\right)^{-1}\mathbf{S}_{ ST }$\;
		$\left(\mathbf{b}_1^{(m)}\right)^T = \mathbf{S}_{ RS }\mathbf{P}^{(m)}$, $\mathbf{b}_2^{(m)} = \left(\mathbf{Q}^{(m)}\right)^{-1} \mathbf{S}_{ ST }$\;
		\For {$k = 1,\ldots,K$}
	        {
		($*$) \quad $c_k^{(m)} \leftarrow b_{1,k}^{(m)}b_{2,k}^{(m)}$\;
		($**$) \quad Compute $\delta_k^{(m)}$ according to \eqref{solution1}\;
		($***$) $\phi_k^{(m+1)} \leftarrow \phi_k^{(m)} + \delta_k^{(m)}$\;
		$X_k^{(m)} = \frac{Z_0}{j}\frac{1+e^{j\phi_k^{(m+1)}}}{1-e^{j\phi_k^{(m+1)}}}$\;
		$\Gamma_{k} \leftarrow \frac{jX_k^{(m)}-Z_0+r_0}{jX_k^{(m)}+Z_0+r_0}$\;
		
		}	

	$\boldsymbol{\Gamma}^{(m+1)}_{k} \leftarrow  \Gamma_{k}$\;
	$\rho \leftarrow \lVert \boldsymbol{\Gamma}^{(m+1)} - \boldsymbol{\Gamma}^{(m)}\rVert$\;
	$m \leftarrow m + 1$.
} 
\label{Alg:S-UNI} 
\end{algorithm}

\subsubsection{\textbf{Sensitivity Analysis}}\label{sensitivity}
The proposed approach for optimizing the tunable impendances of an RIS-aided channel is based on the $S$-parameter representation of the multiport network model for RISs. The available optimization frameworks for multiport network models for RISs are, on the other hand, based on the $Z$-parameter representation \cite{DR2}, \cite{ABR}, \cite{SARIS}, \cite{DR3}. \textcolor{black}{Here, we intend to motivate, from the analytical standpoint, the advantages of using the $S$-parameter representation, in lieu of the $Z$-parameter representation, for RIS-aided channels}.

\textcolor{black}{For both representations, the algorithms available in the literature, i.e., \cite{DR2}, \cite{ABR}, \cite{SARIS}, and the proposed Algorithm \ref{Alg:S-UNI} are iterative, and they update the optimization variables through small increment values that are applied to a linearized version of the objective function, which is obtained by invoking Neumann's series approximation. The key difference between the $Z$-parameter and the $S$-parameter representations is that the tunable load impedances are optimized in the former case, while the reflection coefficients are optimized in the latter case. Even though the tunable impedances and the reflection coefficients at the ports of the RIS are related to one another as given in \eqref{Gamma1}, a small increment applied to an impedance and to a reflection coefficient, which are related but different optimization variables, results in a substantially different impact on the speed of convergence of the algorithms}.

To elaborate, let us consider \eqref{Gamma1} by assuming for simplicity, but no loss of generality, $r_0=0$. We obtain \vspace{-0.1cm}
\begin{equation}
{X_k} = \frac{{{Z_0}}}{j}\frac{{1 + {\Gamma _k}}}{{1 - {\Gamma _k}}}\mathop  = \limits^{\left( a \right)} \frac{{{Z_0}}}{j}\frac{{1 + {e^{j{\phi _k}}}}}{{1 - {e^{j{\phi _k}}}}} \vspace{-0.1cm}
\end{equation}
where $(a)$ follows from $\Gamma_k = e^{j\phi_k}$. Therefore, the derivative of ${X_k}$ with respect to $\phi_k$ can be formulated as \vspace{-0.1cm}
\begin{align}
\frac{{d{X_k}}}{{d{\phi _k}}} &= 2{Z_0}\frac{{{e^{j{\phi _k}}}}}{{{{\left( {1 - {e^{j{\phi _k}}}} \right)}^2}}} \mathop  = \limits^{\left( a \right)} 2{Z_0}\frac{{{\Gamma _k}}}{{{{\left( {1 - {\Gamma _k}} \right)}^2}}}\mathop  = \limits^{\left( b \right)}  - \frac{{X_k^2 + Z_0^2}}{{2{Z_0}}} \vspace{-0.1cm}
\end{align}
where $(a)$ follows from $\Gamma_k = e^{j\phi_k}$ and $(b)$ follows from ${\Gamma _k} = {\left( {j{X_k} + {Z_0}} \right)^{ - 1}}\left( {j{X_k} - {Z_0}} \right)$.

We evince, therefore, the following: (i) if we optimize $X_k^{(m)}$ at the $m$th iteration, a small increment $d X_k$ results in a small update of the tunable reactance at the next iteration, i.e., $X_k^{(m+1)} \leftarrow X_k^{(m)} +d X_k$; but (ii) if we optimize $\phi_k^{(m)}$ at the $m$th iteration, a small increment $d \phi_k$ results in a large update of the tunable reactance at the next iteration, i.e.,  \vspace{-0.1cm}
\begin{equation}
X_k^{(m+1)} \leftarrow X_k^{(m)} + {\left( {2{Z_0}} \right)^{ - 1}}\left( {{{\left( {X_k^{\left( m \right)}} \right)}^2} + Z_0^2} \right)d{\phi _k} \vspace{-0.1cm}
\end{equation}
since $X_k^{(m)}$ and $Z_0$ are large values, even though $d{\phi _k}$ is a small increment. Optimizing an RIS based on the $S$-parameter representation is hence expected to provide a faster convergence rate. This is illustrated with numerical results in Sec. IV.

\subsubsection{\textbf{Step-Size Optimization}} \label{opt_step}
The proposed S-UNI algorithm is based on the assumption that the optimization variables at the $m$th iteration, ${{{\boldsymbol{\delta }}^{\left( m \right)}}}$, are updated by applying the same small increment to all of them. In this section, we propose an approach that overcomes this assumption. To this end, we first establish a connection between the maximization of the signal-to-noise-ratio (SNR) in an RIS-aided channel and the minimization of the mean square error (MSE), i.e., with a minimum mean square error (MMSE) equalizer applied to an RIS-aided channel. The departing point is \eqref{Optprob2} by removing the assumption of equal increments. \textcolor{black}{Specifically, considering the general constraint in \eqref{condition} for applying Neumann's series approximation, the optimization problem can be written as} \vspace{-0.1cm}
\begin{align}\label{Optprob2_UneqIncr}
&\max \limits_{{\boldsymbol{\delta}_d^{(m)}}}  \left| a^{(m)} + \left(\mathbf{c}^{(m)}\right)^T \boldsymbol{\delta}_d^{(m)} \right|^2\\
&\text{s.t.} \quad \text{trace}\left({\left( {{\boldsymbol{\delta }}_d^{\left( m \right)}} \right)^T}{{\boldsymbol{\Theta}}_d^{\left( m \right)}}{\boldsymbol{\delta }}_d^{\left( m \right)}\right) \le \bar \delta_{0}^2, \quad k=1,2, \ldots , K \nonumber \vspace{-0.1cm}
\end{align} 
where ${\boldsymbol{\delta }}_d^{\left( m \right)} = {\rm{diag}}\left( {{{\boldsymbol{\delta }}^{\left( m \right)}}} \right)$, ${{\bf{\Theta }}^{\left( m \right)}} = {{\bf{P}}^{\left( m \right)}}{\left( {{{\bf{P}}^{\left( m \right)}}} \right)^H}$, ${\boldsymbol{\Theta }}_d^{\left( m \right)} = {\rm{diag}}\left({\rm{diag}}\left( {{{\boldsymbol{\Theta }}^{\left( m \right)}}} \right)\right)$, $\mathbf{c}^{(m)} = \left(\mathbf{b}_1^{(m)}\right)^T \odot \mathbf{b}_2^{(m)}$, and $\bar \delta_{0}^2 \ll 1$.

Similar to the S-UNI case, we are interested in solving the sub-problem at the $m$th iteration. For ease of writing, we simplify the notation by dropping the iteration index $m$, and by introducing the shorthands $a = a^{(m)}$, ${\bf{c}}={\bf{c}}^{(m)}$, ${\bf{x}} = {{\boldsymbol{\delta}_d^{(m)}}}$, ${\bf{D}} = {\boldsymbol{\Theta}}_d^{\left( m \right)}$, and $\epsilon = \bar \delta_{0}$. Therefore, we obtain the following: \vspace{-0.1cm}
\begin{align}\label{Optprob2_UneqIncrEasyNot}
&\max \limits_{\bf{x}}  f({\bf{x}}) = \left| a + \mathbf{c}^T \bf{x} \right|^2\\
&\text{s.t.} \quad  {\bf{x}}^T {\bf{D}} {\bf{x}} \le \epsilon^2, \quad k=1,2, \ldots , K \nonumber \vspace{-0.1cm}
\end{align} 

The problem in \eqref{Optprob2_UneqIncrEasyNot} is not convex because it aims to maximize a convex objective function. We tackle it as follows. In \eqref{Optprob2_UneqIncrEasyNot}, the signal $h = a + {{\bf{c}}^T}{\bf{x}}$ is the RIS-aided channel. Consider a transmitted symbol $s \sim \mathcal{CN}(0,\sigma_s^2)$, with $\sigma_s^2$ denoting the transmit power. The received signal is \vspace{-0.1cm}
\begin{equation} \label{Eq:EqChannel}
y = hs + n = \left( {a + {{\bf{c}}^T}{\bf{x}}} \right)s + n \vspace{-0.1cm}
\end{equation}
where $n \sim \mathcal{CN}(0,\sigma_n^2)$ is the additive white Gaussian noise at the receiver, with $\sigma_n^2$ denoting the noise power.

The maximization of $f({\bf{x}})$ in \eqref{Optprob2_UneqIncrEasyNot} is hence equivalent to the maximization of the SNR corresponding to \eqref{Eq:EqChannel}, as follows: \vspace{-0.1cm}
\begin{equation} \label{Eq:SNR}
{\rm{SNR}}\left({\bf{x}}\right) = \left( {{{\sigma _s^2} \mathord{\left/
 {\vphantom {{\sigma _s^2} {\sigma _n^2}}} \right.
 \kern-\nulldelimiterspace} {\sigma _n^2}}} \right){\left| {a + {{\bf{c}}^T}{\bf{x}}} \right|^2} = \left( {{{\sigma _s^2} \mathord{\left/
 {\vphantom {{\sigma _s^2} {\sigma _n^2}}} \right.
 \kern-\nulldelimiterspace} {\sigma _n^2}}} \right)f\left( {\bf{x}} \right) \vspace{-0.1cm}
\end{equation}

Given the received signal in \eqref{Eq:EqChannel}, in addition, the corresponding MMSE equalizer is the complex-valued coefficient $w$ that minimizes the following MSE: \vspace{-0.1cm}
\begin{align} \label{Eq:MSE}
{\rm{MSE}}\left( {w;{\bf{x}}} \right) &= {\mathbb{E}}\left\{ {{{\left| {wy - s} \right|}^2}} \right\}\\
& = \sigma _s^2 + {\left| w \right|^2}\left( {\sigma _s^2{{\left| h \right|}^2} + \sigma _n^2} \right) - 2\sigma _s^2{\mathop{\Re}\nolimits} \left\{ {w{h^*}} \right\} \nonumber \vspace{-0.1cm}
\end{align}
where the expectation is computed with respect to the distributions of the transmitted symbol $s$ and receiver noise $n$.

In an RIS-aided channel, the MSE in \eqref{Eq:MSE} needs to be jointly minimized as a function of $w$ and $\bf{x}$. For the time being, let us assume that $\bf{x}$ is kept fixed. Therefore, the minimization of the MSE in \eqref{Eq:MSE} as a function of $w$ leads to \vspace{-0.1cm}
\begin{equation} \label{Eq:wOpt}
{w_{{\rm{opt}}}}\left( {\bf{x}} \right) = \frac{{\sigma _s^2{h^*}}}{{\sigma _s^2{{\left| h \right|}^2} + \sigma _n^2}} = \frac{{\sigma _s^2{{\left( {a + {{\bf{c}}^T}{\bf{x}}} \right)}^*}}}{{\sigma _s^2{{\left| {a + {{\bf{c}}^T}{\bf{x}}} \right|}^2} + \sigma _n^2}} \vspace{-0.1cm}
\end{equation}

Thus, the corresponding MMSE assuming $\bf{x}$ fixed is \vspace{-0.1cm}
\begin{align} \label{Eq:MSEopt}
{\rm{MSE}}\left( {\bf{x}} \right) &= {\rm{MSE}}\left( {{w_{{\rm{opt}}}}\left( {\bf{x}} \right);{\bf{x}}} \right)\\
&\mathop  = \limits^{\left( a \right)} \frac{{\sigma _s^2\sigma _n^2}}{{\sigma _s^2{{\left| {a + {{\bf{c}}^T}{\bf{x}}} \right|}^2} + \sigma _n^2}}\mathop  = \limits^{\left( b \right)} \frac{{\sigma _s^2}}{{{\rm{SNR}}\left( {\bf{x}} \right) + 1}} \nonumber\\
&\mathop  = \limits^{\left( c \right)} \sigma _s^2\left( {1 - {w_{{\rm{opt}}}}\left( {\bf{x}} \right)\left( {a + {{\bf{c}}^T}{\bf{x}}} \right)} \right) \nonumber \vspace{-0.1cm}
\end{align}
where $(a)$ follows from \eqref{Eq:wOpt}, $(b)$ from \eqref{Eq:SNR}, and $(c)$ from \eqref{Eq:wOpt} expresses the MSE as a function of 
${w_{{\rm{opt}}}}\left( {\bf{x}} \right)$.

From the equality $(b)$ in \eqref{Eq:MSEopt}, we evince that maximizing the SNR (given $\bf{x}$) is equivalent to minimizing the MSE (given $\bf{x}$). Therefore, the proposed approach to tackle the maximization problem in \eqref{Optprob2_UneqIncrEasyNot} consists of solving the following optimization problem, which aims at minimizing the MSE: \vspace{-0.1cm}
\begin{align}\label{Optprob2_MSE}
&\min \limits_{w, \bf{x}} \; \;  \sigma _s^2 + {\left| w \right|^2}\left( {\sigma _s^2{{\left| h \left({\bf{x}}\right) \right|}^2} + \sigma _n^2} \right) - 2\sigma _s^2{\mathop{\Re}\nolimits} \left\{ {w{h\left({\bf{x}}\right)^*}} \right\} \\
&\text{s.t.} \quad \sum\nolimits_{k = 1}^K {x_k^2{D_k}} \le \epsilon^2, \quad k=1,2, \ldots , K \nonumber \vspace{-0.1cm}
\end{align} 
where $D_k$ if the $k$th element on the main diagonal of $\bf{D}$.

The problem in \eqref{Optprob2_MSE} is not convex. To tackle it, we propose an alternating optimization method applied to the two optimization variables $w$ and $\bf{x}$. The approach works as follows:

$(1)$ We start from an initial solution of $\bf{x}$, e.g., by choosing a random but feasible initialization point. This solution at the generic $l$th iteration is denoted by ${\bf{x}}^{(l)}$.

$(2)$ Then, we solve the problem in \eqref{Optprob2_MSE} by considering ${\bf{x}} = {\bf{x}}^{(l)}$ as given. The solution is  ${w_{{\rm{opt}}}}\left( {\bf{x}}^{(l)} \right)$ in \eqref{Eq:wOpt}.

$(3)$ Subsequently, we compute ${\rm{MSE}}\left( {{w_{{\rm{opt}}}}\left( {\bf{x}}^{(l)} \right);{\bf{x}}} \right)$ in \eqref{Eq:MSE}. At the $(l+1)$th iteration, therefore, ${{{\bf{x}}^{(l+1)}}}$ is the solution of the following optimization problem: \vspace{-0.1cm}
\begin{align}\label{Optprob2_AO}
&\min \limits_{\bf{x}} \; \; \sigma _s^2 + {\left| {{w_{{\rm{opt}}}}\left( {{{\bf{x}}^{\left( l \right)}}} \right)} \right|^2}\left( {\sigma _s^2{{\left| {a + {{\bf{c}}^T}{\bf{x}}} \right|}^2} + \sigma _n^2} \right) \nonumber \\ & \hspace{0.75cm} - 2\sigma _s^2{\mathop{\Re}\nolimits} \left\{ {{w_{{\rm{opt}}}}\left( {{{\bf{x}}^{\left( l \right)}}} \right){{\left( {a + {{\bf{c}}^T}{\bf{x}}} \right)}^*}} \right\} \\
&\text{s.t.} \quad \sum\nolimits_{k = 1}^K {x_k^2{D_k}} \le \epsilon^2, \quad k=1,2, \ldots , K \nonumber \vspace{-0.1cm}
\end{align} 

In \eqref{Optprob2_AO}, ${{w_{{\rm{opt}}}}\left( {{{\bf{x}}^{\left( {l} \right)}}} \right)}$ is fixed and is not an optimization variable. Only $\bf{x}$ is an optimization variable. Thus, the objective function in \eqref{Optprob2_AO} is convex in $\bf{x}$, the feasible set is convex, and the problem in \eqref{Optprob2_AO} is convex. For example, its unique solution can be obtained by by utilizing the method of Lagrangian multipliers. Specifically, by computing the gradient of the Lagrangian function corresponding to \eqref{Optprob2_AO} and setting it equal to zero, we obtain the following: \vspace{-0.1cm}
\begin{equation} \label{Eq:AOsolution}
{{\bf{x}}^{\left( {l + 1} \right)}} = {\left( {{{\bf{M}}^{\left( l \right)}} + \mu {{\bf{D}}^{\left( l \right)}}} \right)^{ - 1}}{{\bf{m}}^{\left( l \right)}} \vspace{-0.1cm}
\end{equation}
where the matrix ${{\bf{M}}^{\left( l \right)}}$ and the vector ${{\bf{m}}^{\left( l \right)}}$ are given by \vspace{-0.1cm}
\begin{align}
&{{\bf{M}}^{\left( l \right)}} = \sigma _S^2\left| {{w_{{\rm{opt}}}}\left( {{{\bf{x}}^{\left( l \right)}}} \right)} \right|^2{\mathop{\Re}\nolimits} \left\{ {{{\bf{c}}^*}{{\bf{c}}^T}} \right\} \\
& {{\bf{m}}^{\left( l \right)}} = \sigma _S^2\left| {{w_{{\rm{opt}}}}\left( {{{\bf{x}}^{\left( l \right)}}} \right)} \right|^2{\mathop{\Re}\nolimits} \left\{ {{a^*}{{\bf{c}}^T}} \right\} - \sigma _S^2{\mathop{\Re}\nolimits} \left\{ {{w_{{\rm{opt}}}}\left( {{{\bf{x}}^{\left( l \right)}}} \right){{\bf{c}}^H}} \right\} \nonumber \vspace{-0.1cm}
\end{align}
and $\mu$ is a Lagrange multiplier that is computed ensuring that the constraint in \eqref{Optprob2_AO} is fulfilled with equality.

The algorithm just introduced is referred to as S-OPT. The difference with S-UNI in Algorithm \ref{Alg:S-UNI} is that the RIS elements are optimized jointly instead of being optimized one by one. To this end, the line denoted by ($*$) is deleted, and the line denoted by ($**$) is moved out of the loop that optimizes the RIS elements and is replaced by the proposed alternating optimization algorithm, which iteratively computes $ {w_{{\rm{opt}}}}\left( {\bf{x}}^{(l)} \right)$ in \eqref{Eq:MSE} and ${{\bf{x}}^{\left( {l + 1} \right)}}$ in \eqref{Eq:wOpt} until convergence.

\vspace{-0.25cm}
\subsection{Structural Scattering Aware Optimization}\label{opt_step}
In Sec. II-C, we have unveiled that a key difference between the conventional and the multiport network theory models for RIS-aided channels in \eqref{Eq:HCT} and \eqref{Eq:HE2Ematched}, respectively, is the structural scattering in \eqref{Eq:StSc}. Therefore, conventional optimization algorithms ignore the specular reflection due to \eqref{Eq:StSc} by design. This is the case of the two algorithms S-UNI and S-OPT that we have introduced in this paper as well. To the best of the authors' knowledge, there exist no algorithm for RIS-aided communications that optimizes the signal scattered by the RIS while minimizing the structural scattering, which is an undesired effect. In this sub-section, we propose the first approach to this end, which generalizes the S-OPT algorithm.

Using the same notation as in \eqref{Optprob2_UneqIncrEasyNot}, we formulate the following optimization problem to tackle the structural scattering: \vspace{-0.3cm}
\begin{align}\label{Optprob_StSc}
& \mathop {\max }\limits_{\bf{x}} {f_\omega }\left( {\bf{x}} \right) = {\left| {a + {{\bf{c}}^T}{\bf{x}}} \right|^2} - \omega {\left| {{a_{{\rm{StSc}}}} + {\bf{c}}_{{\rm{StSc}}}^T{\bf{x}}} \right|^2}\\
& {\rm{s}}{\rm{.t}}{\rm{.}}\quad  {{{\bf{x}}^T}{\bf{Dx}}} \le {\epsilon ^2},\quad k = 1,2, \ldots ,K \nonumber \vspace{-0.1cm}
\end{align} 
where the term $\omega {\left| {{a_{{\rm{StSc}}}} + {\bf{c}}_{{\rm{StSc}}}^T{\bf{x}}} \right|^2}$ accounts for the structural scattering. 
\textcolor{black}{More specifically, $a_{\rm{StSc}}$ and $\bf{c}_{\rm{StSc}}$ are equivalent to $a$ and $\bf{c}$, respectively, but they represent a signal that is not evaluated at the location of the intended receiver, as $a$ and $\bf{c}$ are. Instead, they correspond to a signal evaluated at the location of a virtual receiver that is deployed in the direction of specular scattering of the RIS and at the same distance as the intended receiver. In other words, we consider an equivalent two-user system, in which one user is the intended receiver and the other user is located in the specular direction and at the same distance from the RIS as the intended receiver. In mathematical terms, $a_{\rm{StSc}}$ and $\bf{c}_{\rm{StSc}}$ can be formulated as $a_{\rm{StSc}} = {S}^{(\rm{StSc})}_{RT} + \mathbf{S}^{(\rm{StSc})}_{RS} (\mathbf{Q}^{(m)})^{-1}\mathbf{S}^{(\rm{StSc})}_{ST}$ and ${\bf{c}}_{\rm{StSc}} = \left({\bf{b}}_1^{(\rm{StSc})}\right)^T \odot {\bf{b}}_{2}^{(\rm{StSc})}$, where ${S}^{(\rm{StSc})}_{RT}$, $\mathbf{S}^{(\rm{StSc})}_{ST}$ and $\mathbf{S}^{(\rm{StSc})}_{RS}$ denote the scattering sub-matrices corresponding to the virtual receiver, as well as $\left(\mathbf{b}_1^{(\rm{StSc})}\right)^T = \mathbf{S}^{(\rm{StSc})}_{ RS }\mathbf{P}^{(m)}$ and $\mathbf{b}_2^{(\rm{StSc})} = (\mathbf{Q}^{(m)})^{-1} \mathbf{S}^{(\rm{StSc})}_{ ST }$. In free space channels, as mentioned, these scattering matrices depend only on the location of the transmitter and RIS, as the location of the receiver is that of specular reflection. Therefore, the parameters $a_{\rm{StSc}}$ and ${\bf{c}}_{\rm{StSc}}$ can be readily computed. Multipath channels are discussed in Sec. III-D. For them, known channel estimation algorithms can be utilized to estimate the scattering parameters \cite{PanZZHWPRRSZZ22}, \cite{zheng2023impact}}.

Based on the optimization problem formulated in \eqref{Optprob_StSc}, the RIS is optimized to maximize the reradiated signal towards the intended user, while minimizing the reradiated signal towards the specular direction. This is obtained by controlling the positive real-valued weighting parameter $\omega$: the larger $\omega$, the lower the scattering towards the specular direction, i.e., the structural scattering is reduced. The weighting factor $\omega$ cannot, however, be too large to make sure that the scattering towards the intended direction is not deprioritized. \textcolor{black}{By setting $\omega = \infty$, in fact, the optimizer would attempt to completely eliminate the signal scattered towards the specular direction while ignoring the signal scattered towards the intended direction. The most appropriate value for $\omega$ can be found numerically, based on the specified performance requirements and tradeoffs. By varying $\omega \in [0, \infty]$, specifically, we obtain the Pareto frontier, i.e., the set of solutions for which no RIS configuration can be found that either maximizes the desired beam or minimizes the structural scattering without making the other worse}.

To tackle the problem in \eqref{Optprob_StSc}, we utilize the same approach as for S-OPT with the caveat that the equivalence between the maximization of the SNR and the minimization of the MSE is applied, as in \eqref{Optprob2_MSE}, only to the intended receiver. Therefore, the following optimization problem is obtained: \vspace{-0.1cm}
\begin{align}\label{Optprob_StSc_MSE}
&\min \limits_{w, \bf{x}} \; \;  \sigma _s^2 + {\left| w \right|^2}\left( {\sigma _s^2{{\left| h \left({\bf{x}}\right) \right|}^2} + \sigma _n^2} \right) - 2\sigma _s^2{\mathop{\Re}\nolimits} \left\{ {w{h\left({\bf{x}}\right)^*}} \right\}\\
& \hspace{0.75cm}- \omega {\left| {{a_{{\rm{StSc}}}}} \right|^2} - 2\omega {\mathop{\Re}\nolimits} \left\{ {a_{{\rm{StSc}}}^*{\bf{c}}_{{\rm{StSc}}}^T{\bf{x}}} \right\} - \omega {{\left| {{\bf{c}}_{{\rm{StSc}}}^T{\bf{x}}} \right|^2}} \nonumber \\
&\text{s.t.} \quad \sum\nolimits_{k = 1}^K {x_k^2{D_k}} \le \epsilon^2, \quad k=1,2, \ldots , K \nonumber \vspace{-0.1cm}
\end{align} 

The optimization problem in \eqref{Optprob_StSc_MSE} can be solved by applying the same alternating optimization algorithm as for \eqref{Optprob2_MSE}. At the $l$th iteration, specifically, ${w_{{\rm{opt}}}}\left( {\bf{x}}^{(l)} \right)$ is given by \eqref{Eq:wOpt}, since the term corresponding to the structural scattering is independent of $w$. Also, ${{\bf{x}}^{\left( {l + 1} \right)}}$ at the $(l+1)$th iteration is obtained by applying the method of Lagrange multiplifiers to the $\omega$-weighted objective function, which yields the following: \vspace{-0.1cm}
\begin{equation} \label{Eq:AOsolution_StSc}
{{\bf{x}}^{\left( {l + 1} \right)}} = {\left( {{\bf{M}}_\omega ^{\left( l \right)} + {\mu _\omega }{{\bf{D}}^{\left( l \right)}}} \right)^{ - 1}}{\bf{m}}_\omega ^{\left( l \right)} \vspace{-0.1cm}
\end{equation}
where the matrix ${\bf{M}}_\omega ^{\left( l \right)}$ and the vector ${\bf{m}}_\omega ^{\left( l \right)}$ are given by \vspace{-0.1cm}
\begin{align}
& {\bf{M}}_\omega ^{\left( l \right)} = \sigma _S^2{\left| {{w_{{\rm{opt}}}}\left( {{{\bf{x}}^{\left( l \right)}}} \right)} \right|^2}{\mathop{\Re}\nolimits} \left\{ {{{\bf{c}}^*}{{\bf{c}}^T}} \right\} - \omega {\mathop{\Re}\nolimits} \left\{ {{\bf{c}}_{{\rm{StSc}}}^*{\bf{c}}_{{\rm{StSc}}}^T} \right\}\\
& {\bf{m}}_\omega ^{\left( l \right)} = \sigma _S^2{\left| {{w_{{\rm{opt}}}}\left( {{{\bf{x}}^{\left( l \right)}}} \right)} \right|^2}{\mathop{\Re}\nolimits} \left\{ {{a^*}{{\bf{c}}^T}} \right\} - \sigma _S^2{\mathop{\Re}\nolimits} \left\{ {{w_{{\rm{opt}}}}\left( {{{\bf{x}}^{\left( l \right)}}} \right){{\bf{c}}^H}} \right\} \nonumber \\
& \hspace{1cm} - \omega {\mathop{\Re}\nolimits} \left\{ {a_{{\rm{StSc}}}^*{\bf{c}}_{{\rm{StSc}}}^T} \right\}  \vspace{-0.1cm}
\end{align}
and $\mu_{\omega}$ is a Lagrange multiplier that is computed ensuring that the constraint in \eqref{Optprob_StSc_MSE} is fulfilled with equality.

In the rest of this paper, the algorithm that accounts for the structural scattering at the design stage is referred to us S-
OPT$(\omega)$. By definition, it reduces to S-OPT if $\omega=0$.

\vspace{-0.25cm}
\subsection{\textcolor{black}{Generalization to a Finite Feasible Set}}
\textcolor{black}{The proposed approach summarized in Algorithm \ref{Alg:S-UNI} assumes that the reactances of the tunable loads connected to the ports of the RIS can take arbitrary values, i.e., ${X_k} \in [-\infty, \infty]$. In turn, this implies that the phases of the reflection coefficients can take arbitrary values, i.e., $\phi_k \in [0,2\pi]$. In practice, however, the reactances can only take values within some specified sets. This imposes some limitations on the set of phase shifts $\phi_k$. We denote by $\mathcal{Q}$ the set of feasible phase shifts. Algorithm \ref{Alg:S-UNI} can be generalized to consider $\mathcal{Q}$, by simply replacing the line denoted by ($***$) with the following: \vspace{-0.45cm}} 
\begin{align}
\textcolor{black}{\phi_k^{(m+1)} \leftarrow \Pi_\mathcal{Q}\left(\phi_k^{(m)} + \delta_k^{(m)}\right) \vspace{-0.1cm}}  
\end{align}\label{projection_fs}
\textcolor{black}{where $y=\Pi_\mathcal{Q}(x)$ denotes the projection of $x$ onto the feasible set $\mathcal{Q}$ according to the minimum distance criterion, i.e., $y = {\Pi _{\mathcal{Q}}}\left( x \right) = \arg {\min _{\tilde x \in {\mathcal{Q}}}}\left\{ {\left| {x - \tilde x} \right|} \right\}$}.

\vspace{-0.25cm}
\subsection{\textcolor{black}{Generalization to Multipath Channels}}
\textcolor{black}{The proposed approach can be applied in multipath channels as well, based on the analytical model recently introduced in \cite{SARIS}. The multipath channel is modeled through the presence of scattering clusters distributed throughout the environment. Each cluster is constituted by several scattering objects, and it is modeled as a multiport network whose ports are loaded with impedances that depend on the electromagnetic properties of the physical scatterers. The authors of \cite{SARIS} model each scattering object as a loaded thin wire dipole according to the so-called discrete dipole approximation \cite{math10173049}.}

\textcolor{black}{Specifically, under the same modeling assumptions and approximations, the end-to-end channel matrix in \eqref{Eq:HZsimpler} can be readily generalized in multipath channels, by simply replacing the matrix ${{\bf{\tilde Z}}}_{RT}$ with the matrix ${{\bf{\tilde Z}}}_{ROT}$ defined as follows: \vspace{-0.1cm}}
\begin{equation} \label{Eq:MultipathZ}
\textcolor{black}{{{{\bf{\tilde Z}}}_{ROT}} = {{\bf{Z}}_{ROT}} - {{\bf{Z}}_{ROS}}{\left( {{{\bf{Z}}_S} + {{{\bf{\mathord{\buildrel{\lower3pt\hbox{$\scriptscriptstyle\smile$}} 
\over Z} }}}_{SOS}}} \right)^{ - 1}}{{\bf{Z}}_{SOT}} \vspace{-0.1cm}} 
\end{equation}
\textcolor{black}{where ${{{\bf{\mathord{\buildrel{\lower3pt\hbox{$\scriptscriptstyle\smile$}} 
\over Z} }}}_{SOS}} = {{\bf{Z}}_{SS}} + {{\bf{Z}}_{SOS}}$, and the matrices ${{\bf{Z}}_{xOy}}$ account for the presence of the scattering objects besides the transmitter, the receiver, and the RIS. These matrices are defined in \cite{SARIS} and, therefore, are not reported here for brevity.}

\textcolor{black}{Comparing \eqref{Eq:MultipathZ} with \eqref{Eq:HZsimpler}, we evince that they are equivalent from point of view of optimizing the matrix ${{{\bf{Z}}_S}}$ of tunable loads, as the matrix ${{{\bf{\mathord{\buildrel{\lower3pt\hbox{$\scriptscriptstyle\smile$}} 
\over Z} }}}_{SOS}}$ is fixed and known. Thus, Algorithm \ref{Alg:S-UNI} can be applied in multipath channels \textit{mutatis mutandis}.}

\textcolor{black}{It is instructive to expand \eqref{Eq:MultipathZ} in order to identify the different constituents of the end-to-end channel, which include the direct link, first-order scattering components (i.e., anomalous reflection, structural scattering), and higher-order scattering components caused by the interactions between the RIS and the scattering clusters. For brevity, we analyze the scattering matrices in \eqref{Eq:S_ST}-\eqref{Eq:StructScatt} with focus on \eqref{Eq:SZ_RT}, which is of interest in this paper. By inserting the matrices ${{\bf{Z}}_{xOy}}$ defined in \cite{SARIS} into \eqref{Eq:SZ_RT}, we obtain the following ${{\bf{S}}_{ROT}}$ matrix: \vspace{-0.1cm}}
\begin{align} 
& \textcolor{black}{{{\bf{S}}_{ROT}} = \frac{{{{\bf{Z}}_{ROT}}}}{{2{Z_0}}} - \frac{{{{\bf{Z}}_{ROS}}}}{{2{Z_0}}}{\left( {{{\bf{Z}}_0}{\bf{U}} + {{\bf{Z}}_{SS}} + {{\bf{Z}}_{SOS}}} \right)^{ - 1}}{{\bf{Z}}_{SOT}}}\\
& \textcolor{black}{= \frac{1}{{2{Z_0}}}\left( {{{\bf{Z}}_{RT}} - {{\bf{Z}}_{RO}}{{\left( {{{\bf{Z}}_{US}} + {{\bf{Z}}_{OO}}} \right)}^{ - 1}}{{\bf{Z}}_{OT}}} \right)}\\
& \textcolor{black}{- \frac{1}{{2{Z_0}}}\left( {{{\bf{Z}}_{RS}} - {{\bf{Z}}_{RO}}{{\left( {{{\bf{Z}}_{US}} + {{\bf{Z}}_{OO}}} \right)}^{ - 1}}{{\bf{Z}}_{OS}}} \right)}\\
& \textcolor{black}{\times {\left( {{{\bf{Z}}_0}{\bf{U}} + {{\bf{Z}}_{SS}} - {{\bf{Z}}_{SO}}{{\left( {{{\bf{Z}}_{US}} + {{\bf{Z}}_{OO}}} \right)}^{ - 1}}{{\bf{Z}}_{OS}}} \right)^{ - 1}}}\\
& \textcolor{black}{\times \left( {{{\bf{Z}}_{ST}} - {{\bf{Z}}_{SO}}{{\left( {{{\bf{Z}}_{US}} + {{\bf{Z}}_{OO}}} \right)}^{ - 1}}{{\bf{Z}}_{OT}}} \right)}\\
& \textcolor{black}{= \frac{{{{\bf{Z}}_{RT}}}}{{2{Z_0}}}}  \label{Eq:Mult1}\\
 & \textcolor{black}{- \frac{{{{\bf{Z}}_{RO}}{{\left( {{{\bf{Z}}_{US}} + {{\bf{Z}}_{OO}}} \right)}^{ - 1}}{{\bf{Z}}_{OT}}}}{{2{Z_0}}}} \label{Eq:Mult2} \\
& \textcolor{black}{- \frac{{{{\bf{Z}}_{RS}}{{\left( {{{\bf{Z}}_0}{\bf{U}} + {{\bf{Z}}_{SS}} - {{\bf{Z}}_{SO}}{{\left( {{{\bf{Z}}_{US}} + {{\bf{Z}}_{OO}}} \right)}^{ - 1}}{{\bf{Z}}_{OS}}} \right)}^{ - 1}}{{\bf{Z}}_{ST}}}}{{2{Z_0}}}} \label{Eq:Mult3} \\
&  \textcolor{black}{+ \boldsymbol{\Delta}} \label{Eq:Mult4}  \vspace{-0.75cm}
\end{align}
\textcolor{black}{where ${{{\bf{Z}}_{US}}}$ is a diagonal matrix containing the loads of the scattering objects that are set to mimic the scattering behavior of natural objects. In detail, \eqref{Eq:Mult1} corresponds to the direct link, \eqref{Eq:Mult2} corresponds to the first-order scattering component from the scattering objects in the absence of the RIS, and \eqref{Eq:Mult3} corresponds to the first-order structural scattering component from the RIS. Specifically, we note that the term in \eqref{Eq:Mult3} results in a specular reflection as in free space channels, with the only caveat that the presence of scattering objects is equivalent to modifying the matrix of mutual coupling of the RIS from ${{{\bf{Z}}_{SS}}}$ to ${{{\bf{Z}}_{SS}} - {{\bf{Z}}_{SO}}{{\left( {{{\bf{Z}}_{US}} + {{\bf{Z}}_{OO}}} \right)}^{ - 1}}{{\bf{Z}}_{OS}}}$. Finally, the term $\boldsymbol \Delta$ in \eqref{Eq:Mult4} accounts for all the second-order and higher-order scattering components that originate from the interaction of the RIS with the scattering objects. Let us consider, as an example, the term $\boldsymbol{\Delta}_1 = {{\bf{Z}}_{RS}} {\left( {{{\bf{Z}}_0}{\bf{U}} + {{\bf{Z}}_{SS}} - {{\bf{Z}}_{SO}}{{\left( {{{\bf{Z}}_{US}} + {{\bf{Z}}_{OO}}} \right)}^{ - 1}}{{\bf{Z}}_{OS}}} \right)^{ - 1}}$ $\times{{\bf{Z}}_{SO}}{\left( {{{\bf{Z}}_{US}} + {{\bf{Z}}_{OO}}} \right)^{ - 1}}{{\bf{Z}}_{OT}}$. $\boldsymbol{\Delta}_1$ corresponds to the signal that is first emitted by the transmitter, is then scattered by the clusters of scattering objects, and is further reflected by the RIS towards a direction that is specular to that of incidence}.

\textcolor{black}{In general, the analysis of ${{\bf{S}}_{ROT}}$ reveals that, as in free space channels, the waves that impinge upon the RIS are reflected specularly, whether they originate from the transmitter or are the result of prior interactions with the scattering objects. As in free space channels, each scattering component can be identified and can be either nullified or enhanced depending on the considered scenario, by formulating a multi-objective optimization problem similar to that in Sec. III-B}.

\vspace{-0.25cm}
\subsection{Analysis of Computational Complexity and Convergence}
In this sub-section, we analyze the computational complexity and the convergence of the proposed algorithms.

\textbf{Computational complexity} -- The three proposed algorithms are iterative and hence their complexity ultimately depends on the number of iterations to converge. Based on Algorithm \ref{Alg:S-UNI} and the alternating optimization in \eqref{Eq:AOsolution}, we evince the following: (i) the number of iterations for S-UNI is given by the product of the number of iterations of the \texttt{while} loop and the number of RIS elements $K$; and (ii) the number of iterations for S-OPT and S-OPT$(\omega)$ is given by the product of the number of iterations of the \texttt{while} loop and the number of iterations of the alternating optimization algorithm in \eqref{Eq:AOsolution}. In the following, we consider the computational complexity expressed in terms of complex multiplications per iteration.

From Algorithm \ref{Alg:S-UNI}, we evince that the per-iteration complexity of S-UNI is determined by the inversion of the $K \times K$ matrix $\mathbf{Q}^{(m)}$, whose complexity is $\mathcal{O}(K^3)$. It is important to note that the number of times $\mathbf{Q}^{(m)}$ needs to be inverted is equal to the number of iterations of the \texttt{while} loop only. This is similar to the complexity of the algorithm proposed in \cite{DR2}, which is based on the $Z$-parameter representation. As far as S-OPT and S-OPT$(\omega)$ are concerned, the per-iteration complexity is determined by the inversion of the $K \times K$ matrices in \eqref{Eq:AOsolution} and \eqref{Eq:AOsolution_StSc}, respectively, as well as the computation of the Lagrange multipliers. The Lagrange multipliers can be computed by utilizing the bisection method. Assuming that $N_\mu$ iterations are needed for the bisection method to converge, the complexity of S-OPT and S-OPT$(\omega)$ scales with $\mathcal{O}(N_\mu K^3)$.

\begin{figure}[t!]
\centering
\includegraphics[width=0.65\columnwidth]{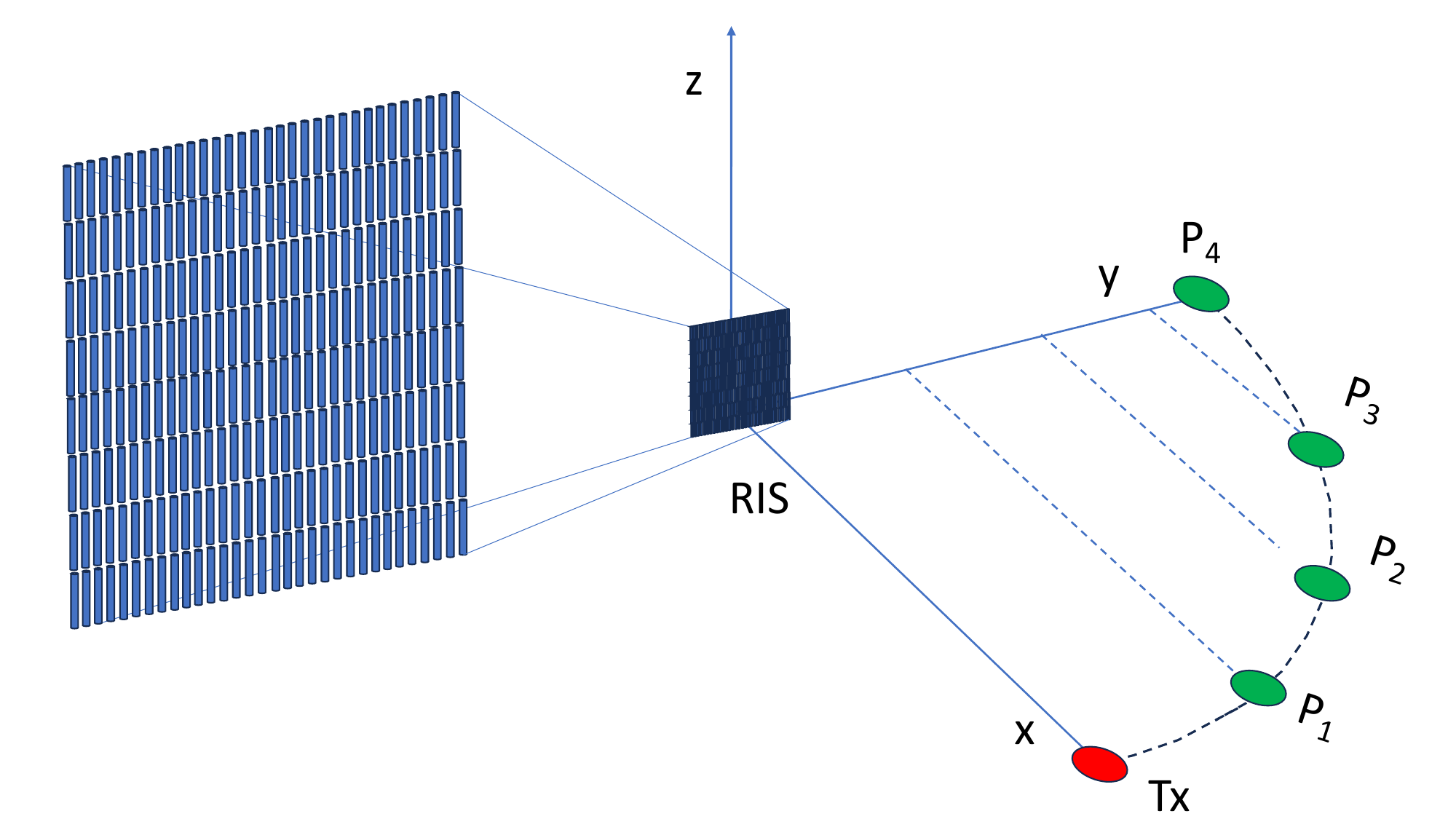}
\caption{Considered scenario. The RIS elements can be either free standing thin wire dipoles (as illustrated in the figure) or rectangular patches on a grounded substrate (detailed next and illustrated in Fig. \ref{fig:7x7Ris}). \textcolor{black}{The scenario in the presence of blocking objects is presented in Fig. \ref{fig:shield_no_shield_Setup} and is elaborated next}.}
\label{Scenario}
\vspace{-0.7cm}
\end{figure}

\textbf{Convergence} -- As far as the convergence of the three proposed algorithms is concerned, they provide optimal solutions at each iteration step\footnote{\textcolor{black}{If the feasible set is a finite set as in Sec. III-C, the solution is not optimal.}}, including the alternating optimization steps for S-OPT and S-OPT$(\omega)$. Therefore, the objective functions monotonically increase (S-UNI) and decrease (S-OPT and S-OPT$(\omega)$) at each iteration step. Since, in addition, the transmitted power is upper bounded, the RIS does not amplify the incident signals, and the MSE is lower bounded, the three proposed algorithms are guaranteed to converge to a local optimum after a sufficient number of iterations.

\vspace{-0.25cm}
\section{Numerical Results}\label{sec:4}
In this section, we illustrate numerical results to evaluate the performance and effectiveness of the proposed algorithms. 
\paragraph{\textcolor{black}{Simulation Setup}}\textcolor{black}{The carrier frequency is $f = 28$ GHz. The RIS is modeled as a uniform planar array centered at the origin of a global reference system and is deployed on the $yz$ plane. The number of RIS elements along the $y$-axis and $z$-axis is denoted by $N_y$ and $N_z$, respectively. The transmitter is located at $(4,0,0)$ m and the receiver is located at $(4 \cos(\psi), 4 \sin(\psi), 0)$ m, i.e., it is on a circle of radius $4$ m and whose center coincides with the center-point of the RIS. We consider four possible locations for the receiver, which are denoted by $\psi_k = \sin^{-1}(k/4)$, with $k = 1,\ldots,4$, and correspond to the locations $P_k$ in Fig. \ref{Scenario}. The transmitting and receiving antennas are modeled as identical $z$-oriented metallic thin wire dipoles with radius $\lambda/500$ and length $0.46 \lambda$, where $\lambda$ denotes the free space wavelength. The length of the dipoles is chosen to have nearly resonant scattering elements whose self-impedance has a low reactance. Two types of RIS elements (unit cells) are considered: (a) free-standing metallic thin wire dipoles and (b) rectangular patch antennas over a grounded substrate. Free space and multipath channels are considered. The direct link between the transmitter and receiver is assumed to be blocked by obstacles and is ignored. This is considered to better emphasize the impact of the RIS and is detailed next.} 
\paragraph{\textcolor{black}{Full-Wave Simulations}}\textcolor{black}{To validate the multiport network model, we utilize the commercial full-wave simulator FEKO\textsuperscript{\textregistered}\footnote{https://altairengineering.fr/feko/.} based on the MoM. Given the considered unit cell structure for the RIS elements, the $S$-matrix and $Z$-matrix of the multiport network model are obtained from the full-wave simulator. In the considered simulation environment, the electromagnetic field at any location is given by the sum of two contributions: (a) the direct link from the transmitter, and (b) the scattered field from the RIS. Therefore, the results based on the full-wave simulations assume a free space scenario. The analysis in multipath channels is discussed next.} 

\textcolor{black}{To focus on the contribution from the RIS, the direct link between the transmitter and receiver is assumed to be obstructed by an obstacle and is then nullified. To make the full-wave simulations computationally affordable, we do not add physical large blocking objects, which obstruct the transmitter-receiver line-of-sight link, in the full-wave simulation environment. We leverage, on the other hand, the superposition principle and utilize the following procedure: (i) we compute the $Z$-matrix in the presence of the RIS; (ii) we compute the $Z$-matrix (a $2 \times 2$ matrix) in the absence of the RIS; and (iii) then, we subtract the off-diagonal entry of the $Z$-matrix obtained in (ii) from the entry corresponding to the transmitter-receiver direct link in the $Z$-matrix obtained in (i).} \textcolor{black}{The theoretical justification of this approach is given in Appendix C. To further validate the proposed approach for modeling the blocking of the direct link in a computationally efficient manner, in addition, we present full-wave simulations in the presence of two physical absorbing screens (shields), which are placed near the transmitter and receiver, that attenuate the direct link. In the following text, these two methods are referred to as ``No Shield'' method and ``With Shield'' method, respectively}.

\textcolor{black}{To validate the effectiveness of the proposed optimization algorithms, the results are obtained by using the following procedure: (i) first, the $Z$-matrix is obtained from the full-wave simulator; (ii) then, the $Z$-matrix is input to the proposed algorithms, and the optimal load impedances are obtained; (iii) finally, the RIS elements are loaded with the obtained optimal impedances and electromagnetic simulations of the scattered field are carried out with the full-wave simulator.}

\vspace{-0.25cm}
\subsection{\textcolor{black}{RIS Made of Thin Wire Dipoles}}\label{sec:RisDipole}
\textcolor{black}{The RIS elements are free-standing metallic thin wire dipoles, which are parallel to the $z$-axis and have the same radius and length as the transmitter and receiver. The inter-distance between adjacent RIS elements is $d_y = \lambda/Q$ along the $y$-direction and $d_z = 3\lambda/4$ along the $z$-direction, where $Q$ is an integer that determines the density of dipoles. We set $N_y = 4 \lambda / d_y = 4Q$ and {$N_z = 8$}, so that the RIS is constituted by {$32Q$} thin wire dipoles, and is a rectangular surface with area equal to {$24 \lambda^2$}. As an example, Fig. \ref{Scenario} shows a deployment with $Q = 8$, i.e., the RIS is constituted by {256} dipoles.} \textcolor{black}{The parasitic resistance of each RIS element is $r_0 = 0.2 \; \Omega$, while the parameter $\delta_0$, which ensures the accuracy of Neumann's series approximation, is set to $\delta_0 = 0.01$ for ensuring a good tradeoff between convergence speed and approximation accuracy}.

\begin{figure}[t!]
	\centering
    \includegraphics[width=0.75\linewidth]{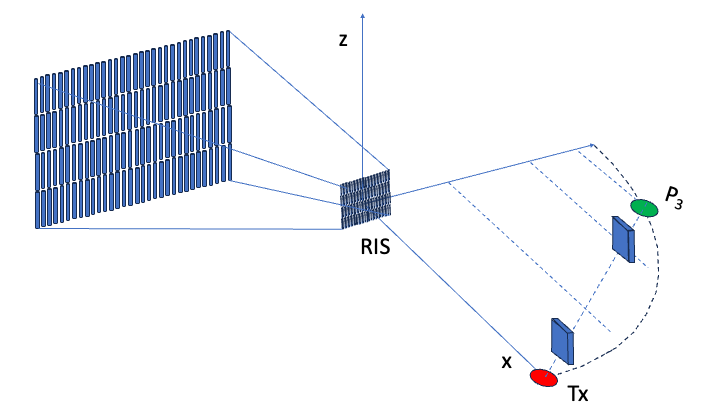} \vspace{-0.1cm} \caption{Considered scenario in the presence of blocking objects (shields) near the transmitter and receiver (position $P_3$).}
    \label{fig:shield_no_shield_Setup} \vspace{-0.5cm}
\end{figure}

\begin{figure}[t!]
	\centering
    \includegraphics[width=0.8\linewidth]{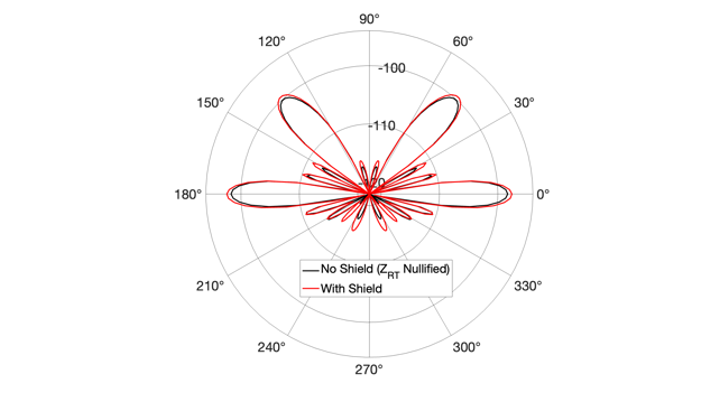} \vspace{-0.1cm} \caption{Scattered field after RIS optimization based on the ``No Shield'' ($Z_{RT}$ is nullified) and ``With Shield'' methods.}
    \label{fig:shield_no_shield_Diagram} \vspace{-1.25cm}
\end{figure}

\paragraph{\textcolor{black}{Comparison Between ``No Shield'' and ``With Shield'' Full-Wave Simulation Methods}} \textcolor{black}{Since the scenario of interest assumes that the direct link is blocked by obstacles and is hence ignored, we commence by comparing the computationally efficient ``No Shield'' method, in which the direct link is canceled out analytically, against the physics-consistent ``With Shield'' method, in which the whole physical structure is simulated in the presence of absorbing objects that obstruct the direct link. The scenario is shown in Fig. \ref{fig:shield_no_shield_Setup}. A free space setting is considered, with two blocking objects deployed near the transmitter and receiver. They are modeled as lossy dielectric screens with a relative permittivity equal to $\varepsilon_r=1.76$ and a dielectric loss tangent equal to $\tan \delta=5.68$, to significantly attenuate the direct link. In the full-wave simulator, the two screens are square cuboids of dimensions $10\lambda\times 10\lambda$ with thickness $t=\sqrt{2}\lambda$. The screens are positioned at a distance of $10\lambda$ from the transmitter and receiver, and they are located normally to the transmitter-receiver line-of-sight path}. 

\textcolor{black}{In Fig. \ref{fig:shield_no_shield_Diagram}, we illustrate the reradiation (scattering) pattern of the RIS, when it is illuminated by a plane wave that originates from the transmitter (normal incidence), and the RIS needs to reflect the incident plane wave towards $P_3$ ($\approx 48.6^{\circ}$), as shown in Fig. \ref{fig:shield_no_shield_Setup}. The figure showcases that the radiation patterns obtained (i) in the presence of the two shields and (ii) in their absence but canceling out the direct link are very similar to one another. This confirms the mathematical foundation of the ``No Shield'' method as summarized in Appendix C, and justifies the use of the ``No Shield'' method in light of its favorable computational efficiency.  In the reminder of this paper, only the ``No Shield'' method is, therefore, considered}.

\paragraph{\textcolor{black}{Free Space Scenario}}\textcolor{black}{We start by analyzing the performance of the proposed algorithms in a free space scenario. The $S$-matrix and $Z$-matrix are obtained by using the full-wave simulator as detailed previously. Also, the feasible set for the phase shifts of the RIS elements is $\mathcal{Q} = [0,2\pi]$. The three algorithms S-UNI, S-OPT, and S-OPT$(\omega)$ are initialized to the optimal solution in the absence of mutual coupling. As a benchmark, the optimization algorithm in \cite{DR2}, which is based on the $Z$-matrix, is considered. This algorithm is referred to as Z-REF. The comparison is given in terms of received power $G= {\left| {{{\bf{{\rm H}}}_{{\rm{e2e}}}}} \right|^2}$ evaluated at the location of the receiver.}

\begin{figure}[t!]
	\centering
\includegraphics[trim={3cm 1cm 4cm 1cm},clip,width=0.8\columnwidth]{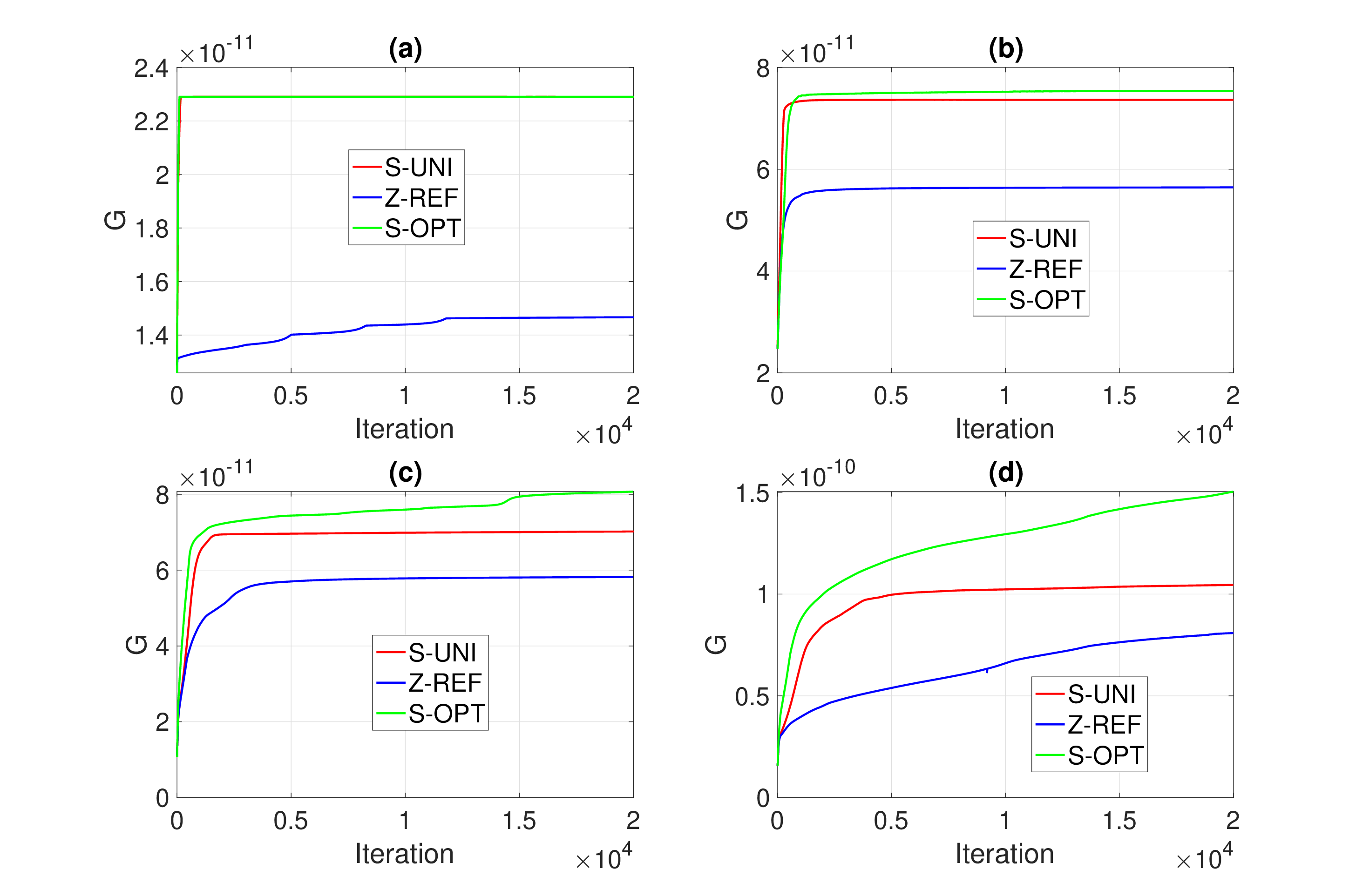}
\caption{Full-wave simulations (the receiver is in $P_4$). Setup: (a) $d_y = \lambda/2$; (b) $d_y = \lambda/4$; (c) $d_y = \lambda/8$; (d) $d_y = \lambda/16$.}
\label{fig1}
\vspace{-1.3cm}
\end{figure}

\begin{figure}[t!]
	\centering
\includegraphics[trim={1.5cm 1cm 3cm 1cm},clip,width=0.8\columnwidth]{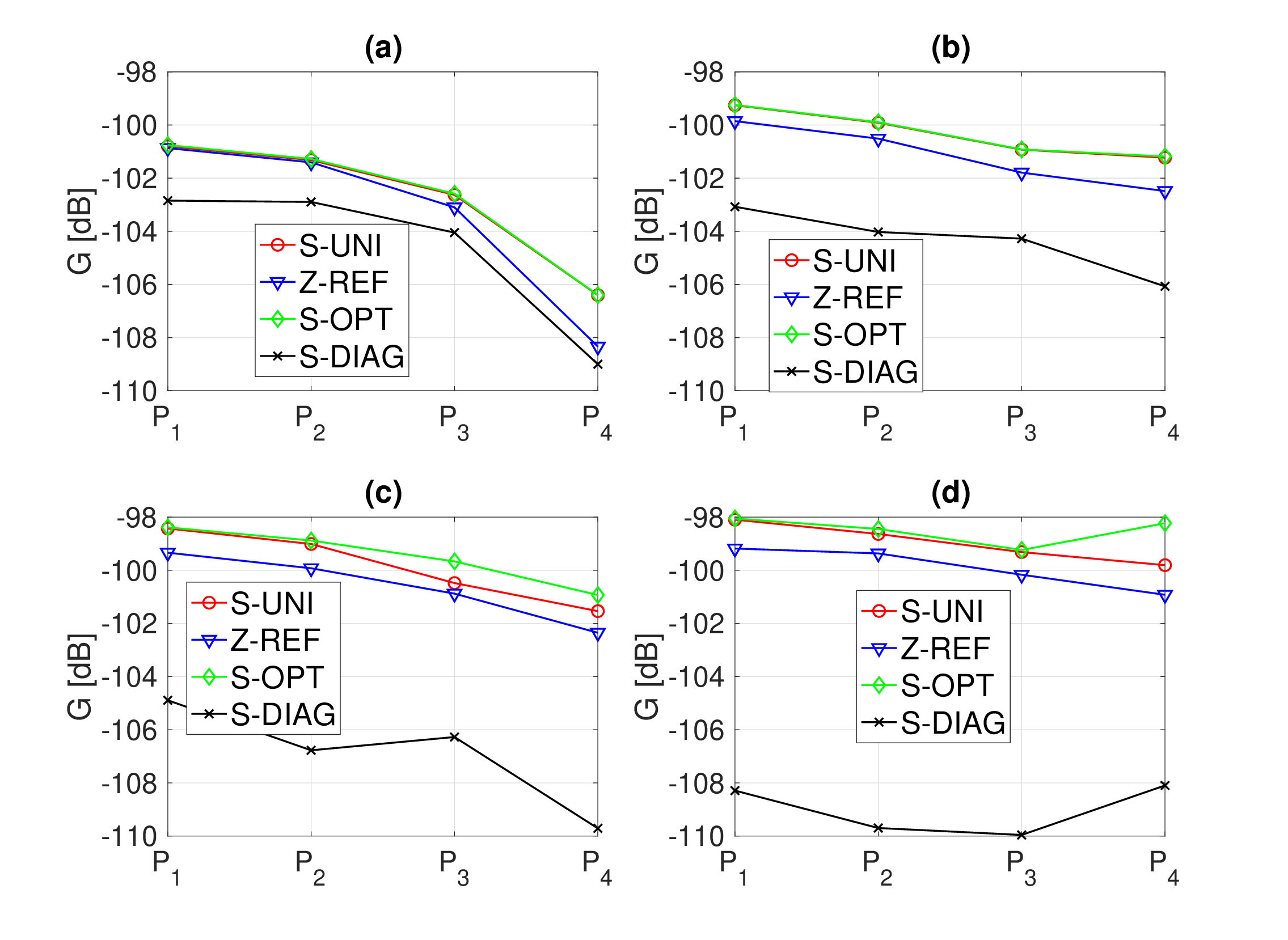}
\vspace{-0.1cm}
\caption{Full-wave simulations (the receiver is in $P_1$-$P_4$). Setup: (a) $d_y = \lambda/2$; (b) $d_y = \lambda/4$; (c) $d_y = \lambda/8$; (d) $d_y = \lambda/16$.}
\label{fig2}
\vspace{-0.25cm}
\end{figure}

In Fig. \ref{fig1}, we show the convergence of the three algorithms based on the scattering parameters obtained from the full-wave simulator. The results confirm the sensitivity analysis in Sec. III-A, i.e. the faster convergence rate of the optimization algorithms applied to the $S$-matrix. Also, we see that the smaller $d_y$, i.e., the more the RIS elements, the better the receiver power, provided that the mutual coupling is exploited at the design stage. Also, we see that the proposed S-OPT algorithm outperforms all the other algorithms.

In Fig. \ref{fig2}, we show the received power (at convergence) as a function of the location of the receiver. Also, we report the received power when the mutual coupling is ignored. This case study is denoted by S-DIAG. We see that ignoring the mutual coupling leads to a large degradation of the received power. 

The numerical results illustrated in Figs. \ref{fig1} and \ref{fig2} based on a full-wave simulator confirm the performance trends first obtained in \cite{DR2} based on a pure mathematical approach. \textcolor{black}{We note, in addition, that the direct link is absent in the considered scenario and that no communication would be possible without the deployment of an RIS. The received power can be further increased by increasing the number of RIS elements.}.

\begin{figure}[t!]
\centering
\includegraphics[width=0.9\columnwidth]{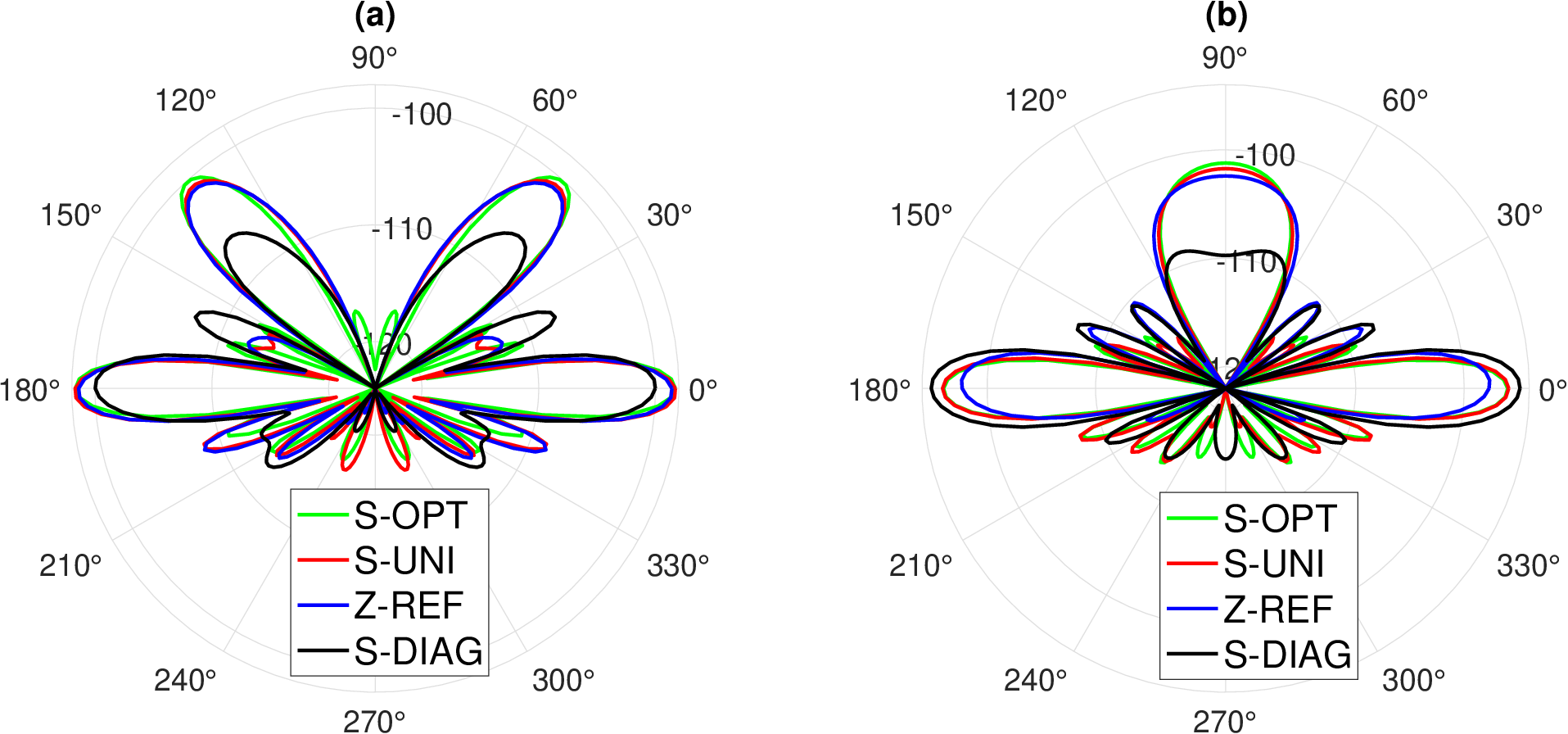}
\vspace{-0.1cm}
\caption{Scattered field in $P_3$ (a) and $P_4$ (b) ($d_y = \lambda/8$).}
\label{Diagram_all}
\vspace{-0.8cm}
\end{figure}
\begin{figure}[t!]
	\centering
\includegraphics[width=0.5\columnwidth]{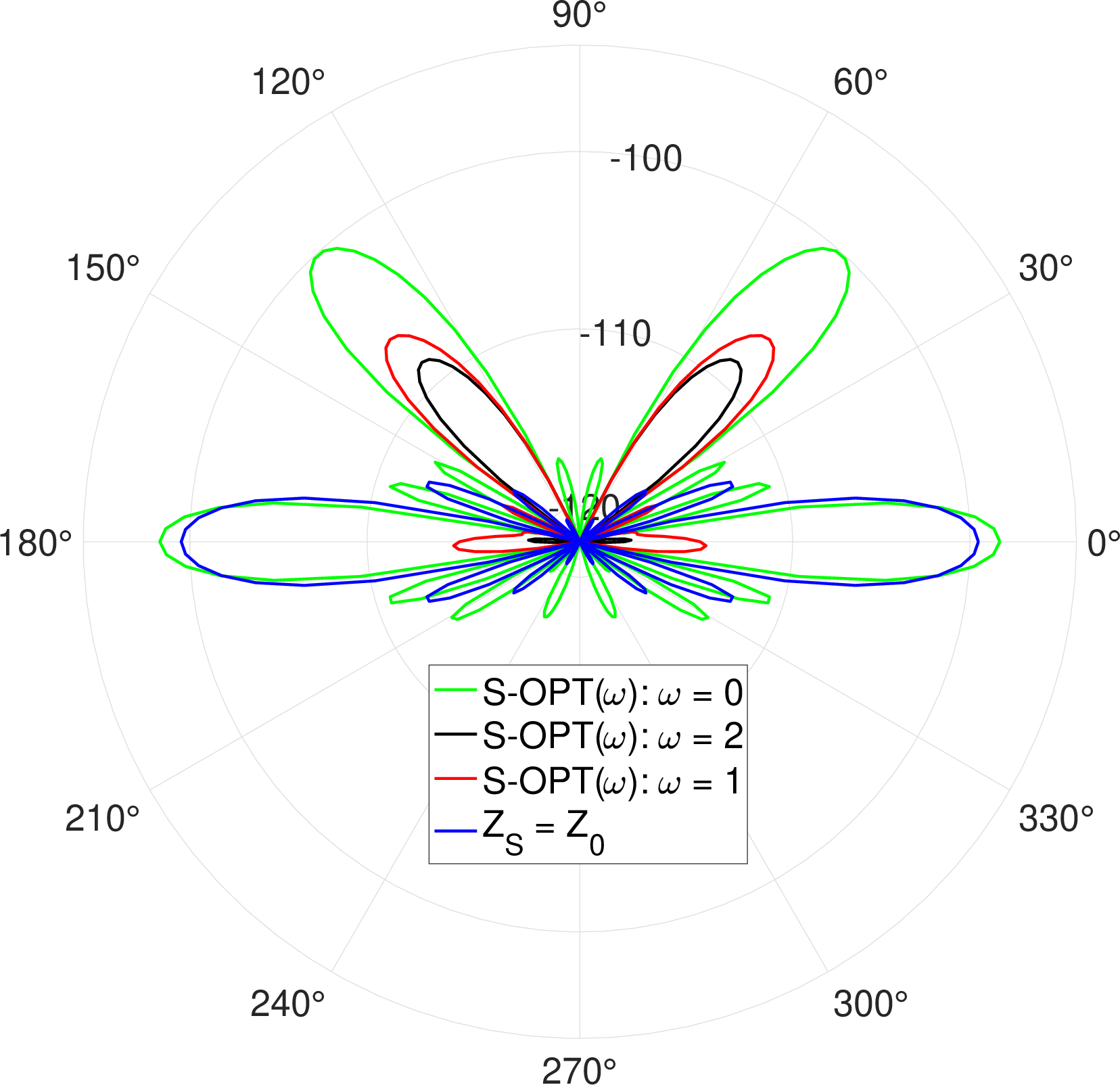}
\vspace{-0.1cm}
\caption{Scattered field in location $P_3$ obtained with S-OPT$(\omega)$ as a function of $\omega$. Setup: $d_y = \lambda/8$.}
	\label{Diagram_omega}
\vspace{-1.0cm}
\end{figure}

In Fig. \ref{Diagram_all}, we analyze the impact of structural scattering. To this end, we illustrate the reradiation (scattering) pattern of the RIS, when it is illuminated by a plane wave that originates from the transmitter (normal incidence), and the RIS reflects the wave towards $P_3$ ($\approx 48.6^{\circ}$) and $P_4$ ($=90^{\circ}$), as shown in Fig. \ref{Scenario}. The figure confirms the presence of a strong specular component towards the transmitter, i.e., towards $0^{\circ}$.

In Fig. \ref{Diagram_omega}, we analyze the effectiveness of the proposed S-OPT$(\omega)$ algorithm, which accounts for the presence of structural scattering at the design stage, and aims at minimizing it while ensuring a sufficient amount of reflected power towards the location of interest. We see that, by appropriately setting the weighing parameter $\omega$, a good trade-off between the structural scattering (specular reflection) and the desired reflection is obtained. Specifically: (i) if $\omega = 0$, i.e., the structural scattering is ignored at the design stage, there is a strong specular reflection; but (ii) if $\omega \ne 0$, i.e., the structural scattering is taken into account at the design stage, the specular reflection is significantly reduced, while a strong non-specular (desired) reflection is obtained. If $\omega = 2$, for example, the specular component is attenuated by 20 dB while the desired beam is attenuated only by 7 dB. For completeness, Fig. \ref{Diagram_omega} reports the reradiation pattern obtained when the ports of the RIS are connected to the reference impedance $Z_0$, i.e., ${\bf{Z}}_S = Z_0 \bf{U}$. As predicted by the proposed multiport network model, only the structural scattering exists but there is no anomalous reflection, since the RIS is a homogeneous surface.

\paragraph{\textcolor{black}{Finite Feasible Set}} \textcolor{black}{To further validate the proposed algorithms, we analyze their performance under a more realistic assumption for the phase shifts of the RIS elements. We still assume a free space scenario, and the $S$-matrix and $Z$-matrix are obtained by using the full-wave simulator.}

\textcolor{black}{Specifically, the feasible set for the allowed phase shifts of the RIS elements is set to $\mathcal{Q}_s=[-36^{\circ}, 36^{\circ}]$. This setup is obtained by considering that the RIS elements are loaded with a varactor diode whose controllable capacitance lies in the range $[C_{\text{min}}, C_{\text{max}}]$. Neglecting the parasitic effects, the corresponding reactances lie in the set $X_C \in [-{j}/(2\pi f C_{\text{max}}), -{j}/{(2\pi f C_{\text{min}})}]$ $\Omega$. Considering, for example, a MACOM's MAVR-011020-1141 gallium arsenide flip chip hyperabrupt varactor diode \cite{Macom}, \cite{Verho2024}, the minimum and maximum values for the capacitance are $C_{\text{min}} = 0.025$ pF and $C_{\text{max}} = 0.19$ pF, respectively, by assuming a reverse bias voltage varying between $11\ V$ and $0\ V$. Hence, at the considered frequency $f = 28$ GHz, we obtain $X_C \in [\approx-227, \approx-30]$ $\Omega$. Then, assuming $Z_0 = 50$ $\Omega$, the phase shifts of the reflection coefficients lie in the range $[\approx-118^{\circ}, \approx-24^{\circ}]$. Also, to ensure that the allowed phase shifts are symmetric around the origin, we connect each varactor diode to a $50 \ \Omega$ transmission line of length $0.1\lambda$, which introduces a fixed phase shift equal to $\approx 71.5^{\circ}$. Accordingly, the set of allowed phase shifts is $\mathcal{Q}_m \approx [-46.7^{\circ}, 46.7^{\circ}]$, which includes the set of phase shifts considered to obtain the simulation results.}

\begin{figure}[t!]
\centering
\includegraphics[width=0.9\columnwidth]{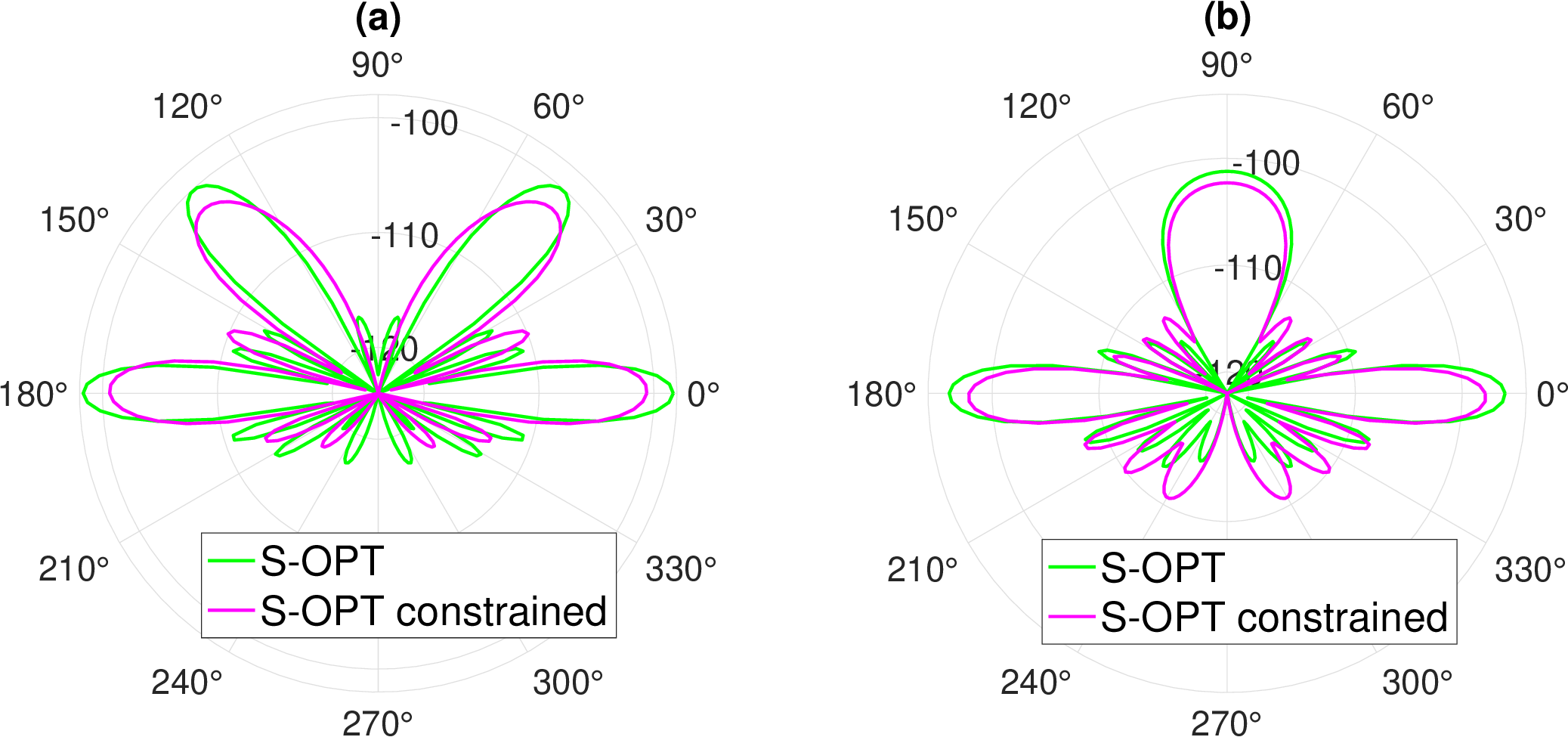}
\vspace{-0.1cm}
\caption{Scattered field in $P_3$ (a) and $P_4$ (b) ($d_y = \lambda/8$).}
\label{Diagram_all2}
\vspace{-0.3cm}
\end{figure}
\begin{figure}[t!]
	\centering
\includegraphics[width=0.5\columnwidth]{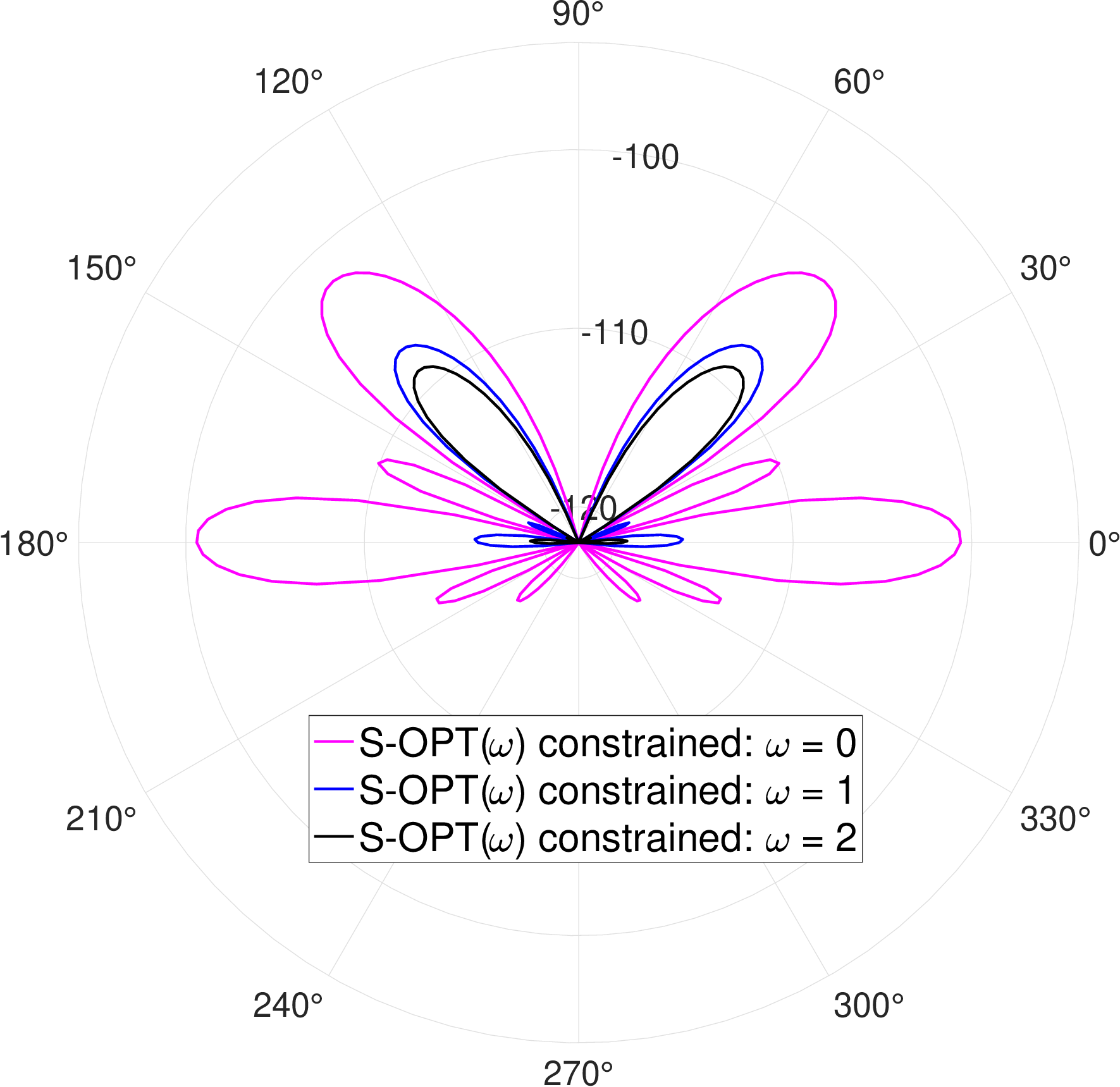}
\vspace{-0.1cm}
\caption{Scattered field in location $P_4$ obtained with S-OPT$(\omega)$ as a function of $\omega$. Setup: $d_y = \lambda/8$.}
	\label{Diagram_omega2}
\vspace{-0.5cm}
\end{figure}
\textcolor{black}{In Fig. \ref{Diagram_all2}, we compare the performance provided by the S-OPT algorithm when $\mathcal{Q}_s=[0, 2 \pi]$ (denoted by S-OPT) and when $\mathcal{Q}_s=[-36^{\circ}, 36^{\circ}]$ (denoted by S-OPT constrained). The same setup as in Fig. \ref{Diagram_all} is considered. It is observed that the performance loss due to the considered finite feasible set is on the order of 1 dB. More importantly, we note that the shape of the radiation pattern is not significantly altered.}

\textcolor{black}{In Fig. \ref{Diagram_omega2}, we study the impact of considering a finite feasible set in terms of mitigating the structural scattering. The same setup as in Fig. \ref{Diagram_omega} is considered. It is observed that the structural scattering can still be reduced, while minimally altering the whole shape of the reradiation pattern of the RIS.}

\paragraph{\textcolor{black}{Multipath Channels}}\textcolor{black}{To validate the proposed algorithms in a realistic wireless scenario, we analyze their performance in a multipath channel. For simplicity, the $S$-matrix and $Z$-matrix are obtained by using the analytical framework in \cite{SARIS}, as described in Sec. III-D. This case study also allows us to investigate whether the analytical models in \cite{DR1} and \cite{10133771} provide results that are consistent with those obtained with the full-wave simulator.
The feasible set for the phase shifts of the RIS elements is assumed to be $\mathcal{Q} = [0,2\pi]$.} 
\begin{figure}[t!]
\centering
\includegraphics[trim={5cm 0cm 5cm 0cm},clip,width=0.8\columnwidth]{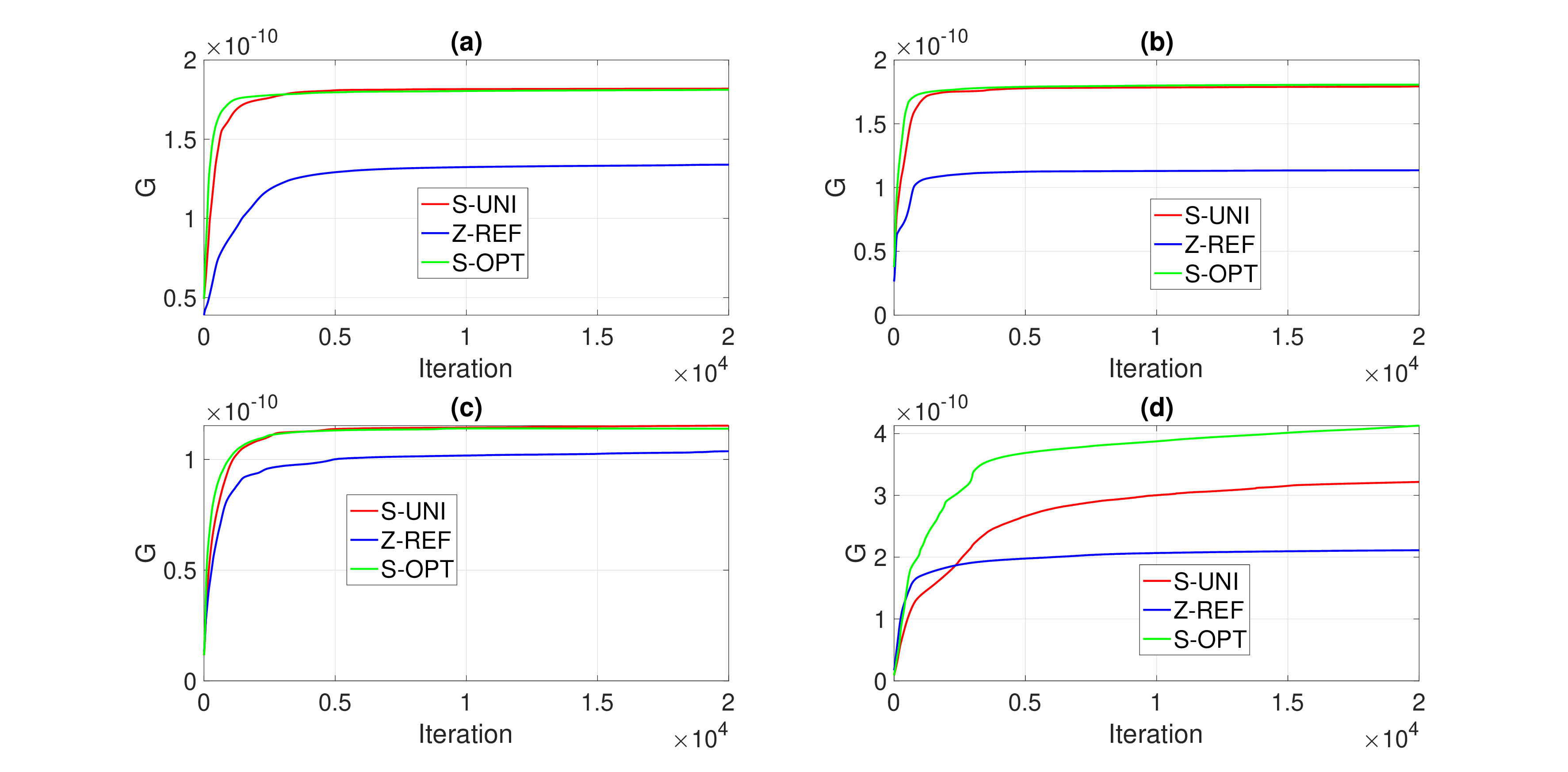}
\caption{Multipath channel with $d_y = \lambda/8$ \cite{DR1}. Setup: the receiver is in (a) $P_1$; (b) $P_2$; (c) $P_3$; (d) $P_4$.}
\label{fig2bis}
\vspace{-1.25cm}
\end{figure}
\begin{figure}[t!]
\centering
\includegraphics[width=0.9\columnwidth]{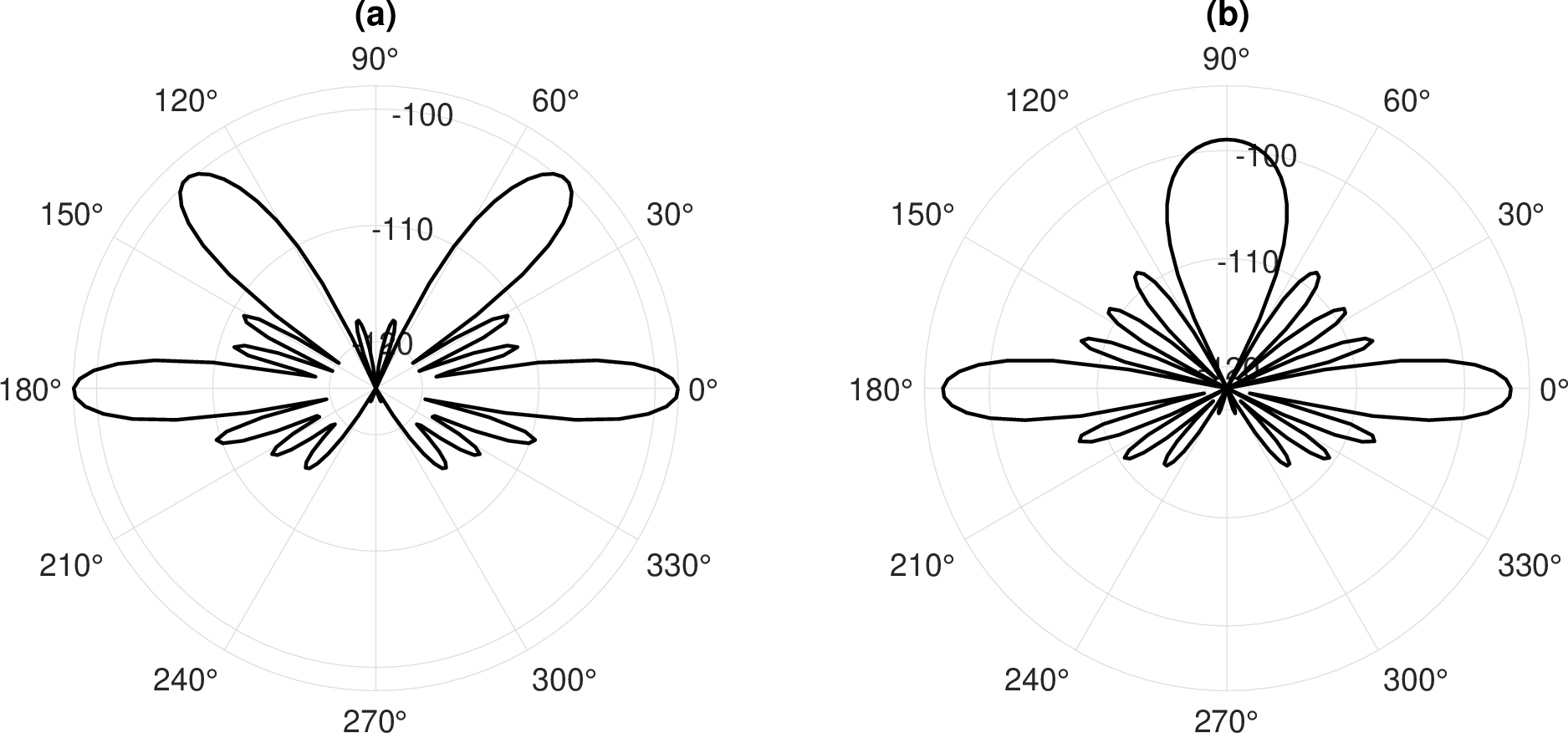}
\vspace{-0.1cm}
\caption{\textcolor{black}{Scattered field in $P_3$ (a) and $P_4$ (b) ($d_y = \lambda/8$) of the S-OPT algorithm in the presence of multipath}.}
\label{Diagram_all3}
\vspace{-0.5cm}
\end{figure}
\textcolor{black}{The multipath channel is modeled as described in Sec. III-D, according to \cite{SARIS}. Specifically, we assume the presence of one multipath cluster, which is constituted by 100 scattering objects that are modeled as short-circuited thin wire dipoles of length $\lambda/2$. The center of the cluster is located at $(2,2,0)$ in the middle of the considered area. The results illustrated in Fig. \ref{fig2bis} are similar to those in Fig. \ref{fig1}, but in this case we consider a fixed $d_y = \lambda/8$ and we show the convergence analysis for $P_1$-$P_4$. Comparing with the results in Fig. \ref{fig1}, we observe a slight increase in the received power at the UE due to the presence of the multipath cluster, which creates additional scattered signals between the transmitter and the receiver, the transmitter and the RIS, and the RIS and the receiver}. \textcolor{black}{In Fig. \ref{Diagram_all3}, in addition, we show the reradiation pattern of the RIS in the presence of multipath. The setting is the same as in Fig. \ref{Diagram_all}, but, for simplicity, only the unconstrained S-OPT algorithm is reported. We see that the presence of multipath slightly deforms the reradiating pattern of the RIS, when compared with that shown in Fig. \ref{Diagram_all}. This is more noticeable when the receiver is in P4. In both cases, a strong specular reflection due to the structural scattering is apparent in the presence of multipath as well, confirming the mathematical analysis in Sec. III-D. The results shown in Figs. \ref{fig2bis} and \ref{Diagram_all3} demonstrate that the conclusions drawn in free space channels remain valid in multipath channels when the transmitter-RIS and the RIS-receiver links are in line-of-sight and are hence dominant.}
\vspace{-0.25cm}
\subsection{\textcolor{black}{RIS Made of Rectangular Patch Antennas}}
\textcolor{black}{Finally, we analyze the accuracy of the multiport network theory model for RIS-aided communications and the effectiveness of the proposed optimization algorithms in steering the beam towards the direction of interest and simultaneously suppressing the structural scattering, by considering an RIS whose RIS elements are loaded rectangular patch antennas.}

\begin{figure}[t!]
	\centering
\includegraphics[width=0.70\columnwidth]{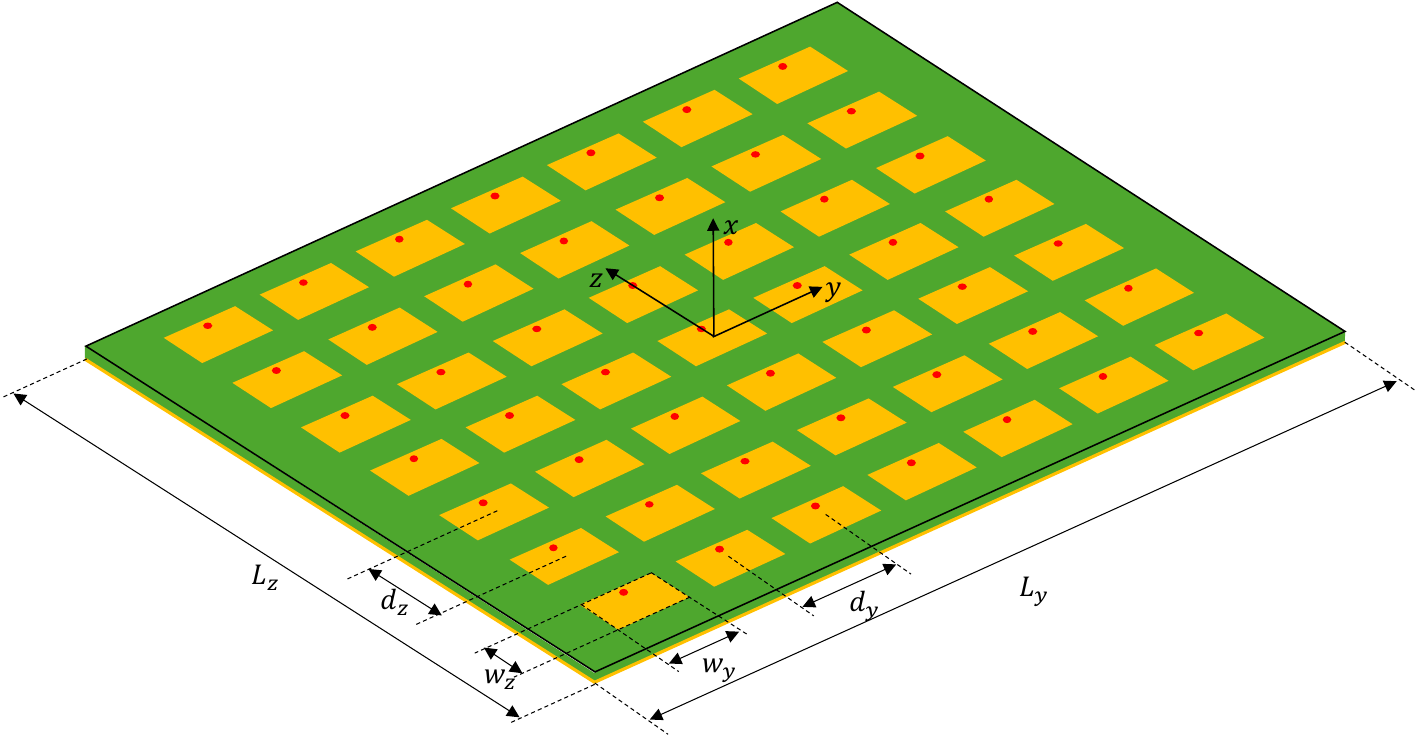}
	\caption{RIS made of $7 \times 7$ rectangular patch antennas.}
	\label{fig:7x7Ris}
	\vspace{-0.3cm}
\end{figure}

\textcolor{black}{The considered RIS is illustrated in Fig. \ref{fig:7x7Ris}. It operates at $f=28$ GHz and is constituted by $7 \times 7$ rectangular patch antennas. The design is based on a Rogers RO3003 grounded substrate with loss tangent equal to $\varepsilon_r=3$ and thickness equal to $t_s=0.51$ mm. The RIS elements are pin-fed resonant patch antennas whose dimensions are $w_y=0.36\lambda$ and $w_z=0.256\lambda$. The RIS elements are spaced by $d_y=d_z=0.49\lambda$, and the total dimensions of the RIS are $L_y=3.85\lambda$ and $L_z=3.58\lambda$, as illustrated in Fig. \ref{fig:7x7Ris}. Similar to the dipole-based RIS, we assume $r_0 = 0.2 \; \Omega$ and $\delta_0 = 0.05$.} 

\begin{figure}[t!]
	\centering
\includegraphics[width=0.6\columnwidth]{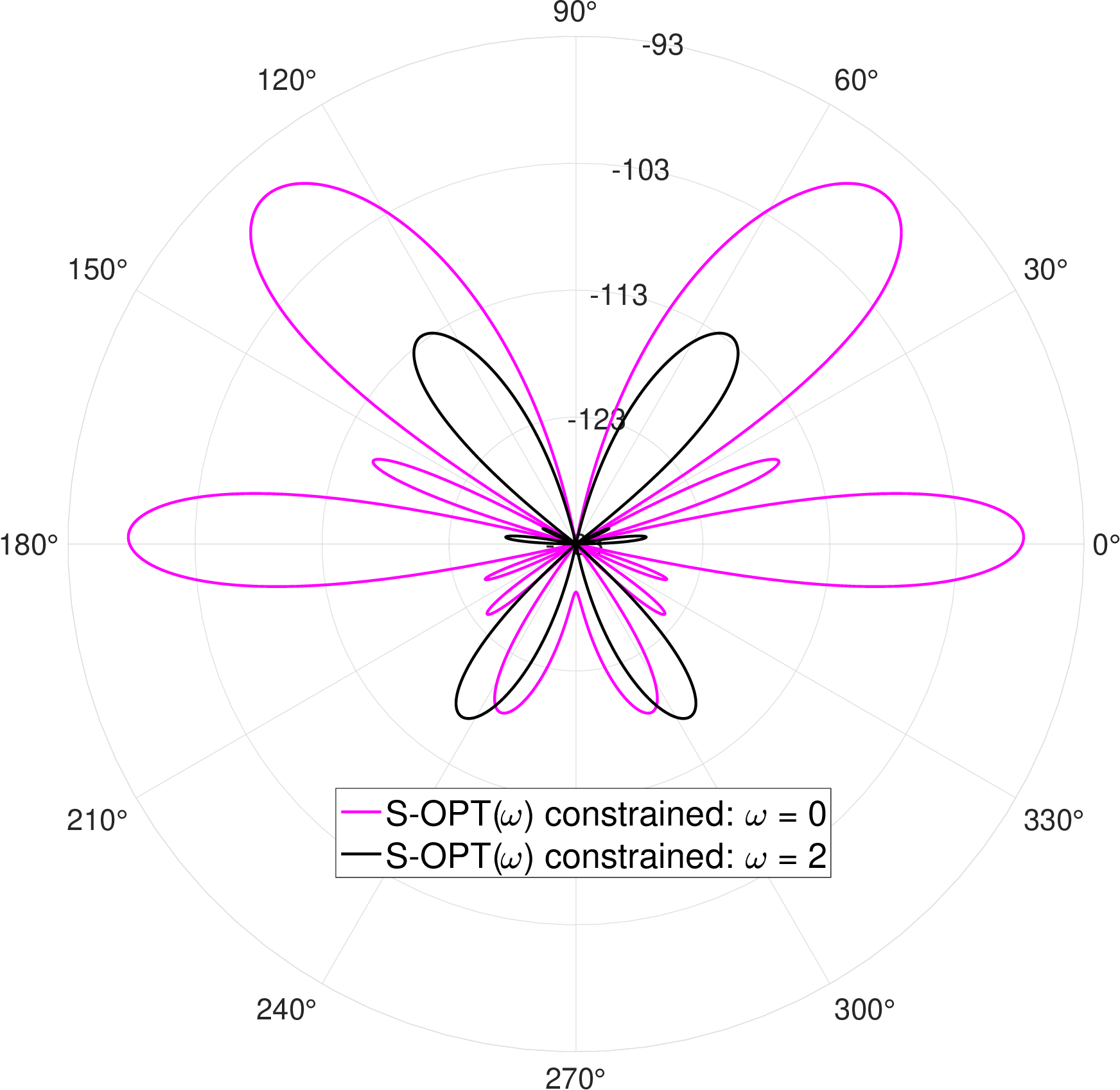}
	\vspace{-0.1cm}
	\caption{Scattered field in location $P_3$ obtained with S-OPT($\omega$) as a function of $\omega$. The RIS in Fig. \ref{fig:7x7Ris} is considered.}
\label{fig:ScatRisPatch}
	\vspace{-1.25cm}
\end{figure}

\textcolor{black}{In Fig. \ref{fig:ScatRisPatch}, we depict the reradiation (scattering) pattern of the RIS, when it is illuminated by a plane wave emitted by the transmitter (normal incidence), and it is designed to reflect the wave towards $P_3$ ($\approx 48.6^{\circ}$) shown in Fig. \ref{Scenario}. The $S$-matrix, $Z$-matrix, and the radiation pattern are obtained by utilizing the full-wave simulator. Similar to the dipole-based RIS, we observe that the considered multiport network model is accurate and the proposed optimization algorithms are effective, as the desired beam is well steered towards the desired direction of reflection while the structural scattering is significantly attenuated when $\omega=2$.}

\vspace{-0.25cm}
\section{Conclusion}
We have put forth a mutiport network model for RIS-aided channels, and have discussed its unique features with respect to conventional scattering models for RISs. \textcolor{black}{Moreover, we have investigated the role of the structural scattering in RIS-aided channels, and have validated its impact with the aid of full-wave simulations}. In addition, we have proposed new optimization algorithms based on the $S$-parameter multiport network model, and we have discussed its advantages with respect to its counterpart based on the $Z$-parameter representation. With the aid of full-wave simulations, finally, we have validated the effectiveness of the proposed algorithms for reducing the structural scattering at the design stage, while optimizing the scattered field towards the direction of interest.

\appendices

\vspace{-0.25cm}
\section*{Appendix A -- End-to-End RIS-Aided Channel}

\subsection{Proof of \eqref{Eq:HE2Eexact}}
From \eqref{Smodel}, we can write the following: \vspace{-0.1cm}
\begin{align}
	\mathbf{b}_T & = \mathbf{S}_{TT} \mathbf{a}_T+ \mathbf{S}_{TS} \mathbf{a}_S+\mathbf{S}_{TR} \mathbf{a}_R  \label{eqs1} \\
	\mathbf{b}_S & = \mathbf{S}_{ST} \mathbf{a}_T+ \mathbf{S}_{SS} \mathbf{a}_S+\mathbf{S}_{SR} \mathbf{a}_R  \label{eqs2} \\
	\mathbf{b}_R & = \mathbf{S}_{RT} \mathbf{a}_T+ \mathbf{S}_{RS} \mathbf{a}_S+\mathbf{S}_{RR} \mathbf{a}_R  \label{eqs3}  \vspace{-0.1cm}
\end{align}

Inserting ${{\bf{b}}_S} = {\bf{\Gamma }}_S^{ - 1}{{\bf{a}}_S}$ from \eqref{eq:atasar} into \eqref{eqs2}, we obtain  \vspace{-0.1cm}
\begin{align} \label{App:aS}
{{\bf{a}}_S} & = {\left( {{\bf{U}} - {{\bf{\Gamma }}_S}{{\bf{S}}_{SS}}} \right)^{ - 1}}{{\bf{\Gamma }}_S}\left( {{{\bf{S}}_{ST}}{{\bf{a}}_T} + {{\bf{S}}_{SR}}{{\bf{a}}_R}} \right)\\
& = {{\bf{\Gamma }}_S}{\left( {{\bf{U}} - {{\bf{\Gamma }}_S}{{\bf{S}}_{SS}}} \right)^{ - 1}}\left( {{{\bf{S}}_{ST}}{{\bf{a}}_T} + {{\bf{S}}_{SR}}{{\bf{a}}_R}} \right) \vspace{-0.1cm}
\end{align}

Substituting \eqref{App:aS} into \eqref{eqs1} and \eqref{eqs3}, we obtain the following reduced system of equations: \vspace{-0.1cm}
\begin{equation} \label{eq1bs}
	\mathbf{b}_T = \mathbf{\tilde{S}}_{TT}\mathbf{a}_T+\mathbf{\tilde{S}}_{TR}\mathbf{a}_R, \quad
	\mathbf{b}_R = \mathbf{\tilde{S}}_{RT}\mathbf{a}_T+\mathbf{\tilde{S}}_{RR}\mathbf{a}_R \vspace{-0.1cm}
\end{equation}
where (with $x,y \in \{T,R\}$) \vspace{-0.1cm}
\begin{equation}
	\mathbf{\tilde{S}}_{xy}=\mathbf{S}_{xy} + \mathbf{S}_{xS} (\mathbf{U}-\mathbf{S}_{SS}\mathbf{\Gamma}_{S})^{-1}\mathbf{\Gamma}_{S}\mathbf{S}_{Sy} \vspace{-0.1cm}
\end{equation}

Inserting ${{\bf{a}}_R} = {\bf{\Gamma }}_R{{\bf{b}}_R}$ from \eqref{eq:atasar} into \eqref{eq1bs}, we obtain  \vspace{-0.1cm}
\begin{equation}
{{\bf{b}}_R} = {\left( {{\bf{U}} - {{{\bf{\tilde S}}}_{RR}}{{\bf{\Gamma }}_R}} \right)^{ - 1}}{{{\bf{\tilde S}}}_{RT}}{{\bf{a}}_T}
	\label{eq:br} \vspace{-0.1cm}
\end{equation}

Inserting ${{\bf{b}}_R}$ in \eqref{eq:br} into \eqref{eq1bs} and using ${{\bf{b}}_T} = {\bf{\Gamma }}_T^{ - 1}\left( {{{\bf{a}}_T} - {{\bf{a}}_g}} \right)$ from \eqref{eq:atasar}, we obtain (${{{{\bf{\bar S}}}_{TT}}}$ is defined in \eqref{Eq:BarSTT}) \vspace{-0.1cm}
\begin{equation}
	{{\bf{a}}_T} = {\left( {{\bf{U}} - {{\bf{\Gamma }}_T}{{{\bf{\bar S}}}_{TT}}} \right)^{ - 1}}{{\bf{a}}_g}  \label{eq:btat} \vspace{-0.1cm}
\end{equation}

Finally, by solving the systems of equations in \eqref{eq:br} and \eqref{eq:btat}, we obtain the end-to-end transfer function \vspace{-0.1cm}
\begin{equation}
{{\bf{b}}_R} = {\left( {{\bf{U}} - {{{\bf{\tilde S}}}_{RR}}{{\bf{\Gamma }}_R}} \right)^{ - 1}}{{{\bf{\tilde S}}}_{RT}}{\left( {{\bf{U}} - {{\bf{\Gamma }}_T}{{{\bf{\bar S}}}_{TT}}} \right)^{ - 1}}{{\bf{a}}_g} \vspace{-0.1cm}
\end{equation}

\vspace{-0.25cm}
\subsection{Proof of \eqref{Eq:HZexact}}
The proof of \eqref{Eq:HZexact} follows along the same lines as the proof of \eqref{Eq:HE2Eexact}. Specifically, the current $\mathbf{I}_S$ is first expressed as a function of $\mathbf{I}_T$ and $\mathbf{I}_R$, by utilizing the second equation in \eqref{eq:Zmat} and ${{\bf{V}}_S} =  - {{\bf{Z}}_S}{{\bf{I}}_S}$ from \eqref{Eq:VoltageCurrent}. Then, the following reduced system of equations is obtained:  \vspace{-0.1cm}
\begin{equation} \label{eq1bsZ}
	\mathbf{V}_T = \mathbf{\tilde{Z}}_{TT}\mathbf{I}_T+\mathbf{\tilde{Z}}_{TR}\mathbf{I}_R , \quad 
	\mathbf{V}_R = \mathbf{\tilde{Z}}_{RT}\mathbf{I}_T+\mathbf{\tilde{Z}}_{RR}\mathbf{I}_R  \vspace{-0.1cm}
\end{equation}
where $\mathbf{\tilde{Z}}_{xy}$ is defined in \eqref{Eq:ZtildeXY}.

The obtained system of equations is solved by using Ohm's laws ${{\bf{I}}_R} =  - {\bf{Z}}_R^{ - 1}{{\bf{V}}_R}$ and ${{\bf{I}}_T} =  - {\bf{Z}}_g^{ - 1}\left( {{{\bf{V}}_T} - {{\bf{V}}_g}} \right)$ from \eqref{Eq:VoltageCurrent}.

\vspace{-0.25cm}
\section*{Appendix B -- Relation Between ${\bf{{H}}}_{{\rm{e2e}}}^{\left( S \right)}$ and ${\bf{{H}}}_{{\rm{e2e}}}^{\left( Z \right)}$}

\setcounter{subsection}{0}
\vspace{-0.1cm}
\subsection{Proof of \eqref{Eq:VR}}
Consider ${{\bf{V}}_R} = {\bf{{H}}}_{{\rm{e2e}}}^{\left( Z \right)}{{\bf{V}}_g}$. We obtain the following: \vspace{-0.1cm}
\begin{align} \label{Eq:VRApp}
\hspace{0.5cm}{{\bf{V}}_R} &= {\bf{H}}_{{\rm{e2e}}}^{\left( Z \right)}{{\bf{V}}_g} \mathop  = \limits^{\left( a \right)} {\bf{H}}_{{\rm{e2e}}}^{\left( Z \right)}\left( {{{\bf{V}}_T} + {{\bf{Z}}_g}{{\bf{I}}_T}} \right) \\
& \mathop  = \limits^{\left( b \right)} {\sqrt {{Z_0}} }{\bf{H}}_{{\rm{e2e}}}^{\left( Z \right)}\left[ {\left( {{{\bf{a}}_T} + {{\bf{b}}_T}} \right) + Z_0^{ - 1}{{\bf{Z}}_g}\left( {{{\bf{a}}_T} - {{\bf{b}}_T}} \right)} \right] \nonumber \\
&\mathop  = \limits^{\left( c \right)} {\sqrt {{Z_0}} }{\bf{H}}_{{\rm{e2e}}}^{\left( Z \right)}  {\left[ {\left( {{{\bf{a}}_g} + {{\bf{\Gamma }}_T}{{\bf{b}}_T}} \right) + {{\bf{b}}_T}} \right]}  \nonumber\\
& \hspace{0.35cm} + {\sqrt {{Z_0}} } Z_0^{ - 1} {\bf{H}}_{{\rm{e2e}}}^{\left( Z \right)} {{\bf{Z}}_g} \left[ {\left( {{{\bf{a}}_g} + {{\bf{\Gamma }}_T}{{\bf{b}}_T}} \right) - {{\bf{b}}_T}} \right] \nonumber\\ 
&\mathop  = \limits^{\left( d \right)} {\sqrt {{Z_0}} }Z_0^{ - 1}{\bf{H}}_{{\rm{e2e}}}^{\left( Z \right)}\left( {{{\bf{Z}}_g} + {Z_0}{\bf{U}}} \right){{\bf{a}}_g}  \nonumber \\  &\mathop  = \limits^{\left( e \right)} 2{\sqrt {{Z_0}} }{\bf{H}}_{{\rm{e2e}}}^{\left( Z \right)}{\left( {{\bf{U}} - {{\bf{\Gamma }}_T}} \right)^{ - 1}}{{\bf{a}}_g} \nonumber
\end{align}
where $(a)$ follows from \eqref{Eq:VoltageCurrent}, $(b)$ follows from \eqref{Eq:ab_VI}, $(c)$ follows from \eqref{eq:atasar}, $(d)$ follows from ${{\bf{\Gamma }}_T} = {\left( {{{\bf{Z}}_g} + {Z_0}{\bf{U}}} \right)^{ - 1}}\left( {{{\bf{Z}}_g} - {Z_0}{\bf{U}}} \right)$, and $(e)$ follows from ${\bf{U}} - {{\bf{\Gamma }}_T} = 2{Z_0}{\left( {{{\bf{Z}}_g} + {Z_0}{\bf{U}}} \right)^{ - 1}}$.

Consider ${{\bf{V}}_R} = {\sqrt {{Z_0}} } \left({{\bf{a}}_R} + {{\bf{b}}_R}\right)$ from \eqref{Eq:ab_VI}. We obtain \vspace{-0.1cm}
\begin{equation} \label{Eq:TEMP1}
{{\bf{V}}_R} \mathop  = \limits^{\left( a \right)} {\sqrt {{Z_0}} } \left( {{\bf{U}} + {{\bf{\Gamma }}_R}} \right){{\bf{b}}_R}\mathop  = \limits^{\left( b \right)}{\sqrt {{Z_0}} } \left( {{\bf{U}} + {{\bf{\Gamma }}_R}} \right){\bf{H}}_{{\rm{e2e}}}^{\left( S \right)}{{\bf{a}}_g} \vspace{-0.1cm}
\end{equation}
where $(a)$ follows from \eqref{eq:atasar} and $(b)$ from ${{\bf{b}}_R} = {\bf{H}}_{{\rm{e2e}}}^{\left( S \right)}{{\bf{a}}_g}$ in \eqref{Eq:HE2Eexact}.

\vspace{-0.25cm}
\subsection{Proof of \eqref{Eq:AgtoVg}}
This follows immediately from the first and the last equality in \eqref{Eq:VRApp}, which gives ${\bf{H}}_{{\rm{e2e}}}^{\left( Z \right)}{{\bf{V}}_g} = 2{\sqrt {{Z_0}} }{\bf{H}}_{{\rm{e2e}}}^{\left( Z \right)}{\left( {{\bf{U}} - {{\bf{\Gamma }}_T}} \right)^{ - 1}}{{\bf{a}}_g}$.

\vspace{-0.25cm}
\subsection{Differences Between \eqref{Eq:HE2Ematched} and \eqref{Eq:HZsimpler}}
For ease of understanding, we consider the SISO case study and assume ${{{Z}}_g} = {Z_0}$ and ${{{Z}}_R} = {Z_0}$. Then, the reduced system of equations in \eqref{eq1bs} and \eqref{eq1bsZ} correspond to a two-port network model. From \eqref{Eq:HZexact}, we obtain the following ${{{H}}}_{{\rm{e2e}}}^{\left( Z \right)}$: \vspace{-0.1cm}
\begin{align}
{{{H}}}_{{\rm{e2e}}}^{\left( Z \right)} & = \frac{{{Z_0}{{\tilde Z}_{RT}}}}{{\left( {{Z_0} + {{\tilde Z}_{RR}}} \right)\left( {{Z_0} + {{\tilde Z}_{TT}}} \right) - {{\tilde Z}_{RT}}{{\tilde Z}_{TR}}}}\\
& \mathop  \approx \limits^{\left( a \right)} \frac{{{Z_0}{{\tilde Z}_{RT}}}}{{\left( {{Z_0} + {{\tilde Z}_{RR}}} \right)\left( {{Z_0} + {{\tilde Z}_{TT}}} \right)}} \vspace{-0.1cm}
\end{align}
where $(a)$ corresponds to the approximation in \eqref{Eq:HZsimpler}.

Then, consider the relation between the $S$-parameter and $Z$-parameter representations in \eqref{Eq:ZtoS}. For a two-port network, we obtain the following end-to-end RIS aided channel: \vspace{-0.1cm}
\begin{equation}
H_{{\rm{e2e}}}^{\left( S \right)} = {{\tilde S}_{RT}} = \frac{{2{Z_0}{{\tilde Z}_{RT}}}}{{\left( {{Z_0} + {{\tilde Z}_{RR}}} \right)\left( {{Z_0} + {{\tilde Z}_{TT}}} \right) - {{\tilde Z}_{RT}}{{\tilde Z}_{TR}}}} \vspace{-0.1cm}
\end{equation}

According to \eqref{Eq:HStoHZ}, we see that $H_{{\rm{e2e}}}^{\left( S \right)} = 2H_{{\rm{e2e}}}^{\left( Z \right)}$, since ${{{{\Gamma }}_T}}={{{{\Gamma }}_R}} = 0$ in the considered case study. Also, the cross-product ${{{\tilde Z}_{RT}}{{\tilde Z}_{TR}}}$ is present not only in the exact expression of $H_{{\rm{e2e}}}^{\left( Z \right)}$, but in the exact expression of $H_{{\rm{e2e}}}^{\left( S \right)}$ as well, even though it is not explicitly observable in the $S$-parameter representation.

\vspace{-0.25cm}
\section*{Appendix C -- Modeling Blocking Objects:\\ The ``No Shield'' Method}

In this section, we discuss and prove the electromagnetic consistency of the ``No Shield'' method, which is based on the superposition principle and has been applied to validate the multiport network model with the aid of full-wave electromagnetic simulations. To ease the mathematical derivation, we consider the case study of two thin wire antennas, which are directed towards the $z$-axis and whose midpoints (feed points) lie on the $z = 0$ plane.

For the two considered coupled antennas in free space, Carter's circuit relations can be written as follows \cite{Carter1932}:
	\begin{align}
		V_1 & = Z_{11}I_1+Z_{12}I_2 \label{eq:1a}\\
		V_2 & = Z_{21}I_1+Z_{22}I_2  \label{eq:2a}
	\end{align}
where $Z_{12}=Z_{21}$ by virtue of reciprocity.
	
First, we consider that antenna 1 is transmitting and antenna 2 is receiving. If antenna 2 is open-circuited at the reference plane and the tangential electric field generated by antenna 1 along the surface of antenna $2$ is ${E}_{21}(z)$, then, according to Lorentz's reciprocity theorem \cite[ch. 3]{harrington1961time}, \cite[ch. 5]{collin1985}, \cite[ch. 7]{balanis2012advanced}, the mutual impedance between the two antennas is
\begin{equation}
		Z_{21}=-\frac{1}{I_1(0)I_2(0)}\int\nolimits_{-h_2/2}^{+h_2/2}{E}_{21}(z) I_2(z) dz
		\label{eq:z21}
\end{equation}
where $I_1(0)$ is the current at the port of antenna $1$ when antenna 2 is open-circuited, $I_2(0)$ is the current at the port of antenna 2 when antenna 1 is open-circuited, which generates  the current $I_2(z)$ along the surface of antenna 2, $h_2$ is the length of antenna 2, and $E_{21}(z)$ is the tangential electric field generated by antenna 1 along the surface of antenna 2 (when the presence of the conductor of antenna 2 is ignored, as required by the reciprocity theorem). Equation \eqref{eq:z21} can be obtained by applying the well-known Carter's induced electromagnetic field method as well \cite{Carter1932}.
	
Let us now consider that the two thin wire antennas radiate in presence of a perfectly conducting scattering object with surface $S_o$. Let us denote the electric and magnetic vector fields scattered by the object as $\underline{E}^s$ and $\underline{H}^s$, respectively. According to the field equivalence principle \cite[par. 7.10]{balanis2012advanced}, the scattered electromagnetic field outside $S_o$ can be thought of as the electromagnetic field produced by equivalent electric surface vector currents $\underline{J}_{es}^o$ along the surface $S_o$ of the scattering object when it radiates in free space, i.e., in the absence of the two thin wire antennas.

In the presence of the considered scattering object, let us consider that antenna 1 is transmitting and antenna 2 is receiving. The total tangential electric field ${E}_{21}$ that is generated by antenna 1, when radiating in presence of the scattering object and being evaluated along the surface of antenna 2, can be expressed as the sum of the tangential electric field ${E}_{21}^i$ produced by antenna 1 in the absence of the scattering object and the scattered tangential electric field ${E}_{21}^s$ produced by the equivalent electric surface vector currents $\underline{J}_{es}^o$ in the absence of antenna 1 and antenna 2, as follows  \cite[par. 3.7 eq. 3-24]{collin1969}:
\begin{equation}
	{E}_{21}(z)={E}_{21}^i(z)+{E}_{21}^s(z) \label{eq:E21}
\end{equation}

Substituting \eqref{eq:E21} in \eqref{eq:z21}, the following expression for the mutual impedance $Z_{21}$ is obtained:
\begin{align}
	Z_{21}&=-\frac{1}{I_1(0)I_2(0)}\int\nolimits_{-h_2/2}^{+h_2/2}{E}_{21}^i(z) I_2(z)dz \\
	& -\frac{1}{I_1(0)I_2(0)}\int\nolimits_{-h_2/2}^{+h_2/2}{E}_{21}^s(z) I_2(z)dz \\
	&= Z_{21}^i + Z_{21}^s
	\label{eq:z21is}
\end{align}

From \eqref{eq:z21is}, we obtain
\begin{equation}
	Z_{21}^s = Z_{21} - Z_{21}^i
	\label{eq:z21s}
\end{equation}

Since $Z_{21}^i$ in \eqref{eq:z21s} corresponds, by definition, to the contribution of the electric field generated by the transmitting antenna 1 when evaluated along the surface of the receiving antenna 2 in the absence of the scattering object, the obtained expression in \eqref{eq:z21s} states that the contribution from the scattering object can be obtained by subtracting the contribution of the whole system (transmitting antenna, receiving antenna, and scattering object) from the contribution generated by the transmitting antenna in the presence of the receiving antenna but in the absence of the scattering object. 

This approach can be generalized to systems with multiple scattering objects, since the corresponding term $Z_{21}^i$ is computed in the absence of all the scattering objects. This validates the electromagnetic foundation of the (computationally efficient) ``No Shield'' method utilized in this paper.

\vspace{-0.25cm}
\bibliographystyle{IEEEtran}
\bibliography{referenceV2,AntennasWithScatterer}

\end{document}